\documentclass[draftcls, 11pt,onecolumn]{IEEEtran}

\IEEEoverridecommandlockouts
\usepackage{graphicx}
\usepackage{cite, amsmath, amsfonts, amssymb, psfrag,times,dsfont}
\usepackage{epsf}
\usepackage{epsfig}
\usepackage{amsfonts}
\usepackage{amstext}
\usepackage{latexsym}
\usepackage{color}
\usepackage{ifthen,bbold}
\usepackage{multirow}
\usepackage{subfigure}

\newcommand{\ZZ}{\mathsf{ZZ}}
\newcommand{\ZS}{\mathsf{ZS}}
\newcommand{\XX}{\mathsf{XX}}
\newcommand{\DZS}{\mathsf{DZS}}
\newcommand{\GZS}{\mathsf{GZS}}
\newcommand{\DZZ}{\mathsf{DZZ}}
\newcommand{\GZZ}{\mathsf{GZZ}}
\newcommand{\Z}{\mathsf{Z}}
\renewcommand{\S}{\mathsf{S}}

\newcommand{\rank}{\mathrm{rank}}

\newcommand{\R}{\mathbb{R}}

\newcommand{\e}{\varepsilon}
\newcommand{\E}{\mathds{E}}

\newcommand{\x}{\mathbf{x}}
\newcommand{\y}{\mathbf{y}}
\newcommand{\z}{\mathbf{z}}
\newcommand{\J}{\mathbf{J}}

\newcommand{\Uoq}{U_1}
\newcommand{\Uoo}{U_1^{(1)}}

\newcommand{\Uto}{U_2^{(1)}}
\newcommand{\Utt}{U_2^{(2)}}
\newcommand{\Utr}{U_2^{(3)}}

\newcommand{\hUoo}{\hat{U}_1^{(1)}}

\newcommand{\hUto}{\hat{U}_2^{(1)}}
\newcommand{\hUtt}{\hat{U}_2^{(2)}}
\newcommand{\hUtr}{\hat{U}_2^{(3)}}

\newcommand{\Voo}{V_1^{(1)}}
\newcommand{\Vot}{V_1^{(2)}}
\newcommand{\Vto}{V_2^{(1)}}
\newcommand{\Vtt}{V_2^{(2)}}
\newcommand{\Vtr}{V_2^{(3)}}
\newcommand{\Vtf}{V_2^{(4)}}
\newcommand{\Vtv}{V_2^{(5)}}

\newcommand{\hVoo}{\hat{V}_1^{(1)}}
\newcommand{\hVot}{\hat{V}_1^{(2)}}
\newcommand{\hVto}{\hat{V}_2^{(1)}}
\newcommand{\hVtt}{\hat{V}_2^{(2)}}
\newcommand{\hVtr}{\hat{V}_2^{(3)}}
\newcommand{\hVtf}{\hat{V}_2^{(4)}}
\newcommand{\hVtv}{\hat{V}_2^{(5)}}

\newcommand{\Tz}{\Upsilon_{0}}
\newcommand{\To}{\Upsilon_{1}}
\newcommand{\Tt}{\Upsilon_{2}}

\newcommand{\Too}{\Upsilon_{1,1}}

\newcommand{\Tto}{\Upsilon_{2,1}}
\newcommand{\Ttt}{\Upsilon_{2,2}}
\newcommand{\Ttr}{\Upsilon_{2,3}}

\newcommand{\Qz}{\Theta_{0}}
\newcommand{\Qo}{\Theta_{1}}
\newcommand{\Qt}{\Theta_{2}}

\newcommand{\Qoo}{\Theta_{1,1}}
\newcommand{\Qot}{\Theta_{1,2}}
\newcommand{\Qto}{\Theta_{2,1}}
\newcommand{\Qtt}{\Theta_{2,2}}
\newcommand{\Qtr}{\Theta_{2,3}}
\newcommand{\Qtf}{\Theta_{2,4}}
\newcommand{\Qtv}{\Theta_{2,5}}

\newcommand{\aoq}{\alpha_{1}}

\newcommand{\ato}{\alpha_{2,1}}
\newcommand{\att}{\alpha_{2,2}}
\newcommand{\atr}{\alpha_{2,3}}
\newcommand{\atf}{\alpha_{2,4}}

\newcommand{\az}{\alpha_{0}}

\newcommand{ \boo }{\beta_{1,1}}
\renewcommand{\bot}{\beta_{1,2}}
\newcommand{ \bz }{\beta_{0}}

\newcommand{\+}[1]{\left(#1\right)^+}
\newcommand{\CG}[2]{\frac{1}{2} \log \left(\frac{#1}{#2}\right)}

\newcommand{\CGN}[1]{\frac{1}{2} \log \left(#1\right)}

\renewcommand{\R}[2]{\mathcal{R}^{#1}_{\textrm{#2}}}

\renewcommand{\l}{^{\ell}}

\newcommand{\SNR}{\mathsf{SNR}}

\newtheorem{lemma}{\textbf{Lemma}}
\newtheorem{theorem}{{Theorem}}

\newtheorem{definition}{{Definition}}
\newtheorem{example}{{Example}}

\newcommand{\N}{\mathcal{N}}

\newcounter{temp}

\title{Approximate Capacity of Gaussian Interference-Relay Networks with Weak Cross Links}

\def\thesubsectiondis{\Roman{section}.\arabic{subsection}}
\def\thesubsection{\thesection.\arabic{subsection}}
\author{S. Mohajer \hspace{0.5in} S N. Diggavi \hspace{0.5in} C. Fragouli \hspace{0.5in} D N C. Tse }

\begin{document}
\maketitle

\begin{abstract}

  In this paper we study a Gaussian relay-interference network, in
  which relay (helper) nodes are to facilitate competing information
  flows over a wireless network. We focus on a two-stage
  relay-interference network where there are weak cross-links, causing
  the networks to behave like a chain of $\Z$ Gaussian channels.  For
  these Gaussian $\ZZ$ and $\ZS$ networks, we establish an approximate
  characterization of the rate region. The outer bounds to the
  capacity region are established using genie-aided techniques that
  yield bounds sharper than the traditional cut-set outer bounds. For
  the inner bound of the $\ZZ$ network, we propose a new interference
  management scheme, termed interference neutralization, which is
  implemented using structured lattice codes.  This technique allows
  for over-the-air interference removal, without the transmitters
  having complete access the interfering signals. For both the $\ZZ$
  and $\ZS$ networks, we establish a new network decomposition
  technique that (approximately) achieves the capacity region. We use
  insights gained from an exact characterization of the corresponding
  linear deterministic version of the problems, in order to establish
  the approximate characterization for Gaussian networks.

\end{abstract}
\section{Introduction}
\label{sec:intro}

The multi-commodity flow problem, where multiple independent unicast 
sessions need to share network resources, can be solved efficiently 
over graphs using linear programming techniques \cite{Schrijver}. 
This is not the case for wireless networks, where the broadcast and
superposition nature
of the wireless medium introduces complex signal interactions between
the competing flows.  The simplest example is the one-hop interference
channel \cite{HK:81}, where two transmitters with independent messages
are attempting to communicate with their respective receivers over the
wireless transmission medium.  Even for this simple one-hop network,
the information-theoretic characterization has been  open for several
decades. To study more general networks, there is a clear need to
understand and develop sophisticated interference management
techniques.

The capacity of the wireless Gaussian interference channel has been
(approximately) characterized, within one bit (see \cite{ETW:08} and
the references therein).  Building on this progress, a
natural next step is to study the approximate capacity region of
small-scale {\em interference-relay networks}, where there are
potentially multiple hops from the sources to destinations through
cooperating relays. Studying even simple two-hop topologies could help
develop techniques and build insight that would enable a (perhaps
approximate) characterization of capacity for more general
networks. We are interested in our work in a {\em universal} type of
approximation, in that it should characterize the capacity to within a
constant number of bits, independently of the signal-to-noise ratio
and the channel parameter values.

\begin{figure}[t!]
\begin{center}
 	\psfrag{s1}[Bc][Bc]{$S_1$}
	\psfrag{s2}[Bc][Bc]{$S_2$}
	\psfrag{r1}[Bc][Bc]{$A$}
	\psfrag{r2}[Bc][Bc]{$B$}
	\psfrag{d1}[Bc][Bc]{$D_1$}
	\psfrag{d2}[ABC][Bc]{$D_2$}
	\psfrag{h111}[Bc][Bc]{$\sqrt{g_{11}}$}
	\psfrag{h211}[Bc][Bc]{$\sqrt{h_{11}}$}
	\psfrag{h122}[Bc][Bc]{$\sqrt{g_{22}}$}
	\psfrag{h222}[Bc][Bc]{$\sqrt{h_{22}}$}
	\psfrag{h112}[Bc][Bc]{$\sqrt{g_{12}}$}
	\psfrag{h212}[Bc][Bc]{$\sqrt{h_{12}}$}
	\psfrag{h121}[Bc][Bc]{$\sqrt{g_{21}}$}
	\psfrag{h221}[Bc][Bc]{$\sqrt{h_{21}}$}
 	\psfrag{x1}[Bc][Bc]{$x_1$}
	\psfrag{x2}[Bc][Bc]{$x_2$}
	\psfrag{y'1}[Bc][Bc]{$y'_1$}
	\psfrag{y'2}[Bc][Bc]{$y'_2$}
	\psfrag{x'1}[Bc][Bc]{$x'_1$}
	\psfrag{x'2}[Bc][Bc]{$x'_2$}
	\psfrag{y1}[Bc][Bc]{$y_1$}
	\psfrag{y2}[Bc][Bc]{$y_2$}
	\psfrag{z11}[Bc][Bc]{$z'_1$}
	\psfrag{z12}[Bc][Bc]{$z'_2$}
	\psfrag{z21}[Bc][Bc]{$z_1$}
	\psfrag{z22}[Bc][Bc]{$z_2$}
	\includegraphics[height=5cm]{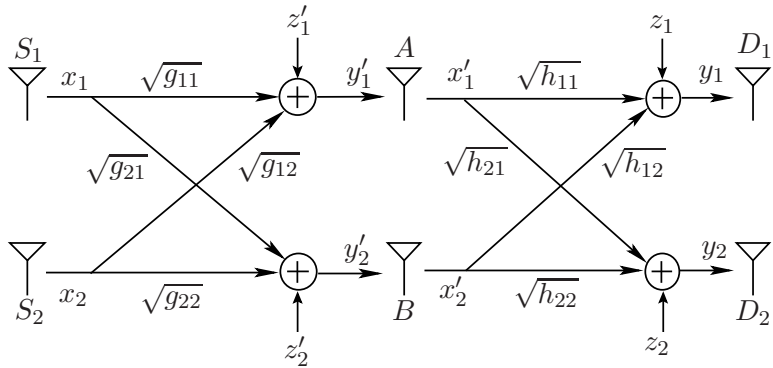}
\end{center}
\vspace{-7mm}
\caption{Two-stage relay-interference network.}
\label{model:GXX}
\vspace{-7mm}
\end{figure}

The focus of this paper is to study the two-stage relay-interference
network illustrated in Figure~\ref{model:GXX}. In particular, we give
an approximate characterization of the capacity region for special
cases of these networks when some of the cross-links are weak. These
are illustrated in Figure~\ref{fig:GZS} and Figure~\ref{fig:GZZ},
which we refer to as the $\ZS$ and $\ZZ$ Gaussian models. We first
study a {\em deterministic} version of these problems by using the
linear deterministic model introduced in \cite{ADT07a}. An exact
capacity region characterization in the deterministic case is then
translated into a universally approximate characterization for the
(noisy) Gaussian network. 
In particular, for $\ZS$ and $\ZZ$
networks we have a capacity region characterization within $2$ bits
(or less), independent of the operating signal-to-noise ratio and the
channel parameters.

In studying these special networks, we discover that many
sophisticated techniques are required to (approximately) characterize
the network capacity region. The main new ingredients that enable this
characterization are as follows: {\sf (i)} a new interference
management technique we term {\em interference neutralization}, in
which interference is canceled over the air, without the relays
necessarily decoding the transmitted messages\footnote{A noise nulling
  technique is proposed in \cite{GJ09} to mitigate correlated noise
  in an amplify-forward relaying strategy for a single unicast
  ``diamond'' parallel relay network.  However, the difference in our
  technique is that we use the structure of the codebooks (without
  necessarily decoding information) to neutralize interference, and
  not noise statistics. Moreover the multiple-unicast nature of the
  problem necessitates strategic partitioning and rate-splitting of
  different components of the messages.}; {\sf (ii)} a {\em
  structured lattice code} that enables interference neutralization
over Gaussian networks; {\sf (iii)} A {\em network decomposition}
technique which enables appropriate rate-splitting of the message and
power allocation for the different message components; {\sf (iv)}
genie-aided outer bounding techniques that enable bounds that are
tighter than the information-theoretic cut-set outer bounds.

A way to interpret the achievability results for the $\ZS$ networks is
that the relays perform a partial-decoding of strategically split
messages from the sources, and then cooperate to deliver the required
messages to the destination, again through strategically splitting the
messages. The power allocated to each of the sub-messages is
determined using the insight derived from the deterministic model,
that messages that are not intended be decoded arrive at the
noise-level. The achievability for the $\ZZ$ network is slightly more
sophisticated in that one of the relays is required to only decode a
function of the sub-messages. The function is chosen such that its
signal in combination with the transmission of the other relay causes
the unwanted interference to be cancelled (neutralized) at the
destination. This interference neutralization is enabled in the
Gaussian channel using the group property of a structured lattice
code.

Work in the literature over the past decade has examined scaling laws
for multiple independent flows over wireless networks, see for example
\cite{GK00,OLT07,Massimo09}. The goal there is to characterize the
order of the wireless network capacity as the network size grows.  In
contrast, in our work, instead of seeking order arguments and scaling
laws, we try to characterize the capacity (perhaps within a universal
constant of a few bits) for specific topologies. The interference
channel is a special case of such networks, where there is only
one-hop communication between the sources and destinations. There has
been a surge of recent work on this topic including cooperating
destinations \cite{PV09} and use of feedback in
inducing cooperation at the transmitters \cite{SuhTse09}. The
deterministic approach developed in \cite{ADT07a} has been
successfully applied to the interference channel in
\cite{BT08}. The fundamental role of interference alignment in
$K$-user interference channel (still a one-hop network) has been
demonstrated in \cite{CJ08,MMK08}. 

The paper is organized as follows. Section~\ref{sec:model} introduces
our notation and the basic network models we study.
Section~\ref{sec:examples} illustrates the transmission techniques
used in this paper through simple deterministic examples.  The main
results are given in Section~\ref{sec:MainRes}, along with a proof
outline for the Gaussian networks in Section \ref{sec:pr-outline}. The
achievability and converse for the deterministic $\ZS$ network is
given in Section~\ref{sec:dzs}, and many of these ideas are translated
into the precise proof for the corresponding Gaussian $\ZS$ network in
Appendix~\ref{sec:gzs}.  Section~\ref{sec:dzz} follows a similar
program for the $\ZZ$ network, by first identifying the capacity
region for the deterministic version.  This allows illustration of
ideas such as interference neutralization, as well as genie-aided
outer bounding techniques. The precise translation of these results
into Gaussian $\ZZ$ networks is given in Appendix ~\ref{sec:gzz}.
Section \ref{sec:disc} concludes the paper with a short discussion. 


\section{Problem Statement}
\label{sec:model}

A well accepted model for wireless communication is a linear Gaussian
model. In this, 
the received signal $y_i(t)$ at time $t$, is related to the
transmitted signals $\{x_j[t]\}$ as
\begin{equation}
\label{eq:GaussModel}
y_i [t] = \sum_{j} h_{ij} x_j[t] + z_i[t],
\end{equation}
where $z_i(t)$ is i.i.d. (unit-variance) Gaussian noise, and $h_{ij}$
represents the fading channel from transmitter $i$ to receiver $j$.

\subsection{The Deterministic Model}
\label{subsec:det-model}

In \cite{ADT07a}, a deterministic model was proposed, 
to capture the essence
of wireless interaction described in \eqref{eq:GaussModel}. The advantage
of the deterministic model is its simplicity, which allows exact characterizations; its purpose is to build insights for the
 noisy wireless network in \eqref{eq:GaussModel}. The
deterministic model of \cite{ADT07a} simplifies the wireless
interaction model by eliminating the noise and discretizing the
channel gains through a binary expansion of $q$ bits. Therefore, the
received signal $Y_i$, which is a binary vector of size $q$, is modeled
as
\begin{equation}
\label{eq:DetModel}
Y_i [t] = \sum_{j} N_{ij} X_j[t],
\end{equation}
where $N_{ij}$ is a $q\times q$ binary matrix representing the
(discretized) channel transformation between nodes $j$ and $i$ and
$X_j$ is a $q \times 1 $ vector that contains the (discretized) transmitted signal. We will  drop the time index $t$ when it does not play a role for simplicity.
All operations in 
(\ref{eq:DetModel}) are done over the binary field, $\mathds{F}_2$.
We use the terminology {\em deterministic wireless network} when the
signal interaction model is governed by (\ref{eq:DetModel}).  The
model in \eqref{eq:DetModel} is an approximate representation of a
Gaussian fading channel, which attempts to capture the attenuation
effect of the signal caused by the channel gain. This can be
interpreted as the number of significant bits of a binary
representation of the input, $x_j$, that is above the noise level.
More precisely, typically the model in \eqref{eq:DetModel} assigns
$N_{ij}=\J^{q-n_{ij}}$, where $\J$ is a shift matrix, i.e.,
\begin{eqnarray}
\footnotesize{\J=\left(
\begin{array}{ccccc}
0 & 0 & 0 & \cdots & 0\\
1 & 0 & 0 & \cdots & 0\\
0 & 1 & 0 & \cdots & 0\\
\vdots & \ddots & \ddots & \ddots & \ddots\\
0 & \cdots & 0 & 1 & 0
\end{array}
\right)_{q\times q}.}
\label{def:shift-matrix}
\end{eqnarray}
For real channel gain $h_{ij}$ in the Gaussian model 
\eqref{eq:GaussModel}, we calculate $n_{ij}$ as $n_{ij}=\lceil \frac{1}{2} \log |h_{ij}|^2 \rceil$. 
The parameter $q$ is chosen such that $q \geq
\max_{i,j} \lceil \frac{1}{2} \log |h_{ij}|^2 \rceil$.

\begin{figure}[th]
\begin{center}
 	\psfrag{s1}[Bc][Bc]{\begin{small}$S_1$\end{small}}
	\psfrag{s2}[Bc][Bc]{\begin{small}$S_2$\end{small}}
	\psfrag{r1}[Bc][Bc]{\begin{small}$A$\end{small}}
	\psfrag{r2}[Bc][Bc]{\begin{small}$B$\end{small}}
	\psfrag{d1}[Bc][Bc]{\begin{small}$D_1$\end{small}}
	\psfrag{d2}[Bc][Bc]{\begin{small}$D_2$\end{small}}
\includegraphics[width=7cm]{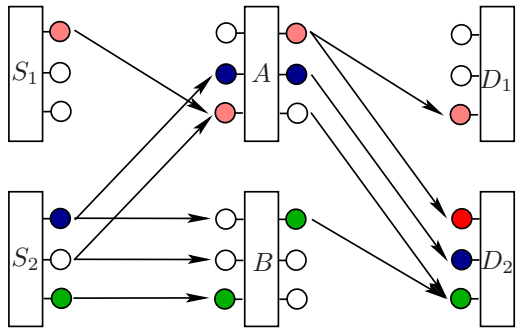}
\end{center}
\vspace{-7mm}
\caption{A deterministic network.}
\label{IF_sep}
\vspace{-5mm}
\end{figure}

An example of a deterministic network is illustrated in Figure~\ref{IF_sep}.
Each node contains several channel inputs and outputs, which are called \emph{sub-node} or \emph{level} 
through out this paper. Source $S_1$ can only send one bit to node 
$A$ and no bit to node $B$; source $S_2$ can send its two MSB to both $A$ and $B$, and its LSB to node $B$.
The transmitted bits from nodes $S_1$ and $S_2$ interfere on the LSB that node $A$ receives.

In a deterministic network, given a cut that separates 
nodes $\mathcal{U}$ from node $\mathcal{V}$,
the cut-value equals the rank of the transfer matrix between the nodes
in $\mathcal{U}$ and $\mathcal{V}$. For example, in Figure~\ref{IF_sep},
the cut that separates nodes $\mathcal{U}=\{S_1,S_2,A,B\}$ and $\mathcal{V}=\{D_1,D_2\}$ equals
\begin{eqnarray}
\rank
\footnotesize{\left(
\begin{array}{cccc}
1 & 0 & 0  & 0\\
1 & 0 & 0  & 0\\
0 & 1 & 0  & 0\\
0 & 0 & 1  & 1
\end{array}
\right)}=3.
\label{def:rank}
\end{eqnarray}
The rows of this transfer matrix correspond to the four transmitted inputs by nodes $A$ and $B$, 
while the  columns to the four receives outputs at nodes $D_1$ and $D_2$.


\subsection{Interference-relay network model}
\label{subsec:IntRel}

Our goal in this paper is to derive approximate capacity
characterizations for a class of $2$-user relay-interference networks
shown in Figure~\ref{model:GXX}, which we call the $\XX$ network.
We start by describing our notation for  Gaussian channels.

Two transmitters, $S_1$ and $S_2$,  encode their messages
$W_1$ and $W_2$ of rates $R_1$ and $R_2$, respectively, and broadcast
the obtained signals to the relay nodes, $A$ and $B$. Denote the
transmitted signals by $x_1$ and $x_2$, and
the received signals at the 
 relays by $y'_1$ and $y'_2$.
Then
\begin{align}
\label{eq:Glayer1}
\begin{array}{l}
y'_1[t]=\sqrt{g_{11}} x_1[t] + \sqrt{g_{12}} x_2[t] + z'_1[t],\\ 
y'_2[t]=\sqrt{g_{21}} x_1[t] + \sqrt{g_{22}} x_2[t] + z'_2[t], 
\end{array}
\end{align}
where $z'_1,z'_2$ are unit-variance Gaussian noises, independent of
each other and of $x_1,x_2$.

The relay nodes perform any (causal) processing on their received
signal sequences $\{y'_1[t]\}$ and $\{y'_2[t]\}$ respectively, to
obtain their transmitting signal sequences, $\{x'_1(t)\}$ and
$\{x'_2(t)\}$. The received signals at the destination nodes can be
written as
\begin{align}
\label{eq:Glayer2}
\begin{array}{l}
y_1[t] =\sqrt{h_{11}} x'_1[t] + \sqrt{h_{12}} x'_2[t] + z_1[t]\\ 
y_2[t] =\sqrt{h_{21}} x'_1[t]+ \sqrt{h_{22}} x'_2[t] + z_2[t],
\end{array}
\end{align}
where the $z'_1$, $z'_2$, $z_1$, and $z_2$ are independent zero-mean
unit-variance noises, which are also independent of $x_1$ and
$x_2$. There is a power constraint for each transmitted signal, that
is, $\E[x_1^2]\leq 1$, $\E[x_2^2]\leq 1$, $\E[x'^2_1]\leq 1$ and
$\E[x'^2_2]\leq 1$.

Each destination node $D_i$, $i=1,2$, is interested in decoding its
message $W_i$, using its received signals $\{y_i[t]\}$. We define a
rate pair $(R_1,R_2)$ to be admissible if there exist a transmission
scheme under which $D_1$ and $D_2$ can decode $W_1$ and $W_2$,
respectively, with arbitrary small (average) error probability in the
standard manner \cite{CT91}.  This would allow two end-to-end reliable
unicast sessions at rates $(R_1,R_2)$ for the source/destination pairs
$(S_1,D_1)$ and $(S_2,D_2)$.

A useful tool to examine the network problem defined above is to
study its deterministic version, based on the model developed in
\eqref{eq:DetModel}. Using the deterministic approach, we can rewrite
\eqref{eq:Glayer1}-\eqref{eq:Glayer2} as
\begin{align}
\label{eq:Dlayer1}
\begin{array}{l}
Y'_1[t] =M_{11} X_1[t] + M_{12} X_2[t]\\ 
Y'_2[t] =M_{21} X_1[t] + M_{22} X_2[t], 
\end{array}
\end{align}
and
\begin{align}
\label{eq:Dlayer2}
\begin{array}{l}
Y_1[t] =N_{11} X'_1[t] + N_{12} X'_2[t]\\
Y_2[t] =N_{21} X'_1[t] + N_{22} X'_2[t],
\end{array}
\end{align}
where the matrices $\{M_{ij}\}$ and $\{N_{ij}\}$ approximately model
the channels in \eqref{eq:Glayer1}-\eqref{eq:Glayer2}, \emph{i.e.,}
$M_{ij}=\J^{q-m_{ij}},N_{ij}=\J^{q-n_{ij}}$. The matrix $\J$ is defined as
in \eqref{def:shift-matrix}, while $m_{ij}=\lceil \frac{1}{2} \log
|g_{ij}|^2 \rceil$ and $n_{ij}=\lceil \frac{1}{2} \log |h_{ij}|^2
\rceil$.

It is worth mentioning that though this network looks like cascaded
interference channels, there is an important difference.  Unlike the
interference channel, the messages sent by the relays at the second
layer of transmission need not independent, {\em i.e.,} we can try to
induce cooperation at the relays to transmit information to the final
destinations. This distinction makes this network more interesting
than a simple cascade of interference channels.

In this paper, we focus on two specific realizations of the network,
namely, the $\ZS$ and the $\ZZ$ networks, which further simplify the
connectivity models of \eqref{eq:Glayer1}-\eqref{eq:Glayer2}. We describe 
these two networks in the following, 
and give an approximate characterization
of their admissible rate region in Section~\ref{sec:MainRes}.
 
\noindent\emph{Notation alert:} Throughout this paper, we use the
lowercase letters $x$ and $y$ for the signals transmitted by the
sources and received signals at the destinations in the Gaussian
networks. The received and transmitting signals by the relays are
denoted by $x'$ and $y'$. Similarly, uppercase letters will be used
for the deterministic networks.

\subsection{The $\ZS$ Network}
\label{subsec:ZSnetDef}

\begin{figure}[h!]
 \centering
     \subfigure[The Gaussian $\ZS$ network]{
 	\psfrag{s1}[Bc][Bc]{$S_1$}
	\psfrag{s2}[Bc][Bc]{$S_2$}
	\psfrag{r1}[Bc][Bc]{$A$}
	\psfrag{r2}[Bc][Bc]{$B$}
	\psfrag{d1}[Bc][Bc]{$D_1$}
	\psfrag{d2}[Bc][Bc]{$D_2$}
	\psfrag{a}[Bc][Bc]{$\sqrt{g_{11}}$}
	\psfrag{b}[Bc][Bc]{$\sqrt{h_{11}}$}
	\psfrag{c}[Bc][Bc]{$\sqrt{g_{22}}$}
	\psfrag{d}[Bc][Bc]{$\sqrt{h_{22}}$}
	\psfrag{f}[Bc][Bc]{$\sqrt{g_{12}}$}
	\psfrag{g}[Bc][Bc]{$\sqrt{h_{21}}$}
 	\psfrag{X_1}[Bc][Bc]{$x_1$}
	\psfrag{X_2}[Bc][Bc]{$x_2$}
	\psfrag{Y'_1}[Bc][Bc]{$y'_1$}
	\psfrag{Y'_2}[Bc][Bc]{$y'_2$}
	\psfrag{X'_1}[Bc][Bc]{$x'_1$}
	\psfrag{X'_2}[Bc][Bc]{$x'_2$}
	\psfrag{Y_1}[Bc][Bc]{$y_1$}
	\psfrag{Y_2}[Bc][Bc]{$y_2$}
	\psfrag{Z'_1}[Bc][Bc]{$z'_1$}
	\psfrag{Z'_2}[Bc][Bc]{$z'_2$}
	\psfrag{Z_1}[Bc][Bc]{$z_1$}
	\psfrag{Z_2}[Bc][Bc]{$z_2$}
	\psfrag{t_1}[Bc][Bc]{$\ $}
	\psfrag{t_2}[Bc][Bc]{$\ $}
\includegraphics[width=9cm]{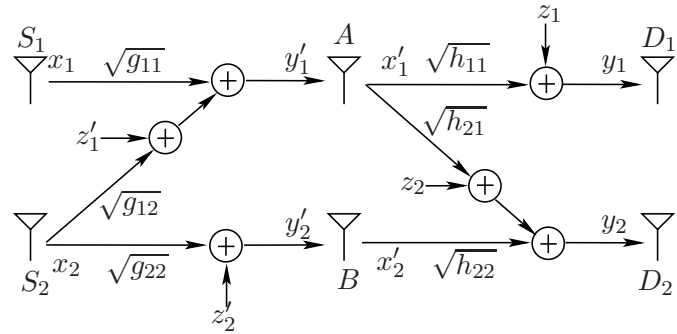}
\label{fig:GZS}
}
\hspace{.5in}
    \subfigure[The deterministic $\ZS$ network]{
     \psfrag{s1}[Bc][Bc]{$S_1$}
	\psfrag{s2}[Bc][Bc]{$S_2$}
	\psfrag{r1}[Bc][Bc]{$A$}
	\psfrag{r2}[Bc][Bc]{$B$}
	\psfrag{d1}[Bc][Bc]{$D_1$}
	\psfrag{d2}[Bc][Bc]{$D_2$}
	\psfrag{a}[Bc][Bc]{$m_{11}$}
	\psfrag{b}[Bc][Bc]{$n_{11}$}
	\psfrag{c}[Bc][Bc]{$m_{22}$}
	\psfrag{d}[Bc][Bc]{$n_{22}$}
	\psfrag{e}[Bc][Bc]{$m_{21}$}
	\psfrag{f}[Bc][Bc]{$m_{12}$}
	\psfrag{g}[Bc][Bc]{$n_{21}$}
	\psfrag{h}[Bc][Bc]{$m_{12}$}
\includegraphics[width=9cm]{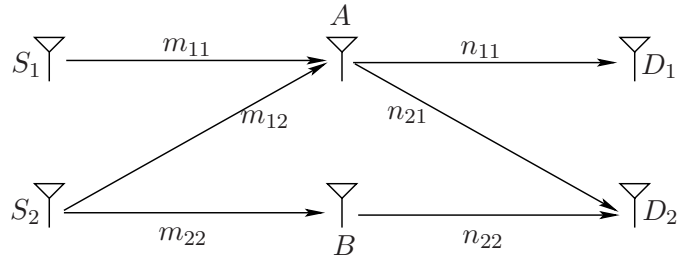}
\label{fig:DZS}
}
\vspace{-5mm}
\caption{The $\ZS$ network.}
\vspace{-5mm}
\label{model:ZS}
\end{figure}

The $\ZS$ network is a special case of the interference-relay network
defined in \eqref{eq:Glayer1}-\eqref{eq:Glayer2}.  In the $\ZS$
network one cross link in each layer has a negligible gain, and
therefore does not cause interference, as illustrated in
Figure~\ref{fig:GZS}.  In particular, we assume $g_{21}=h_{12}=0$ in
the Gaussian network, and $m_{21}=n_{12}=0$ in the deterministic
network.  The resulting Gaussian $\ZS$ network is shown in
Figure~\ref{fig:GZS}, and the deterministic model for this network is
given in Figure~\ref{fig:DZS}.

\subsection{The $\ZZ$ Network}
\label{subsec:ZZnetDef}
\begin{figure}[h!]
 \centering
     \subfigure[The Gaussian $\ZZ$ network]{
 	\psfrag{s1}[Bc][Bc]{$S_1$}
	\psfrag{s2}[Bc][Bc]{$S_2$}
	\psfrag{r1}[Bc][Bc]{$A$}
	\psfrag{r2}[Bc][Bc]{$B$}
	\psfrag{d1}[Bc][Bc]{$D_1$}
	\psfrag{d2}[Bc][Bc]{$D_2$}
	\psfrag{a}[Bc][Bc]{$\sqrt{g_{11}}$}
	\psfrag{b}[Bc][Bc]{$\sqrt{h_{11}}$}
	\psfrag{c}[Bc][Bc]{$\sqrt{g_{22}}$}
	\psfrag{d}[Bc][Bc]{$\sqrt{h_{22}}$}
	\psfrag{f}[Bc][Bc]{$\sqrt{g_{12}}$}
	\psfrag{g}[Bc][Bc]{$\sqrt{h_{12}}$}
 	\psfrag{X_1}[Bc][Bc]{$x_1$}
	\psfrag{X_2}[Bc][Bc]{$x_2$}
	\psfrag{Y'_1}[Bc][Bc]{$y'_1$}
	\psfrag{Y'_2}[Bc][Bc]{$y'_2$}
	\psfrag{X'_1}[Bc][Bc]{$x'_1$}
	\psfrag{X'_2}[Bc][Bc]{$x'_2$}
	\psfrag{Y_1}[Bc][Bc]{$y_1$}
	\psfrag{Y_2}[Bc][Bc]{$y_2$}
	\psfrag{Z'_1}[Bc][Bc]{$z'_1$}
	\psfrag{Z'_2}[Bc][Bc]{$z'_2$}
	\psfrag{Z_1}[Bc][Bc]{$z_1$}
	\psfrag{Z_2}[Bc][Bc]{$z_2$}
	\psfrag{t_1}[Bc][Bc]{$t_1$}
	\psfrag{t_2}[Bc][Bc]{$t_2$}
\includegraphics[width=9cm]{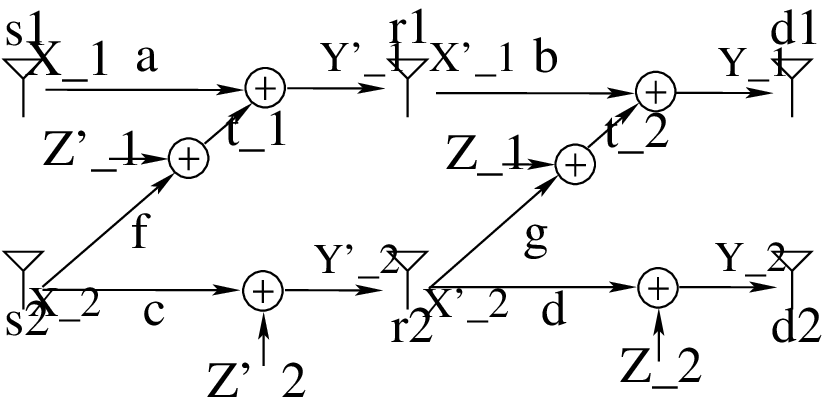}
\label{fig:GZZ}
}
\hspace{.5in}
    \subfigure[The deterministic $\ZZ$ network]{
     \psfrag{s1}[Bc][Bc]{$S_1$}
	\psfrag{s2}[Bc][Bc]{$S_2$}
	\psfrag{r1}[Bc][Bc]{$A$}
	\psfrag{r2}[Bc][Bc]{$B$}
	\psfrag{d1}[Bc][Bc]{$D_1$}
	\psfrag{d2}[Bc][Bc]{$D_2$}
	\psfrag{a}[Bc][Bc]{$m_{11}$}
	\psfrag{b}[Bc][Bc]{$n_{11}$}
	\psfrag{c}[Bc][Bc]{$m_{22}$}
	\psfrag{d}[Bc][Bc]{$n_{22}$}
	\psfrag{e}[Bc][Bc]{$m_{21}$}
	\psfrag{f}[Bc][Bc]{$m_{12}$}
	\psfrag{g}[Bc][Bc]{$n_{21}$}
	\psfrag{h}[Bc][Bc]{$m_{12}$}
	\psfrag{X_1}[Bc][Bc]{$X_1$}
	\psfrag{X_2}[Bc][Bc]{$X_2$}
	\psfrag{Y'_1}[Bc][Bc]{$Y'_1$}
	\psfrag{Y'_2}[Bc][Bc]{$Y'_2$}
	\psfrag{X'_1}[Bc][Bc]{$X'_1$}
	\psfrag{X'_2}[Bc][Bc]{$X'_2$}
	\psfrag{Y_1}[Bc][Bc]{$Y_1$}
	\psfrag{Y_2}[Bc][Bc]{$Y_2$}
\includegraphics[width=9cm]{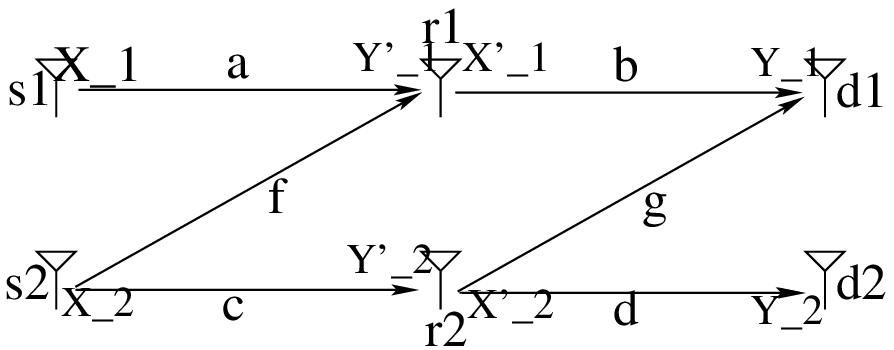}
\label{fig:DZZ}
}
\vspace{-5mm}
\caption{The $\ZZ$ network.}
\vspace{-5mm}
\label{fig:ZZ}
\end{figure}

The $\ZZ$ network is another special configuration interference-relay
network, wherein one cross link in each layer has zero gain. However,
the difference is that, here the missing links are in parallel. In
particular, we assume $g_{21}=h_{21}=0$ and $m_{21}=n_{21}=0$ in the
Gaussian and deterministic networks, respectively. The Gaussian and
corresponding deterministic $\ZZ$ networks are shown in
Figure~\ref{fig:ZZ}.

\section{Examples illustrating transmission techniques}
\label{sec:examples}

In this section, we illustrate through examples some of the main interference
management techniques we will use to (approximately) achieve the
capacity of our relay-interference networks.  For simplicity in
demonstrating the ideas, we focus on deterministic networks throughout
the examples.  However, similar techniques will be used later for
Gaussian channels as well.

We also present a simple Gaussian example at the end of this section, 
to illustrate the message splitting idea used in many places through out this work. 

\begin{example}[Network Decomposition for the $\ZS$ Network]
\label{ex:NetPart}
  A deterministic $\ZS$ network can be always decomposed into two
  subnode-disjoint networks, where the first partition consists of a set
  of sub-nodes of $S_1$, $A$ and $D_1$, and looks like a line network.
  The second partition is, however, a diamond network, with a broadcast channel 
  from $S_2$ to $A$ and $B$ in the first layer, and a multiple access channel 
  from $A$ and $B$ to $D_2$ in the second layer. This diamond network can be used
  to send information from $S_2$ to $D_2$. Since these two networks
  are sub-node disjoint, there would be no interfering signal, and
  each of them can be analyzed separately. This is more illustrated in
  Figure~\ref{fig:zs-parts}.
\begin{figure}[h]
\begin{center}
 	\includegraphics[width=7cm]{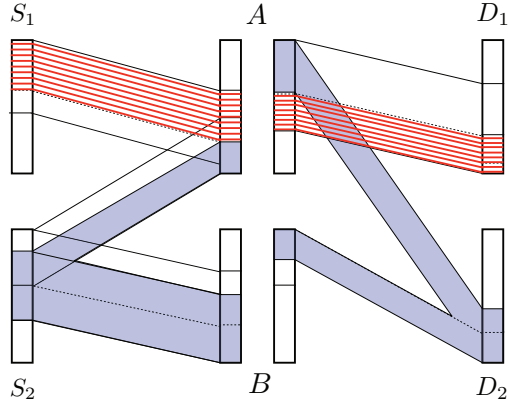}
\end{center}
\vspace{-4mm}
\caption{Network partitioning for a deterministic $\ZS$ network.}
\label{fig:zs-parts}
\vspace{-4mm}
\end{figure}

In a Gaussian $\ZS$ network the network decomposition can be done using
message splitting, superposition coding and proper power
allocation. We will use this technique to achieve an approximate
capacity for the Gaussian $\ZS$ network.
\end{example}

\begin{example}[Interference Neutralization]
\label{ex:DetIntNeut}
  This technique can be used in networks which contain more than one
  disjoint path from $S_i$ to $D_j$ for $i\neq j$, where $D_j$ is not
  interested in decoding the message sent by the source node $S_i$,
  and therefore it receives the interference through more than one
  link. The proposed technique is to tune these interfering signals
  such that they neutralize each other at the destination node. In
  words, the interfering signal should be received at the same power
  level and with different sign such that the effective interference,
  obtained by adding them, occupies a smaller number of degrees of
  freedom. To best of our knowledge, this technique is new and was
  been introduced in \cite{MDFT:Al08}.

\begin{figure}[h]
\begin{center}
 	\psfrag{s1}[Bc][Bc]{$S_1$}
	\psfrag{s2}[Bc][Bc]{$S_2$}
	\psfrag{r1}[Bc][Bc]{$A$}
	\psfrag{r2}[Bc][Bc]{$B$}
	\psfrag{d1}[Bc][Bc]{$D_1$}
	\psfrag{d2}[Bc][Bc]{$D_2$}	
	\psfrag{+}[Bc][Bc]{$+$}	
	\psfrag{-}[Bc][Bc]{$-$}	
\includegraphics[width=7cm]{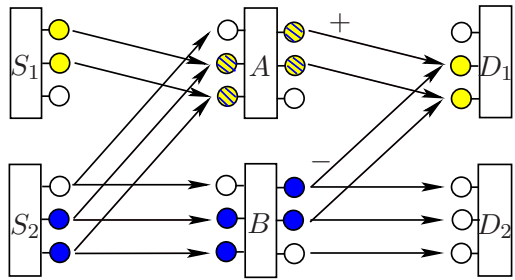}
\end{center}
\vspace{-4mm}
\caption{Interference Neutralization; $(R_1,R_2)=(2,2)$ is achievable.}
\label{fig:neutralization}
\vspace{-4mm}
\end{figure}

\begin{figure}[h]
\begin{center}
 	\includegraphics[width=7cm]{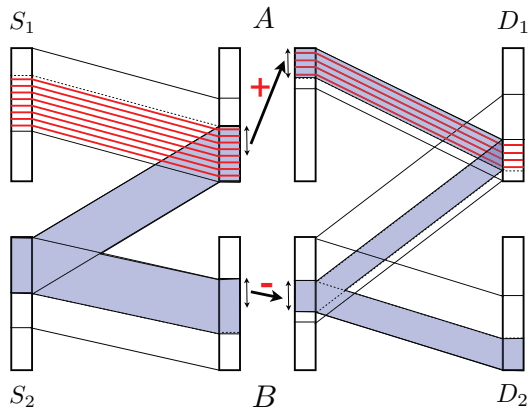}
\end{center}
\vspace{-4mm}
\caption{Interference neutralization.}
\label{fig:neutralizationGenEx}
\vspace{-6mm}
\end{figure}

Figure~\ref{fig:neutralization} shows a network in which interference
neutralization is essential to achieve the desired rate pair
$(R_1,R_2)=(2,3)$. Here $D_1$ has only two degrees of freedom, and
receives information bits from both $A$ and $B$ over these
sub-nodes. However, notice that there are two disjoint paths
($S_2,A,D_1$) and ($S_2,B,D_1$), which connect $S_2$ to $D_1$. As it
is shown in Figure~\ref{fig:neutralization}, using a proper mapping
(permutation) at the relay nodes, one can make the interference
neutralized at the destination node $D_1$, and provide two
non-interfered links from $S_1$ to $D_2$. Note that this permutation
does not effect the admissible rate of the other unicast from $S_2$ to
$D_2$, the cost we pay, is to permute the received bits at $D_2$.
A more general illustration of this phenomenon is given in Figure
\ref{fig:neutralizationGenEx}.
\label{ex:IN}
\end{example}

\begin{example}[Use of Lattice Codes to Implement Interference
  Neutralization over Gaussian $\ZZ$ Network]
\label{ex:LatticeIntNeut}
  The idea of interference neutralization illustrated in
  Example~\ref{ex:IN} can be also used in Gaussian networks. In this
  case a group structured code, such as lattice code, is required to
  play the role of composition and decomposition of the signal and
  interference in two layers of the network.  Consider the Gaussian
  $\ZZ$ network in Figure~\ref{fig:GZZ}.
  We can use message splitting and interference neutralization to
  improve the achievable rate pairs of this network.

  Let the second source split its message into two parts as
  $W_2=(W_2^{(N)}, W_2^{(P)} )$, namely, the functional (neutralization) and
  private parts, of rates $R_{2,N}=R_1$ and $R_{2,P}=R_2-R_1$.  Both
  transmitters use a common lattice code to encode $W_1$ and
  $W_2^{(N)}$, and map them into $\x_1^{(N)}$ and $\x_2^{(N)}$,
  respectively. The other message $W_2^{(P)}$ can be encoded to
  $\x_2^{(P)}$ using a random Gaussian code. We assume that both the
  lattice code and the random Gaussian code have average power equal to
  $1$. Then, the transmitting signals would be a linear combination of
  the codewords with a proper power allocation, \emph{i.e.},
\begin{align}
 \x_1=\sqrt{\alpha_N} \x_1^{(N)}, \qquad \x_2= \sqrt{\beta_N} \x_2^{(N)} +\sqrt{\beta_P} \x_2^{(P)},  
\end{align}
where the power allocation coefficients satisfy $\alpha_N\leq 1$ and 
$\beta_N+\beta_P\leq 1$.
The transmitters choose the power allocated to $\x_1^{(N)}$ to
$\x_2^{(N)}$ in a way that they get received at $A$ with the same
power. In this way, their summation would be again a lattice code and
can be decoded at $A$ by treating $\x_1^{(P)}$ as noise.  A similar
strategy will be used for signaling at the relay for transmission in
the second layer of the network.  The only difference is that instead
of sending $\x_2^{(N)}$, the relay node $B$ sends $-\x_2^{(N)}$. Then,
the lattice point observed at $D_1$ would be exactly $\x_1^{(N)}$ and
it can find $W_1$. The other decoder can simply first reverse
$-\x_2^{(N)}$ to $\x_2^{(N)}$, and then decode it. This idea is
illustrated in Figure~\ref{fig:lattice}.

\begin{figure}[h]
\begin{center}
 	\psfrag{s1}[Bc][Bc]{$S_1$}
	\psfrag{s2}[Bc][Bc]{$S_2$}
	\psfrag{A}[Bc][Bc]{$A$}
	\psfrag{B}[Bc][Bc]{$B$}
	\psfrag{d1}[Bc][Bc]{$D_1$}
	\psfrag{d2}[Bc][Bc]{$D_2$}
 	\psfrag{a}[Bc][Bc]{$\sqrt{\alpha_N} \x_1^{(N)}$}
	\psfrag{b}[Bc][Bc]{$\sqrt{\beta_N} \x_2^{(N)}$}
	\psfrag{c}[Bc][Bc]{$\sqrt{g_{11}}$}
	\psfrag{d}[Bc][Bc]{$\sqrt{g_{12}}$}
	\psfrag{e}[Bc][Bc]{$\sqrt{g_{22}}$}
	\psfrag{f}[Bc][Bc]{$\sqrt{h_{11}}$}
	\psfrag{g}[Bc][Bc]{$\sqrt{h_{12}}$}
	\psfrag{h}[Bc][Bc]{$\sqrt{h_{22}}$}
	\includegraphics[height=6cm]{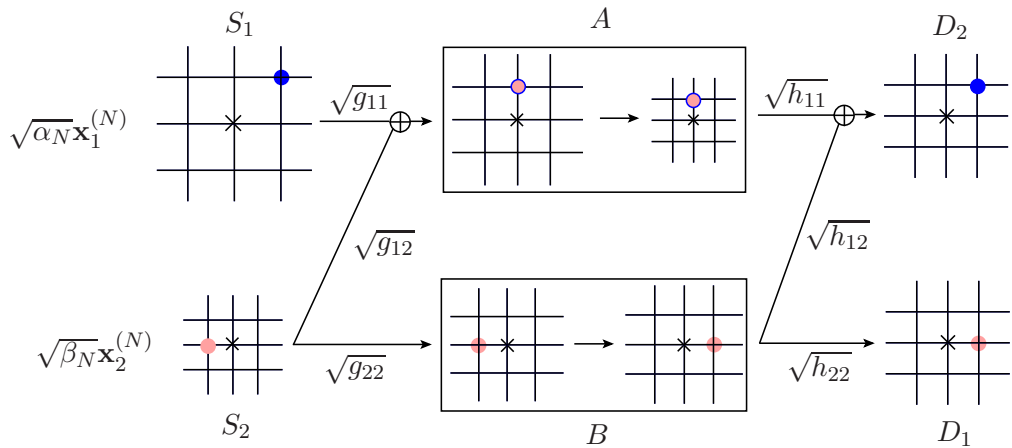}
\end{center}
\vspace{-4mm}
\caption{Using lattice codes for interference neutralization over
  a Gaussian $\ZZ$ network. The origin is specified by a cross ``$\times$''. 
  Power allocated to the messages at the
  transmitters are chosen such that the two lattice points corresponding 
  to $\x_1^{(N)}$ and $\x_2^{(N)}$ get received at $B$ at the
  same power level, and their summation becomes a point on the 
   scaled lattice. The same strategy is used by the
  relays. The relay node $B$ also reverses its transmitting lattice
  point in order to neutralize the interference caused in the first layer
  of the network.}
\label{fig:lattice}
\vspace{-6mm}
\end{figure}
\end{example}

\begin{example}
\label{ex:GZ}

Consider the Gaussian $\Z$ network shown in Figure~\ref{fig:GZ}, with channel gains 
$g_{11} \geq 1$, $g_{12}\geq 1$, and  $g_{22}\geq 1$.
\begin{figure}[th]
\begin{center}
 	\psfrag{s1}[Bc][Bc]{$F_1$}
	\psfrag{s2}[Bc][Bc]{$F_2$}
	\psfrag{r1}[Bc][Bc]{$G_1$}
	\psfrag{r2}[Bc][Bc]{$G_2$}
	\psfrag{a}[Bc][Bc]{$\sqrt{g_{11}}$}
	\psfrag{f}[Bc][Bc]{$\sqrt{g_{12}}$}
	\psfrag{c}[Bc][Bc]{$\sqrt{g_{22}}$}
	\psfrag{Z'_1}[Bc][Bc]{$z_1$}
	\psfrag{Z'_2}[Bc][Bc]{$z_2$}
	\psfrag{X_1}[Bc][Bc]{$x_1$}
	\psfrag{X_2}[Bc][Bc]{$x_2$}
	\psfrag{Y'_1}[Bc][Bc]{$y_1$}
	\psfrag{Y'_2}[Bc][Bc]{$y_2$}
	\psfrag{w1}[Bc][Bc]{$W_1$}
	\psfrag{w2}[Bc][Bc]{$W_2$}
	\psfrag{w3}[Bc][Bc]{$\hat{W}_1$}
	\psfrag{w4}[Bc][Bc]{$\hat{W}_2$}
	\includegraphics[height=4cm]{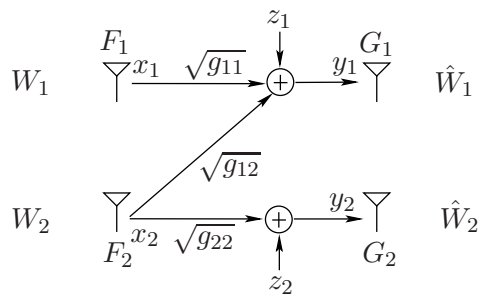}
\end{center}
\vspace{-4mm}
\caption{A Gaussian $\Z$ network.}
\label{fig:GZ}
\vspace{-4mm}
\end{figure}

The source nodes $F_i$ wishes to encode and send 
message $W_i$ to the destination node $G_i$, for $i=1,2$. Denoting the rate of message $W_i$ 
by $R_i$, an approximate capacity characterization for this network is given by 
network is given by 
\begin{align}
\R{\Z}{}=\Big\{ (R_1,R_2): & R_1 \leq \CGN{1+g_{11}}\nonumber\\
			  & R_2 \leq \CGN{1+g_{22}}\label{ex:eq:GZ-out}\\
			  & R_1+R_2 \leq \CGN{1+g_{11}+g_{12}}+\CGN{1+\frac{g_{22}}{g_{12}}}
\Big\}. \nonumber
\end{align}
It is easy to show that any achievable rate pair belongs to $\R{\Z}{}$, and hence $\R{\Z}{}$ establishes an outer bound for the capacity region. 
Moreover, one can show that the rate pair $(R_1-\frac{1}{2},R_2-\frac{1}{2})$ is achievable provided that $(R_1,R_2)\in\R{\Z}{}$. 
The encoding strategy to achieve such rate pair involves message splitting and proper power allocation. We will discuss this in more details in 
Appendix~\ref{sec:app-lm}. 

\end{example}

\section{Main Results}
\label{sec:MainRes}
In this section we present the main results of this paper, which is
the approximate capacity characterization of the Gaussian $\ZS$ and
$\ZZ$ interference-relay networks. In order to obtain such an
approximate characterization, we have a complete characterization of
the deterministic versions of the $\ZS$ and $\ZZ$ networks. The coding
strategies for the Gaussian problems are outlined in Section
\ref{sec:pr-outline}. The detailed analysis of these strategies and
the corresponding outer bounds which lead to Theorems \ref{thm:gzs}
and \ref{thm:gzz} are given in Appendices \ref{sec:gzs} and
\ref{sec:gzz}, respectively. Most of the insights are obtained by
analyzing the deterministic versions of these problems, and the exact
characterizations are summarized in Theorems \ref{thm:dzs} and
\ref{thm:dzz} respectively. We prove these results in Sections
\ref{sec:dzs} and \ref{sec:dzz}, respectively. The achievability and
outer bound results for the Gaussian cases are directly inspired by
these results.

\subsection{The $\ZS$ Network}

The $\ZS$ network illustrated in Figure~\ref{fig:DZS} and the
corresponding Gaussian $\ZS$ network is given in Figure~\ref{fig:GZS}.
Theorems \ref{thm:dzs} and \ref{thm:gzs} give the exact and
approximate (within $2$ bits) characterizations of their capacity
regions.

\begin{theorem}[The capacity region of deterministic $\ZS$ network]

  The capacity region of the deterministic $\ZS$ network is specified
  by $\R{\DZS}{}$, where $\R{\DZS}{}$ is the set of all rate pairs
  $(R_1,R_2)$ that satisfy 
  \setcounter{temp}{\value{equation}}
  \setcounter{equation}{0} \let\thee\theequation
  \renewcommand{\theequation}{$\DZS$-\arabic{equation}}
\begin{align}
 R_1 &\leq m_{11},\label{dzs:s1}\\
 R_2 &\leq \max(m_{12}, m_{22}),\label{dzs:s2}\\
 R_1+R_2 &\leq \max(m_{11},m_{12}) + (m_{22}-m_{12})^+,\label{dzs:s1s2}\\
 R_2 &\leq m_{12}+n_{22},\label{dzs:s2r2}\\
 R_1+R_2 &\leq m_{22}+\max(n_{11},n_{21}),\label{dzs:s1s2r1}\\
 R_1+R_2 &\leq \max(m_{11},m_{12})+n_{22},\label{dzs:s1s2r2}\\
 R_1 &\leq n_{11},\label{dzs:d1}\\
 R_2 &\leq \max(n_{21},n_{22}),\label{dzs:d2}\\
 R_2 &\leq m_{22}+n_{21}, \label{dzs:r2d2}\\
 R_1+R_2 &\leq \max(n_{21},n_{22}) + (n_{11}-n_{21})^+.\label{dzs:d1d2}
\end{align}
\setcounter{equation}{\value{temp}} \let\theequation\thee 
\label{thm:dzs}
\end{theorem}

\begin{theorem}[An Approximate capacity region of Gaussian $\ZS$ network]
  Let $\R{\GZS}{}$ be the set of all rate pairs $(R_1,R_2)$ which
  satisfy \eqref{gzs:s1}--\eqref{gzs:d1d2} given below. Then
  $\R{\GZS}{}$ is an outer bound for the capacity region of the
  Gaussian $\ZS$ network. Moreover, for any $(R_1,R_2)\in
  \R{\GZS}{}$, there exists a transmission scheme with rates
  $(R'_1,R'_2)=(R_1-\delta_1,R_2-\delta_2)$, where $\delta_1=1$ and
  $\delta_2=1.5$ are universal constants, independent of the channel gain,
  and required rates.  \setcounter{temp}{\value{equation}}
  \setcounter{equation}{0} \let\thee\theequation
  \renewcommand{\theequation}{$\GZS$-\arabic{equation}}
\begin{align}
R_1 &\leq \frac{1}{2}\log(1+ g_{11})\label{gzs:s1}\\
R_2 &\leq \frac{1}{2}\log(1+ g_{12} + g_{22})\label{gzs:s2}\\
R_1 + R_2 &\leq  \frac{1}{2}\log(1+ g_{11} + g_{12}) + \frac{1}{2}\log \left(1+\frac{g_{22}}{g_{12}}\right)\label{gzs:s1s2}\\
R_2 &\leq \frac{1}{2}\log(1+ g_{12}) + \frac{1}{2}\log(1+ h_{22})\label{gzs:s2r2}\\
R_1+R_2  &\leq  \frac{1}{2} \log(1+g_{22}) + \frac{1}{2}\log(1+h_{11} +h_{21}) \label{gzs:s1s2r1}\\
R_1+R_2  &\leq  \frac{1}{2} \log(1+g_{11}+g_{12}) + \frac{1}{2}\log(1+h_{22} ) \label{gzs:s1s2r2}\\
R_1 &\leq \frac{1}{2}\log(1+ h_{11})\label{gzs:d1}\\
R_2 &\leq \frac{1}{2}\log(1+ h_{21} + h_{22}+ 2 \sqrt{h_{21}h_{22}})\label{gzs:d2}\\
R_2 &\leq  \frac{1}{2}\log(1+ g_{22}) + \frac{1}{2}\log(1+ h_{21})\label{gzs:r2d2}\\
R_1 + R_2 &\leq  \frac{1}{2}\log(1+ h_{21} + h_{22}+2\sqrt{h_{21}h_{22}})+ \frac{1}{2}\log \left(1+\frac{h_{11}}{h_{21}}\right).  
  \label{gzs:d1d2}
\end{align}
\setcounter{equation}{\value{temp}} \let\theequation\thee 
\label{thm:gzs}
\end{theorem}

The outer bound for the results above are fairly standard arguments
based on reducing a multi-letter mutual information into single-letter
forms by appropriately using decodability requirements at the
different destinations. The details of these are given in Section
\ref{subsec:dzs:convrse} and Appendix~\ref{subsec:gzz-conv}, respectively.

The coding strategy achieving these regions is based on two
ideas. One is that of a {\em network decomposition} illustrated in
Section \ref{sec:examples}, Example \ref{ex:NetPart} for the
deterministic network. The insight from the network decomposition
leads to the idea of strategic {\em rate-splitting} and power
allocation in the Gaussian channel. For the Gaussian coding scheme, we
need to strategically partition the messages and allocate powers in
order for the relays to partially decode appropriate messages and setup
cooperation. The details of this strategy are outlined in Section
\ref{sec:pr-outline}.

\subsection{The $\ZZ$ Network}

The $\ZZ$ network illustrated in Figure~\ref{fig:DZZ} and the
corresponding Gaussian $\ZZ$ network is given in Figure~\ref{fig:GZZ}.
Although superficially the $\ZS$ and $\ZZ$ networks may look similar,
the subtle difference in the network connectivity, makes the two
problems completely different, both in terms of capacity
characterization, as well as transmission schemes.  It will be shown
that a new interference management scheme, which we term as interference
neutralization, is needed to (approximately) achieve the capacity of
this network.  The most intuitive description for interference
neutralization is to cancel interference over air without processing
at the destinations.  This scheme can be used whenever there are more
than one path for interference to get received at a destination. We will
explain it in more detail in Sections~\ref{sec:pr-outline} and
\ref{sec:dzz}.  

Theorems \ref{thm:dzz} and \ref{thm:gzz} give the
exact and approximate (within $2$ bits) characterizations 
for the capacity region of the deterministic and the Gaussian $\ZZ$
networks, respectively. Another new ingredien used here is needed a
genie-aided outer bound that gives the (noisy) cross link of the first
(or correspondingly second) layer to the destination (or
correspondingly to the relay). This genie-aided bound allows us to
develop outer bounds that are apparantly tighter than the
information-theoretic cut-set bounds by utilizing the decoding
structure needed.

\begin{theorem}[The capacity region of deterministic $\ZZ$ network]
The capacity region of the deterministic $\ZZ$ network is given by 
$\R{\DZZ}{}$, where $\R{\DZZ}{}$ is the set of  all rate pairs 
$(R_1,R_2)$ which satisfy
\setcounter{temp}{\value{equation}}
\setcounter{equation}{0}
\let\thee\theequation
\renewcommand{\theequation}{$\DZZ$-\arabic{equation}}
\begin{align}
R_1 &\leq m_{11},\label{dzz:s1}\\
R_2 &\leq m_{22},\label{dzz:s2}\\
R_1 &\leq n_{11},\label{dzz:d1}\\
 R_2 &\leq n_{22},\label{dzz:d2}\\
 R_1+r_2 &\leq \max(m_{11},m_{12})  + (m_{22}-m_{12})^+ + n_{12},\label{dzz:s1s2}\\
 R_1+R_2 &\leq \max(n_{11},n_{12})+(n_{22}-n_{12})^+  +m_{12}.\label{dzz:d1d2}
\end{align}
\setcounter{equation}{\value{temp}}
\let\theequation\thee
\label{thm:dzz}
\end{theorem}

\begin{theorem}[An approximate capacity region of Gaussian $\ZZ$ network]
  Let $\R{\GZZ}{}$ be the set of all rate pairs $(R_1,R_2)$ which
  satisfy \eqref{gzz:s1-o}--\eqref{gzz:d1d2-o} given below.  
  \setcounter{temp}{\value{equation}} \setcounter{equation}{0}
  \let\thee\theequation
  \renewcommand{\theequation}{$\GZZ$\arabic{equation}}
\begin{align}
R_1 &\leq \frac{1}{2}\log(1+ g_{11})\label{gzz:s1-o}\\
R_2 &\leq \frac{1}{2}\log(1+ g_{22})\label{gzz:s2-o}\\
R_1 &\leq \frac{1}{2}\log(1+ h_{11})\label{gzz:d1-o}\\
R_2 &\leq \frac{1}{2}\log(1+ h_{22})\label{gzz:d2-o}\\
R_1+R_2  &\leq  \frac{1}{2} \log(1+g_{11}+g_{12}) + \frac{1}{2}\log\left(1+\frac{g_{22}}{g_{12}}\right) 
  + \frac{1}{2}\log(1+h_{12}),\label{gzz:s1s2-o}\\
R_1+R_2  &\leq  \frac{1}{2} \log(1+h_{11}+h_{12}) + \frac{1}{2}\log\left(1+\frac{h_{22}}{h_{12}}\right) 
  + \frac{1}{2}\log(1+g_{12})  \label{gzz:d1d2-o}
\end{align}
Then,
  any admissible rate pair $(R_1,R_2)$  for the Gaussian $\ZZ$
  networks belongs to $\R{\GZZ}{}$. Moreover, for any rate pair
  $(R_1,R_2)\in \R{\GZZ}{}$, there exists an encoding scheme with 
  rates  $(R'_1,R'_2)=(R_1-\frac{7}{4},R_2-\frac{7}{4})$.  

\label{thm:gzz}
\end{theorem}

\section{Gaussian coding strategies}
\label{sec:pr-outline}
This section is devoted to providing the basic ideas of the coding
schemes used in the Gaussian $\ZS$ and $\ZZ$ networks. We also develop
an outline of how to analyze these coding strategies. 

\subsection{The Gaussian $\ZS$ network: Achievability}

\begin{figure}[th]
\begin{center}
 	\psfrag{s1}[Bc][Bc]{$F_1$}
	\psfrag{s2}[Bc][Bc]{$F_2$}
	\psfrag{r1}[Bc][Bc]{$G_1$}
	\psfrag{r2}[Bc][Bc]{$G_2$}
	\psfrag{a}[Bc][Bc]{$\sqrt{g_{11}}$}
	\psfrag{f}[Bc][Bc]{$\sqrt{g_{12}}$}
	\psfrag{c}[Bc][Bc]{$\sqrt{g_{22}}$}
	\psfrag{Z'_1}[Bc][Bc]{$z_1$}
	\psfrag{Z'_2}[Bc][Bc]{$z_2$}
	\psfrag{X_1}[Bc][Bc]{$x_1$}
	\psfrag{X_2}[Bc][Bc]{$x_2$}
	\psfrag{Y'_1}[Bc][Bc]{$y_1$}
	\psfrag{Y'_2}[Bc][Bc]{$y_2$}
	\psfrag{w1}[Bc][Bc]{\begin{small}$\Uoo$\end{small}}
	\psfrag{w2}[Bc][Bc]{\begin{small}$\left(\Uto,\Utt,\Utr\right)$\end{small}}
	\psfrag{w3}[Bc][Bc]{\begin{small}$\left(\hUoo, \hUto, \hUtt \right)$\end{small}}
	\psfrag{w4}[Bc][Bc]{\begin{small}$\left(\hUto,\hUtr\right)$\end{small}}
	\includegraphics[height=4cm]{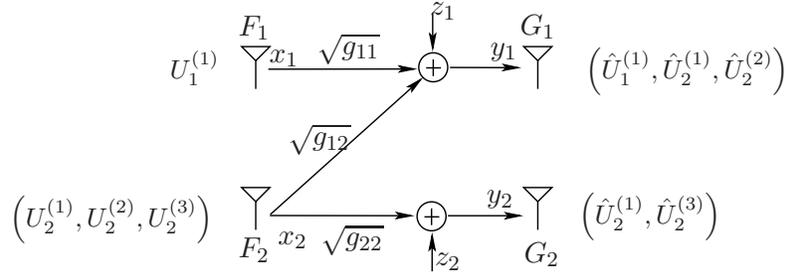}
\end{center}
\vspace{-4mm}
\caption{The $\Z$ interference channel with particular message
  requirements, captures the proposed coding scheme for the first layer of the
  Gaussian $\ZS$ network.}
\label{fig:GZS:Z}
\vspace{-4mm}
\end{figure}

The coding strategy for the Gaussian $\ZS$ network is essentially a
partial-decode-and-forward strategy, along with a strategic
rate-splitting of the messages. Let the messages to be sent from
$S_1,S_2$ be denoted by $W_1,W_2$ respectively (see
Figure~\ref{fig:GZS}). We will break the $\ZS$ network into two
cascaded interference channels, where we require particular messages
to be decoded at the relays and forwarded to the destinations. The 
first stage is a $\Z$ interference channel, where the message $W_2$ 
is split into three
parts: $\left(\Uto,\Utt,\Utr\right)$. The intention of this strategic
split is to allow the the node $G_1$ (which is relay $A$ in the
original $\ZS$ network) to decode $\left(\Uoo, \Uto, \Utt \right)$ and
node $G_2$ (which is relay $B$ in the original $\ZS$ network), to
decode $\left(\Uto,\Utr\right)$. This is illustrated in Figure
\ref{fig:GZS:Z}. Here, $\Uto$ plays the role of a common message which can 
be decoded at both receivers, whereas $\Utt$ and $\Utr$ are the private messages 
for $G_1$ and $G_2$ respectively.

\begin{figure}[th]
\begin{center}
  \psfrag{s1}[Bc][Bc]{$F_1$} \psfrag{s2}[Bc][Bc]{$F_2$}
  \psfrag{r1}[Bc][Bc]{$G_1$} \psfrag{r2}[Bc][Bc]{$G_2$}
  \psfrag{a}[Bc][Bc]{$\sqrt{h_{11}}$}
  \psfrag{f}[Bc][Bc]{$\sqrt{h_{21}}$}
  \psfrag{c}[Bc][Bc]{$\sqrt{h_{22}}$} \psfrag{Z'_1}[Bc][Bc]{$z_1$}
  \psfrag{Z'_2}[Bc][Bc]{$z_2$} \psfrag{X_1}[Bc][Bc]{$x_1$}
  \psfrag{X_2}[Bc][Bc]{$x_2$} \psfrag{Y'_1}[Bc][Bc]{$y_1$}
  \psfrag{Y'_2}[Bc][Bc]{$y_2$}
  \psfrag{w1}[Bc][Bc]{\begin{small}$\left(\Voo,\Vot,\Vto,\Vtt,\Vtr,\Vtf\right)$\end{small}}
  \psfrag{w2}[Bc][Bc]{\begin{small}$\left(\Vto,\Vtt,\Vtv\right)$\end{small}}
  \psfrag{w3}[Bc][Bc]{\begin{small}$\left(\hVoo,\hVot, \hVto, \hVtr
      \right)$\end{small}}
  \psfrag{w4}[Bc][Bc]{\begin{small}$\left(\hVoo, \hVto,\hVtt,
        \hVtr,\hVtf,\hVtv\right)$\end{small}}
	\includegraphics[height=4cm]{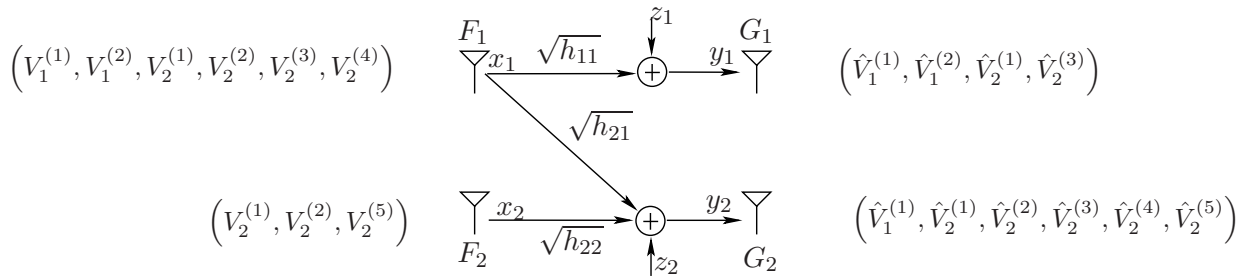}
\end{center}
\vspace{-4mm}
\caption{The $\S$ interference channel with particular message
  requirements, depicting the proposed coding strategy for the second layer of
  the Gaussian $\ZS$ network.}
\label{fig:GZS:S}
\vspace{-4mm}
\end{figure}

The next stage of the $\ZS$ network is a $\S$
interference channel depicted in Figure \ref{fig:GZS:S}. Here we take
the messages delivered and decoded by the $\Z$ interference channel of
the first stage and further process them to ensure delivery of the
desired messages to the destination. In particular, we further split
the decoded messages from the first stage into several parts and
require delivery of messages as shown in Figure \ref{fig:GZS:S}. This
splitting and delivery of appropriate pieces, finally ensures that
$W_1$ and $W_2$ are decodable at the destinations. This is the encoding
strategy in the $\ZS$ network. In the following lemmas, we give the
rates at which messages at each stage can be delivered. Putting
together Lemmas \ref{lm:GZS:Z} and \ref{lm:GZS:S}, we get the desired
result given in Theorem \ref{thm:gzs}. The proofs of these lemmas
follow fairly standard arguments, and are given in Appendix
\ref{sec:app-lm}.

A formal statement of the argument above is given below.

\begin{lemma}
  Consider the Gaussian $\Z$ interference network with channel gains $(g_{11},g_{12},
  g_{22})$, and decoding requirements as shown in
  Figure~\ref{fig:GZS:Z}.  Denoting the rate of the sub-message $U_i^{(j)}$ by $\Upsilon_{i,j}$,
  any rate tuple $(\Too,\Tto,\Ttt,\Ttr)$ which satisfies
\begin{align}
\Too &\leq \+{\CGN{1+g_{11}}-\frac{1}{2}},\label{eq:GZ-1}\\
\Ttt &\leq  \+{\CGN{1+\frac{g_{12}}{g_{22}} }-\frac{1}{2}},\label{eq:GZ-2}\\
\Tto+\Ttt &\leq  \+{\CGN{1+g_{12}}-\frac{1}{2}},\label{eq:GZ-3}\\
\Too +\Tto+\Ttt & \leq \+{\CGN{1+g_{11}+g_{12}}-\frac{1}{2}},\label{eq:GZ-4}\\
\Ttr &\leq \+{\CGN{1+\frac{g_{22}}{g_{12}} }-\frac{1}{2}},\label{eq:GZ-5}\\
\Tto+\Ttr & \leq \+{\CGN{1+g_{22}}-\frac{1}{2}},\label{eq:GZ-6}
\end{align}
 is achievable.
 \label{lm:GZS:Z}
\end{lemma}

The next lemma gives an achievable rate region for the second layer of the $\ZS$ network,
which is a $\S$ interference network depicted in Figure~\ref{fig:GZS:S}.

\begin{lemma}
  Consider the Gaussian $\S$ interference network with channel gains $(h_{11},h_{21},
  h_{22})$, and decoding requirements as shown in
  Figure~\ref{fig:GZS:S}, where $\Theta_{i,j}$ denotes the rate of 
  message $V_i^{(j)}$. Any rate tuple
  $(\Qoo,\Qot,\Qto,\Qtt,\Qtr,\Qtf,\Qtv)$ which satisfies
\begin{align}
\Qoo+\Qot+\Qto+\Qtr &\leq \+{\CGN{1+h_{11}} -\frac{1}{2}},\label{eq:GS-1}\\
\Qot &\leq \+{\CGN{1+\frac{h_{11}}{h_{12}} }-\frac{1}{2}},\label{eq:GS-2}\\
\Qtf  &\leq \+{\CGN{1+\frac{h_{21}}{h_{11}}}-\frac{1}{2}},\label{eq:GS-3}\\
\Qoo+\Qtr+\Qtf  &\leq \+{\CGN{1+h_{21}}-\frac{1}{2}},\label{eq:GS-4}\\
\Qtv &\leq \+{\CGN{1+h_{22}}-\frac{1}{2}},\label{eq:GS-5}\\
\Qoo+\Qto+\Qtt+\Qtr+\Qtf+\Qtv  &\leq \+{\CGN{1+h_{21}+h_{22}}-\frac{1}{2}},\label{eq:GS-6}
\end{align}
 is achievable.
 \label{lm:GZS:S}
\end{lemma}

\subsection{The Gaussian $\ZZ$ network: Achievability}
The encoding scheme needed for the $\ZZ$ network is slightly more
sophisticated than the $\ZS$ network. An
additional component to strategic message splitting is that of
interference neutralization.  This was illustrated in examples
\ref{ex:DetIntNeut} and \ref{ex:LatticeIntNeut} in Section
\ref{sec:examples}. This along with message splitting 
inspired by the network decomposition illustrated in example
\ref{ex:NetPart} of Section \ref{sec:examples}, form the basis of the
encoding scheme for the $\ZZ$ network.

More formally, the interference that has to be neutralized, will be
combined with the main message in the first layer according to some
partial-invertible function. In the second layer the inverse of
the function is applied on this combination and the other interference
received through the cross link. The remaining parts of the
interference has to be either decoded or treated as noise. The
neutralization is implemented using lattice codes and the
rate-splitting along with appropriate power allocation is also used.

We formally define a partial-invertible function and a
$\Z$-\emph{neutralization} network in the following. The Gaussian
$\ZZ$ network is essentially a cascade of two $\Z$-neutralization
networks. An achievable rate region for the $\Z$-\emph{neutralization}
network is given in Lemma~\ref{lm:GZZ:Z}. This rate region will be
later used to obtain an achievable rate region for the Gaussian $\ZZ$
network. We will analyze the performance of the Gaussian encoding/decoding schemes 
 in Appendix~\ref{subsec:gzz-ach}.

\begin{definition}
  Let $\mathcal{U}$ and $\mathcal{V}$ be two finite sets.  A function
  $\phi(\cdot,\cdot)$ defined on $\mathcal{U}\times\mathcal{V}$ is
  called \emph{partial-invertible}, if and only if having $\phi(u,v)$
  and $u$, one can always reconstruct $v$ for any $u\in\mathcal{U}$
  and $v\in\mathcal{V}$. Similarly, $u$ can be obtained from
  $\phi(u,v)$ and $v$.
\end{definition}
An intuitive way of thinking about a partial-invertible $\phi(u,v)$ is
the following. An arbitrary function defined on a finite sets
$\mathcal{U}$ and $\mathcal{V}$ creates a table with rows corresponding
to the elements of $\mathcal{U}$ and columns corresponding to the
elements of $\mathcal{V}$, the each cell of the table consists the
value assigned to its row and column by the function. A function will
be partial-invertible, if and only if no two cells in the same column
or row of its table be identical.

Note that summation over real numbers, and multiplication over
non-zero numbers are two examples of partial-invertible
functions. However, it is clear multiplication over real numbers is
not partial-invertible, since $w=\phi(1,0)=\phi(2,0)$, and therefore
having $w$ and $v=0$, $u$ can be anything.

\begin{definition}
  Consider the $\Z$ network shown in Fig~\ref{fig:GZZ:Z}, which
  consists of a Gaussian broadcast channel from $F_2$ to the receivers
  and a Gaussian multiple access channel from $F_1$ and $F_2$ to
  $G_1$.
\begin{figure}[th]
\begin{center}
 	\psfrag{s1}[Bc][Bc]{$F_1$}
	\psfrag{s2}[Bc][Bc]{$F_2$}
	\psfrag{r1}[Bc][Bc]{$G_1$}
	\psfrag{r2}[Bc][Bc]{$G_2$}
	\psfrag{a}[Bc][Bc]{$\sqrt{g_{11}}$}
	\psfrag{f}[Bc][Bc]{$\sqrt{g_{12}}$}
	\psfrag{c}[Bc][Bc]{$\sqrt{g_{22}}$}
	\psfrag{Z'_1}[Bc][Bc]{$z_1$}
	\psfrag{Z'_2}[Bc][Bc]{$z_2$}
	\psfrag{X_1}[Bc][Bc]{$x_1$}
	\psfrag{X_2}[Bc][Bc]{$x_2$}
	\psfrag{Y'_1}[Bc][Bc]{$y_1$}
	\psfrag{Y'_2}[Bc][Bc]{$y_2$}
	\psfrag{w1}[Bc][Bc]{$\left(U_1^{(0)},U_1^{(1)} \right)$}
	\psfrag{w2}[Bc][Bc]{$\left(U_2^{(0)} , U_2^{(1)}\right)$}
	\psfrag{w3}[Bc][Bc]{$\left(\hat{U}_1^{(1)}, \hat{\phi}(U_1^{(0)}, U_2^{(0)}) \right)$}
	\psfrag{w4}[Bc][Bc]{$\left(\hat{U}_2^{(0)} , \hat{U}_2^{(1)}\right)$}
	\includegraphics[height=4cm]{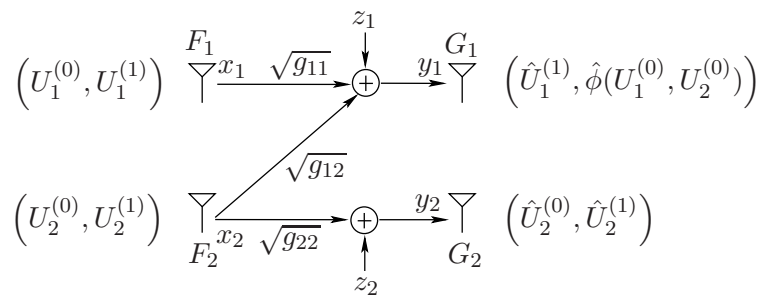}
\end{center}
\vspace{-4mm}
\caption{The Gaussian $Z$ channel. }
\label{fig:GZZ:Z}
\vspace{-4mm}
\end{figure}
A $\Z$-\emph{neutralization} network is a $\Z$ network, wherein
the first source node has two messages $(U_1^{(0)},U_1^{(1)} )$ of
rates $\Tz$ and $\To$, respectively.  Similarly the second
source observes two independent messages $(U_2^{(0)} , U_2^{(1)})$ of
rates $\Tz$ and $\Tt$.

The second receiver is interested in decoding $U_2^{(0)}$ and
$U_2^{(1)}$, while the first destination wishes to decode
$\phi(U_1^{(0)}, U_2^{(0)})$ and $U_1^{(1)}$, where
$\phi(\cdot,\cdot)$ can be any arbitrary partial-invertible function.
A rate tuple $(\Tz,\To,\Tt)$ is called achievable if the
receivers can decode their messages with arbitrary small error
probability.
\label{def:GZZ:Z}
\end{definition}

\begin{lemma}
Consider the $\Z$-neutralization network defined Definition~\ref{def:GZZ:Z}  with 
channel gains $(g_{11},g_{12},g_{22})$ (see Figure~\ref{fig:GZZ:Z}). Let
\begin{align}
\lambda\triangleq\min\{g_{11},g_{12},g_{22}\},
\end{align}
and
\begin{align}
\mu\triangleq\max\left\{g_{11},g_{12},g_{22},\frac{g_{11}g_{22}}{g_{12}}\right\}.
\end{align}
Any rate tuple $(\Tz,\To,\Tt)$ satisfying
\begin{align}
\Tz &\leq \+{\CGN{\lambda}-\frac{1}{2}},\label{gzz:z:1}\\
\Tz+\To &\leq \+{\CGN{g_{11}}-1},\label{gzz:z:2}\\
\Tz +\Tt &\leq \+{\CGN{g_{22}}-1},\label{gzz:z:3}\\
\Tz +\To + \Tt &\leq \+{\CGN{\mu}-\frac{3}{2}},\label{gzz:z:4}
\end{align}
is achievable.
\label{lm:GZZ:Z}
\end{lemma}

As mentioned before, we
strategically split the messages and require functional
reconstructions for some of them at the relay nodes to
facilitate neutralization at the destinations. More precisely, in the
first layer of the network, each source node splits its message into
two parts, namely, ``functional'' and private parts, $W_1=\left(
  U_1^{(0)}, U_1^{(1)}\right)$ and $W_2=\left( U_2^{(0)},
  U_2^{(1)}\right)$. The ``functional'' parts $U_1^{(0)}, U_2^{(0)}$
both have the same rates $\Tz$.  Both transmitters use
a common lattice code to encode their functional sub-messages. Now the
first layer encodes the message such that the first receiver 
(which is relay $A$ in the original $\ZZ$ network) can decode
$U_1^{(1)}$ and $\phi(U_1^{(0)},U_2^{(0)})$, and the second one (relay $B$ 
 in the original $\ZZ$ network) can decode
$U_2^{(0)}$ and $U_2^{(1)}$. Lemma~\ref{lm:GZZ:Z} gives the rates at
which these can be sent reliably. The second stage  operates in a
manner similar to the first stage, by splitting the messages into
functional and private parts. The first sender (relay $A$ in the original network) 
uses $U_1^{(1)}$ and $\phi(U_1^{(0)},U_2^{(0)})$ as the
private and functional parts and the other one (relay $B$) uses $U_2^{(1)}$ 
and $U_2^{(0)}$ as the private and functional parts. 

The functional parts are sent appropriately, using a common lattice code 
in both stages. Let $\x_{1}^{(N)}$ and $\x_{2}^{(N)}$ be the lattice codewords, 
corresponding to $U_1^{(0)}$ and $U_2^{(0)}$, respectively. The power allocation 
In the first layer it is done so that two lattice points get received at $A$ at the
same power (see Figure~\ref{fig:lattice}). The group structure of the lattice code implies that the summation 
of two received lattice point, $\tilde{\x}^{(N)}=\x_1^{(N)}+\x_{2}^{(N)}$ is 
still a valid codeword, and can be decoded by $A$.  
The function $\phi(\cdot,\cdot)$ is in fact the decoded message from $\tilde{\x}^{(N)}$. 
In the second stage, relay node $B$, sends the inverse of the the received lattice point, 
that is $\x_{2}^{'(N)}=-\x_{2}^{(N)}$, while $A$ forwards the sum lattice point, 
$\x_{1}^{'(N)}=\tilde{\x}^{(N)}$. Again these lattice points are scaled properly so that 
they get received at $D_1$ at the same power. Thus, their summation would be a lattice point 
and equals $\x_{1}^{'(N)} + \x_{2}^{'(N)} = (\x_1^{(N)}+\x_{2}^{(N)}) -\x_{2}^{(N)}= \x_1^{(N)}$, 
which will be decoded to $U_1^{(0)}$. The other destination $D_2$, receives $-\x_{2}^{(N)}$, 
finds its inverse $\x_{2}^{(N)}$, and finally decodes it to $U_2^{(0)}$. 
This idea is illustrated in Example~\ref{ex:LatticeIntNeut}, and the precise details of this
argument are given in Appendix \ref{subsec:gzz-ach}.

\section{The Deterministic $\ZS$ Network}
\label{sec:dzs}

In this section we prove Theorem~\ref{thm:dzs}. We study this problem
in two parts. First we present the converse proof, which shows any
achievable rate pair belongs to $\R{\DZS}{}$. Then for any rate pair
in this region, we propose an encoding scheme which is able to
transmit messages up to the desired rates.

\subsection{The Outer Bound}
\label{subsec:dzs:convrse}

In this section we show that any achievable rate pair
$(R_1,R_2)$ for the deterministic $\ZS$ network belongs to
$\R{\DZS}{}$. 
Assume there
exists a coding scheme with block length $\ell$ which can be used to
communicate at rates $R_1$ and $R_2$ over the network.
We use fold face matrices to denote $\ell$ copy of 
them, as the transfer matrix applied over a codeword of length $\ell$,
\emph{e.g.,} $\mathbf{M}_{11}= I_{\ell} \otimes M_{11}$.

All of the bounds in the theorem except \eqref{dzs:s1s2} and \eqref{dzs:d1d2}
can be obtained straight-forwardly
 using  the generalized cut-set bound in \cite{ADT07b}, which 
shows that in a linear finite-field network, the maximum reliable 
rate can be transmitted through a cut is upper bounded by the rank 
of the transition matrix of the cut. Here, we only 
present the proof of \eqref{dzs:s1s2r1} to illustrate this idea. Then we 
prove the two remaining bounds, which are tighter than the cut-set bound.

\paragraph*{\eqref{dzs:s1s2r1} $R_1+ R_2 \leq m_{22}+\max(n_{11},n_{21})$}

This bound corresponds to the cut $\Omega_s=\{S_1, S_2,A\}$ and
  $\Omega_d=\{ B, D_1, D_2\}$. The transition matrix from the input of 
the cut $X_{\Omega_s}=(X_2,X'_1)$ to its output $Y_{\Omega_d}=(Y'_2,Y_1,Y_2)$
can be written as
\begin{align}
\left[\begin{array}{c}
Y^{'\ell}_2\\
Y^{\ell}_1\\
Y^{\ell}_2
\end{array}\right] 
= 
\underbrace{
\left[\begin{array}{cc}
\mathbf{M}_{22} & \mathbf{0}\\
\mathbf{0} & \mathbf{N}_{11}\\
\mathbf{0} & \mathbf{N}_{21}
\end{array}\right] }_{\mathbf{G}_{\Omega_s,\Omega_d}}
\left[\begin{array}{c}
X^{\ell}_2\\
X^{'\ell}_1
\end{array}\right]
+
\left[\begin{array}{c}
\mathbf{0}\\
\mathbf{0} \\
\mathbf{N}_{22}
\end{array}\right]
X^{'\ell}_2. 
\end{align}

Therefore, from \cite{ADT07b} we have
\begin{align}
 \ell( R_1 + R_2) &\leq  \rank(\mathbf{G}_{\Omega_s,\Omega_d})
= \rank(\mathbf{M}_{22}) 
  + \rank \left(
\left[\begin{array}{c}
\mathbf{N}_{11}\\
\mathbf{N}_{21}
\end{array}\right]
\right)
=
\ell m_{22}+\ell \max(n_{11},n_{21}).
\end{align}

As mentioned before, we skip the proof of those bounds which follow from the generalized 
cut-set bound. In the  following  we present the proof of the two remaining inequalities 
which are tighter that the cut-set bound. 

\paragraph*{ \eqref{dzs:s1s2} $R_1+R_2 \leq \max(m_{11},m_{12})+(m_{22}-m_{12})^+$}

In order to prove this bound, we can start with 
\begin{align}
 \ell( R_1+R_2) &\leq I(X^{\ell}_1, X^{\ell}_2 ;  Y^{\ell}_1, Y^{\ell}_2 )\nonumber\\
  &\leq I( X^{\ell}_1, X^{\ell}_2 ;  Y^{'\ell}_1, Y^{'\ell}_2  )\label{dzs:s1s2-MC}\\
  & = I( X^{\ell}_1, X^{\ell}_2 ;  Y^{'\ell}_1 ) + I( X^{\ell}_1, X^{\ell}_2 ;  Y^{'\ell}_2 | Y^{'\ell}_1)\nonumber\\
  &\leq  I( X^{\ell}_1, X^{\ell}_2 ;  Y^{'\ell}_1 ) 
      + H(Y^{'\ell}_2  | Y^{'\ell}_1) - H(Y^{'\ell}_2 | X^{\ell}_1, X^{\ell}_2 , Y^{'\ell}_1),\label{fn}
\end{align}
where in \eqref{dzs:s1s2-MC} we used the data-processing inequality for the Markov chain 
\begin{align}
(X_1^{\ell},X_2^{\ell}) \leftrightarrow (Y_1^{'\ell},Y_2^{'\ell}) \leftrightarrow (X_1^{'\ell},X_2^{'\ell})\leftrightarrow (Y_1^{\ell},Y_2^{\ell}),
\label{eq:MC-layer}
\end{align}
and  (\ref{fn}) holds since $Y^{'\ell}_2$ is function of $X^{\ell}_2$. Now, it is clear that 
\begin{align}
 I( X^{\ell}_1, X^{\ell}_2 ;  Y^{\ell}_1 ) \leq \rank \left(\left[\begin{array}{cc}
\mathbf{M}_{11} & \mathbf{M}_{21}
\end{array}\right]\right) = \ell \max(m_{11},m_{12}). 
\label{dzs:s1s2-0}
\end{align}
In order to bound the second term, we can write
\begin{align}
  H(Y^{'\ell}_2 | Y^{'\ell}_1)&=  H(Y^{'\ell}_2 | Y^{'\ell}_1, X^{'\ell}_1, Y^{\ell}_1)\label{dzs:s1s2-1}\\
  & \leq H(Y^{'\ell}_2, W_2  | Y^{'\ell}_1, X^{'\ell}_1, Y^{\ell}_1) \nonumber\\
  &=  H(Y^{'\ell}_2  | Y^{'\ell}_1, X^{'\ell}_1, Y^{\ell}_1, W_2) +  H( W_2  | Y{'\ell}_1, X^{'\ell}_1, Y^{\ell}_1)\nonumber\\
  &\leq H(Y^{'\ell}_2  | Y^{'\ell}_1, X^{'\ell}_1, Y^{\ell}_1, W_2, X_1^{\ell}) +\ell \e_{\ell}\label{dzs:s1s2-2}\\
  &\leq H(Y^{'\ell}_2  | Y^{'\ell}_1 - \mathbf{M}_{11} X_1^{\ell}) +\ell \e_{\ell}\nonumber\\
  &\leq H(M_{22} X_2^{\ell}  | \mathbf{M}_{12} X_2^{\ell}) +\ell \e_{\ell}\nonumber\\
  &\leq \ell \rank \left( \left[ \begin{array}{c}
        M_{12}\\
        M_{22}
  \end{array}\right]\right) - \ell \rank \left(M_{12}\right)+\ell \e_{\ell}\nonumber\\
&= \ell (m_{22}-m_{12})^+ +\ell \e_{\ell},\label{dzs:s1s2-3}
\end{align}  
where \eqref{dzs:s1s2-1} holds since $X^{'\ell}_1$ is also a function of
$Y^{'\ell}_1$, and $Y_1^{\ell}$ is also a deterministic
function of $X^{'\ell}_1$.. We used Fano's inequality in \eqref{dzs:s1s2-2},
where $W_1$ should be decodable based on $Y_1^{\ell}$. 
Summing up \eqref{dzs:s1s2-0} and \eqref{dzs:s1s2-3}, we get the desired bound. 

Note that the cut-set bound for the cut $\Omega_s=\{S_1,S_2\}$ and $\Omega_d=\{A, B, D_1, D_2\}$ gives us
\begin{align}
\ell(R_1+R_2) \leq \rank \left(
\left[\begin{array}{cc}
\mathbf{M}_{11} & \mathbf{M}_{12}\\
\mathbf{0} & \mathbf{M}_{22}
\end{array}\right]
\right)
= \ell \max(m_{11}+m_{22} , m_{12}),
\end{align}
in which the RHS can be arbitrarily larger than the RHS of the presented bound. The reason for this difference is the following. 
It is inherently assumed in deriving the cut-set bound  that the receivers can cooperate to decode the messages of rates $R_1$
and $R_2$, and no decodability requirement is posed for individual receivers. However, the setup of this problem 
impose an extra constraint, that is $B$ alone should be able to decode  $W_2$. Incorporating this decodability requirement shrinks the 
set of admissible rates, and gives us a tighter bound.

\paragraph*{\eqref{dzs:d1d2} $R_1+R_2 \leq \max(n_{21},n_{11}) + (n_{11}-n_{21})^+$}

  The last inequality captures the maximum flow of information from the relays 
  to the destinations, such that $D_1$ and $D_2$ be able to decode $W_1$ and $W_2$, respectively. 
We again start with 
\begin{align}
 \ell( R_1+R_2) &\leq I(X^{\ell}_1, X^{\ell}_2 ;  Y^{\ell}_1, Y^{\ell}_2 )
  = H(   Y^{\ell}_1, Y^{\ell}_2  )
  = H( Y^{\ell}_2 ) + H(Y^{\ell}_1  | Y^{\ell}_2).\label{dzs:d1d2-0}
\end{align}
The first term can be easily bounded by 
\begin{align}
  H(Y_2^{\ell}) \leq \rank \left(\left[\begin{array}{cc} \mathbf{N}_{21} &
        \mathbf{N}_{22} \end{array}\right]\right)=
  \ell \max(n_{21},n_{22}).\label{dzs:d1d2-1}
\end{align}
In order to bound the second term, we use the fact that $W_2$ can be
decoded from $Y_2^{\ell}$. Therefore,
\begin{align}
  H(Y^{\ell}_1 | Y^{\ell}_2 ) &\leq H(Y^{\ell}_1, W_2| Y^{\ell}_2 )\nonumber\\
  &= H(Y^{\ell}_1| Y^{\ell}_2, W_2 )  +  H(W_2| Y^{\ell}_2 )\nonumber\\
  &\leq  H(Y^{\ell}_1 | Y^{\ell}_2, W_2 ) + \ell \e_{\ell}\label{dzs:d1d2-2}\\
  &=  H(Y^{\ell}_1 | Y^{\ell}_2, W_2, X_2^{\ell}, Y^{'\ell}_2, X^{'\ell}_2 ) + \ell \e_{\ell}\nonumber\\
  &\leq  H(Y^{\ell}_1 | Y^{\ell}_2- \mathbf{N}_{22} X^{'\ell}_2 ) + \ell \e_{\ell}\nonumber\\
  &=  H(N_{11} X^{'\ell}_1 | \mathbf{N}_{21} X^{'\ell}_1 ) + \ell \e_{\ell}\nonumber\\
  &\leq \ell \rank \left( \left[ \begin{array}{c}
        N_{11}\\
        N_{21}
  \end{array}\right]\right) - \ell \rank \left(N_{21}\right)+ \ell \e_{\ell}\nonumber\\
&= \ell (n_{11}-n_{21})^+ + \ell \e_{\ell}\label{dzs:d1d2-3}.
\end{align}
In \eqref{dzs:d1d2-2} we used the Fano's inequality, as well as the fact that   $X_2^{\ell}$, $Y^{'\ell}_2$, 
and $X^{'\ell}_2$ are known having $W_2$. The bound is obtained by replacing \eqref{dzs:d1d2-1} and
\eqref{dzs:d1d2-3} in \eqref{dzs:d1d2-0}.

  It is worth mentioning that this bound is tighter than the cut-set bound for the cut 
  $\Omega_s=\{S_1,S_2, A, B\}$ and $\Omega_d=\{D_1,  D_2\}$, which is 
  \begin{align}
  R_1+R_2 \leq \max(n_{11}+n_{22},n_{12}).
  \end{align}

\subsection{The Achievability Part}
\label{subsec:DZS-Ach}

\paragraph*{Network Decomposition:}
The achievability scheme presented here is based on decomposition of
the deterministic $\ZS$ network into two node-disjoint networks. In
fact, such partitioning depends on the demanded rate pair
$(R_1,R_2)\in\R{\DZS}{}$.  The resulting family of separations
immediately suggests a simple coding scheme. We will show that this
separation is optimal, and does not cause any loss in the admissible
rate region of the network.

Before introducing the network decomposition, we define an equivalence class
for the sub-nodes (levels) in a network.

\begin{definition}
  In a $\Z$ (or $\S$) deterministic network, two sub-nodes $a$ and $b$ are
  called \emph{related} sub-nodes, and denoted by $a\sim b$ if any of the following conditions hold:
\begin{itemize}
\item $a=b$;
\item $a$ is connected to $b$; 
\item $b$ is connected to $a$;
\item there exists a sub-node $c$ such that $c$ broadcasts to both $a$ and $b$;
\item there exists a sub-node $d$ where both $a$ and $b$ are connected to. 
\end{itemize}
Note that this relation is reflective, symmetric, and
transitive. Therefore, it forms equivalence classes for the sub-nodes.
\label{def:node-related}
\end{definition}

We denote by $\N_1$ and $\N_2$ the partitions of the network. Assume we
wish transmitting at rate $R_1=r\leq \min(m_{11},n_{11})$ from $S_1$ to
$D_1$. The first part of the network $\N_1$, includes the top
$(m_{11}-m_{12})^+$ levels as well as the lowest
$(r-(m_{11}-m_{12})^+)^+$ levels of $S_1$. It also includes all the
related sub-nodes of $S_2$, and the receiver levels of $A$ and
$B$. Similarly, in the second layer of the network, $\N_1$ includes
the lowest $(n_{11}-n_{21})^+$ levels as well as the top
$(r-(n_{11}-n_{21})^+)^+$ nodes of the transmitter part of $A$. All related 
sub-nodes of the transmitter part of $B$, as well as $D_1$
and $D_2$ also belong to $\N_1$. The second part of the network $\N_2$, is 
formed by all the remaining nodes. 

We will use $\N_1$ for transmitting data from $S_1$ to $D_1$. Similarly $\N_2$
is only used to communicate from $S_2$ to $D_2$.  Therefore, we have two
uni-cast networks, and each pair of transmitter-receiver can
communicate up to the capacity of their own partition, which is the
min-cut of the partition \cite{ADT07a}.

It is worth mentioning that any two ``related'' sub-nodes belong to the
same partition. Therefore, these two networks are node-disjoint, and do not cause 
interference for each other. This allows us to derive the capacity of each 
network separately, and argue that $(R_1,R_2)$ can be achieved simultaneously
for the original network, if $R_1$ and $R_2$ are achievable for partitions $\N_1$ 
and $\N_2$.

\paragraph*{Encoding Scheme}

A transmission from $S_1$ and $S_2$ to $D_1$ and $D_2$ is performed as
follows. $S_1$ transmits only on its sub-nodes which belong to
$\N_1$, and keeps its other sub-nodes silent. Similarly, $S_2$ encodes its
message on the sub-nodes included in $\N_2$, and sends zero on the
other levels. Therefore, the \emph{effective} communication over each
partition is a simple uni-cast.
\begin{figure}[hbtp]
\centering
\subfigure[Effective channel for $(S_1,D_1)$.]{\label{fig:zs-eff-1}
	\psfrag{s1}[Bc][Bc]{$S_1$}
	\psfrag{r1}[Bc][Bc]{$A$}
	\psfrag{d1}[Bc][Bc]{$D_1$}
	\psfrag{a}[Bc][Bc]{$r$}
	\psfrag{b}[Bc][Bc]{$r$}
	\psfrag{X_1}[Bc][Bc]{$X_1$}
	\psfrag{Y'_1}[Bc][Bc]{$Y^{'(1)}_1$}
	\psfrag{X'_1}[Bc][Bc]{$X^{'(1)}_1$}
	\psfrag{Y_1}[Bc][Bc]{$Y_1$}
\includegraphics[width=85mm]{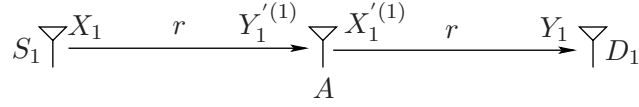}}
\hspace{3mm}
\subfigure[Effective channel for $(S_2,D_2)$.]{\label{fig:zs-eff-2}
	\psfrag{s2}[Bc][Bc]{$S_2$}
	\psfrag{r1}[Bc][Bc]{$A$}
	\psfrag{r2}[Bc][Bc]{$B$}
	\psfrag{d2}[Bc][Bc]{$D_2$}
	\psfrag{c}[Bc][Bc]{$m'_{22}(r)$}
	\psfrag{d}[Bc][Bc]{$n'_{22}(r)$}
	\psfrag{f}[Bc][Bc]{$m'_{12}(r)$}
	\psfrag{g}[Bc][Bc]{$n'_{21}(r)$}
	\psfrag{X_2}[Bc][Bc]{$X_2$}
	\psfrag{Y'_1}[Bc][Bc]{$Y^{'(2)}_1$}
	\psfrag{X'_1}[Bc][Bc]{$X^{'(2)}_1$}
	\psfrag{Y'_2}[Bc][Bc]{$Y^{'(2)}_2$}
	\psfrag{X'_2}[Bc][Bc]{$X^{'(2)}_2$}
	\psfrag{Y_2}[Bc][Bc]{$Y_2$}
\includegraphics[width=85mm]{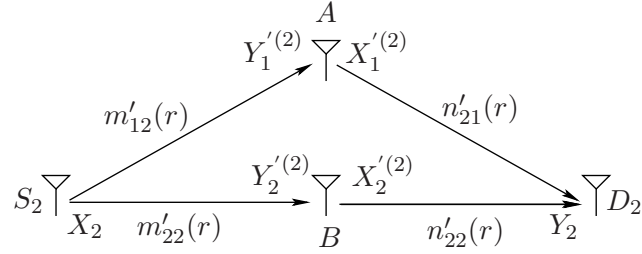}}
\caption{The effective separated $\ZS$ network.}
\label{fig:ZS-parts}
\end{figure}

Fig.~\ref{fig:ZS-parts} shows the effective parts of the network. It
is easy to see that the diamond network in Figure~\ref{fig:zs-eff-2}
is also a linear shift deterministic networks, with channel gains

\begin{align}
  m'_{12}(r)&=\min(\max(m_{11},m_{12}) -r, m_{12}),\\
  m'_{22}(r)&=\min(\max(m_{11},m_{12}) +(m_{22}-m_{12})^+ -r, m_{22}),\\
  n'_{21}(r)&=\min(\max(n_{11},n_{21}) -r, n_{21}),\\
  n'_{22}(r)&=\min(\max(n_{11},n_{21}) +(n_{22}-n_{21})^+ -r, n_{22}).
\end{align}

\paragraph*{Achievable Rate Region}
The cut values of $\N_1$ can be easily computed as
\begin{align*}
  \Omega=\{S_1\}: & & (m_{11}-m_{12})^+ + (r-(m_{11}-m_{12})^+)^+ = \max\{(m_{11}-m_{12})^+,r\}\geq r\\
  \Omega=\{S_1,A\}: & & (r-(n_{11}-n_{12})^+)^+ + (n_{11}-n_{12})^+ =
  \max\{(n_{11}-n_{12})^+,r\}\geq r.
\end{align*}
Therefore any rate in $\R{\DZS}{1}(r)=\{R_1: R_1\leq r\}$ can be
conveyed from $S_1$ to $D_1$ through $\N_1$.

The capacity of $\N_2$ can be found using the generalized max-flow min-cut
theorem \cite{ADT07a}. Hence, the rate region of the second partition $\N_2$  would
be
\begin{align}
\R{\DZS}{2}(r) =\{R_2: 
& R_2 \leq \max(m'_{12}(r), m'_{22}(r)),\\
&R_2 \leq m'_{22}(r)+n'_{21}(r),\\
&R_2 \leq m'_{12}(r)+n'_{22}(r),\\
&R_2 \leq \max(n'_{21}(r),n'_{22}(r)) \}.
\end{align}
Therefore, by  using this decomposition, any rate pair in the set
$\R{\DZS}{1}(r) \times \R{\DZS}{2}(r)=\{(R_1,R_2):R_1\in
\R{\DZS}{1}(r),R_2\in\R{\DZS}{2}(r)\}$ can be achieved.  It remains to
prove the following lemma.

\begin{lemma}
  For any deterministic $\ZS$ network,
\begin{align}
\R{\DZS}{} \subseteq \bigcup_{r\leq \min(m_{11},n_{11})}  \left(\R{\DZS}{1}(r) \times \R{\DZS}{2}(r)\right). 
\end{align}
\label{prop:dzs}
\end{lemma}
We will prove this lemma in Appendix~\ref{sec:app-lm}.

\section{The Deterministic $\ZZ$ Network}
\label{sec:dzz}

In this section we prove Theorem~\ref{thm:dzz}. This is done in two
parts, that provide the converse and achievability proofs.

\subsection{The Outer Bound}
\label{subsec:dzz:convrse}

In the following we will show that any achievable rate pair
$(R_1,R_2)$ satisfies constraints \eqref{dzz:s1}-\eqref{dzz:d1d2}.  
The individual rate bounds can be directly obtained by the generalized 
cut-set bound introduced in \cite{ADT07b}, where the maximum flow of 
information through a cut in a linear deterministic network is upper bounded 
by the rank of the transition matrix from the sender part of the cut to its receiver part. 
Hence, we skip the proofs of \eqref{dzz:s1}-\eqref{dzz:d2}. 

The sum-rate bounds in \eqref{dzz:s1s2}-\eqref{dzz:d1d2} are, however, 
genie-aided bounds which are tighter that the cut-set bounds. In the following, 
we focus on these two bounds, and present their proofs in detail.  
 Again we assume that there 
exists a coding scheme with block length $\ell$ which can be used to
communicate at rates $R_1$ and $R_2$ over the network.

\paragraph*{ \eqref{dzz:s1s2} $ R_1+R_2 \leq \max(m_{11},m_{12})  + (m_{22}-m_{12})^+ + n_{12}$}

  In order to prove this inequality we focus on the flow of information 
  from the sources to the relays.  The key idea here is to provide $A$ 
  with the information
  sent by $B$ to $D_1$ as side information. In such condition, the
  information $A$ has received about $W_1$ is stronger than the information
  available at $D_1$, and therefore $A$ can decode $W_1$ since $D_1$ can as well.  
  Once $W_1$ is decoded at $A$, it can determine the transmitted codeword from $S_1$. 
  By removing the
  interference from $S_1$, $A$ can also partially decode $W_2$. 
  
  More  precisely, we can write
\begin{align}
 \ell( R_1+R_2 ) &\leq I( X^{\ell}_1, X^{\ell}_2; Y^{'\ell}_1, Y^{'\ell}_2) = H(Y^{'\ell}_1, Y^{'\ell}_2) \leq  H(Y^{'\ell}_1, Y^{'\ell}_2, \Gamma^{\ell}_2) \nonumber\\
  &=  H(Y^{'\ell}_1, \Gamma^{\ell}_2) + H(Y^{'\ell}_2  | Y^{'\ell}_1, \Gamma^{\ell}_2) \nonumber\\
  &\leq H(Y^{'\ell}_1) + H( \Gamma^{\ell}_2) + H(Y^{'\ell}_2 | Y^{'\ell}_1,
  \Gamma^{\ell}_2), \label{zz:fn}
\end{align}
where $\Gamma^{\ell}_2=\mathbf{N}_{12} X^{'\ell}_2$ is the part of the signal received at
$D_2$ from $B$ as in Figure~\ref{fig:DZZ}.  The first two terms are easily bounded by 
$\ell \max(m_{11},m_{12})$ and $\ell n_{12}$, respectively. Deriving an upper bound for the last term is 
more involved. 


Similar to $\Gamma\l_2$, we define
$\Gamma^{\ell}_1=\mathbf{M}_{12}X^{\ell}_2$, where we have
\begin{align}
H(\Gamma^{\ell}_{1}| Y^{'\ell}_1, \Gamma\l_{2}) &= H(  Y^{'\ell}_1-\mathbf{M}_{11} X^{\ell}_{1} | Y^{'\ell}_1, \Gamma\l_{2}) \label{zz:inv}\\
&\leq  H(X^{\ell}_{1}| Y^{'\ell}_1, \Gamma\l_{2}) \nonumber\\
&\leq H(W_1 | Y^{'\ell}_1, \Gamma\l_{2})\nonumber\\
&= H(W_1 | Y^{'\ell}_1, X^{'\ell}_{1}, \Gamma\l_{2})\nonumber\\
&\leq H(W_1 | \mathbf{N}_{11} X^{'\ell}_{1}+ \Gamma\l_{2} )\nonumber\\
&= H(W_1 | Y\l_1) \leq \ell\e_{\ell}, \label{zz:fano1}
\end{align}
where $\e_{\ell} \rightarrow 0$ as $\ell$ grows. We have used the
invertibility property of the deterministic multiple access channel in (\ref{zz:inv}), and
(\ref{zz:fano1}) follows from the Fano's inequality, and the fact that
$D_1$ can decode the message sent by $S_1$.  Therefore, we have
$H(\Gamma\l_{1}| Y^{'\ell}_1, \Gamma\l_{2})\leq \ell \e_{\ell}$. Hence,
\begin{align}
H(Y^{'\ell}_2 | Y^{'\ell}_1, \Gamma\l_{2}) &\leq H(Y^{'\ell}_2 , \Gamma\l_{1} | Y^{'\ell}_1, \Gamma\l_{2})\nonumber\\
& =  H(Y^{'\ell}_2 | \Gamma\l_{1}, Y^{'\ell}_1, \Gamma_{2}) + H(\Gamma\l_{1}| Y^{'\ell}_1, \Gamma\l_{2}) \nonumber\\
&\leq H(Y^{'\ell}_2 | \Gamma\l_{1}) + \ell\e_{\ell}\nonumber\\
&= H(\mathbf{M}_{22} X^{\ell}_2 | \mathbf{M}_{12} X^{\ell}_2) + \ell\e_{\ell}\nonumber\\
&\leq \ell (m_{22}-m_{12})^+ + \ell \e_{\ell}.
\end{align}
Replacing the upper bounds for each term in \eqref{zz:fn}, we get
\begin{align}
R_1+R_2 
\leq \max(m_{11},m_{12}) + n_{12} + (m_{22}-m_{12})^+.
\end{align}

It is worth mentioning that the cut-set bound for $\Omega_s=\{S_1, S_2\}$ and
  $\Omega_d=\{A, B, D_1, D_2\}$ gives us
  \begin{align}
  R_1+R_2 \leq \max(m_{11}+m_{22},m_{12}),
  \end{align}
which is looser than the genie-aided bound. 

\paragraph*{ \eqref{dzz:d1d2} $ R_1+R_2 \leq \max(n_{11},n_{12})+(n_{22}-n_{12})^+  +m_{12}$}

  The last inequality captures the maximum flow of information from 
  the relays to the destinations. 
 Intuitively,  this inequality says that the number of
  interfering bits can get neutralized at $D_1$ cannot exceed the minimum of $m_{12}$ and
  $n_{12}$.
 In order to make this intuition formal, we provide $\Gamma\l_1$, the partial information
  about $W_2$ which is available at $A$, as side information for
  $D_1$. We then have 
\begin{align}
\ell(R_1+R_2) & \leq I( Y\l_1, Y\l_2 ; X\l_1 , X\l_2) = H( Y\l_1, Y\l_2 ) \nonumber\\
&\leq H(Y\l_1, Y\l_2, \Gamma\l_{1})\nonumber\\
&\leq H(Y\l_1) + H(\Gamma\l_{1}) + H(Y\l_2| Y\l_1, \Gamma\l_{1} ).\nonumber
\end{align}
Again, we can simply upper bound the first two terms by the rank of the corresponding matrices. 
In order to bound the last term, similar to the proof of (\ref{dzz:s1s2}), we use the following bounding
technique.
\begin{align}
H(\Gamma\l_{2} | Y\l_1 , \Gamma^{\ell}_{1}) &= H( Y\l_1- \mathbf{N}_{11} X^{'\ell}_1 | Y\l_1 , \Gamma\l_{1} )\nonumber\\
&\leq  H(  X^{'\ell}_1 | Y\l_1 , \Gamma\l_{1} )\nonumber\\
&\leq H(  Y^{'\ell}_1 | Y\l_1 , \Gamma\l_{1} ) \nonumber\\
&= H(  \mathbf{M}_{11}X\l_{1} + \Gamma\l_{1} | Y\l_1 , \Gamma\l_{1} ) \nonumber \\
&\leq H(  X\l_{1}  | Y\l_1 , \Gamma\l_{1} )\nonumber\\
&\leq H(  X\l_{1}  | Y\l_1  )\nonumber\\
&\leq  H(W_1| Y\l_1)\leq \ell\e_{\ell}, \label{zz:fano2}
\end{align}
where (\ref{zz:fano2}) follows from the Fano's inequality. This inequality can be used as
\begin{align}
H(Y\l_2| Y\l_1, \Gamma\l_{1} ) &\leq H(Y\l_2, \Gamma\l_{2}| Y\l_1, \Gamma\l_{1})\nonumber\\
&= H(Y\l_2| \Gamma\l_{2}, Y\l_1, \Gamma\l_{1}) +H(\Gamma\l_{2} | Y\l_1 , \Gamma\l_{1})\nonumber\\
&\leq H(Y\l_2| \Gamma\l_{2}) + \ell \e_{\ell}\nonumber\\
&= H(\mathbf{N}_{22} X^{'\ell}_2 | \mathbf{N}_{12} X^{'\ell}_2) + \ell\e_{\ell}\nonumber\\
&\leq \ell (n_{22}-n_{12})^+ +\ell\e_{\ell}.
\end{align}
Therefore, we have
\begin{align}
R_1+R_2 \leq \max(n_{11},n_{12}) +m_{12}+(n_{22}-n_{12})^+.
\end{align}

Again, it is easy to show that this bound is tighter than the cut-set bound for
 $\Omega_s=\{S_1,S_2, A, B\}$ and $\Omega_d=\{D_1,  D_2\}$,
 \begin{align}
 R_1+R_2\leq \max(n_{11}+n_{22},n_{12}).
 \end{align}

This completes the proof of the converse part of Theorem~\ref{thm:dzz}.

\subsection{The Achievability Proof}
\label{subsec:dzz:ach}

In this part we will show that all rate pairs satisfying inequalities
(\ref{dzz:s1})-(\ref{dzz:d1d2}) are achievable. In particular, we
introduce a coding scheme which achieves such rates. Our coding strategy 
provides the interference neutralization at the destination. This is performed
by splitting the messages into two parts, namely \emph{private} and 
\emph{functional} parts. The private sub-messages can be decoded at the 
relays, and forwarded to the destinations. The functional sub-message 
of the second source can be also decoded at $B$. However, $A$ only receives 
a combination (\texttt{xor}) of the functional sub-messages, and cannot 
decode them. It only forwards such combination on proper (power) levels 
such that the interference caused by the functional sub-message of $S_2$
get neutralized over the second layer of the network, and $D_1$ can decode 
the sub-message of its interest.

Our analysis is based on characterizing the number of \emph{pure} and \emph{combined} bits 
can be sent through each layer of the network. In the following we focus on 
one layer of the network, and obtain an achievable rate region for these numbers. 
Next, we use this region to build the encoding scheme for the $\ZZ$ network, and obtain 
an achievable rate region, which matches with the outer bound.

\begin{figure}[th]
\begin{center}
\hspace{-5mm}
 	\psfrag{s1}[Bc][Bc]{\footnotesize{$F_1$}}
	\psfrag{s2}[Bc][Bc]{\footnotesize{$F_2$}}
	\psfrag{r1}[Bc][Bc]{\footnotesize{$G_1$}}
	\psfrag{r2}[Bc][Bc]{\footnotesize{$G_2$}}
	\psfrag{w1}[Bc][Bc]{$\{W_{1,N}(1),\dots,W_{1,N}(\Upsilon_0)\}$}
	\psfrag{w2}[Bc][Bc]{$\{W_{1,P}(1),\dots,W_{1,P}(\Upsilon_1)\}$}
	\psfrag{w3}[Bc][Bc]{$\{W_{2,N}(1),\dots,W_{2,N}(\Upsilon_0)\}$}
	\psfrag{w4}[Bc][Bc]{$\{W_{2,P}(1),\dots,W_{2,P}(\Upsilon_2)\}$}
	\psfrag{w5}[Bc][Bc]{$\{W_{1,N}(1)\oplus W_{2,N}(1),\dots,W_{1,N}(\Upsilon_0)\oplus W_{2,N}(\Upsilon_0)\}$}
	\psfrag{w6}[Bc][Bc]{$\{W_{1,P}(1),\dots,W_{1,P}(\Upsilon_1)\}$}
	\psfrag{w7}[Bc][Bc]{$\{W_{2,N}(1),\dots,W_{2,N}(\Upsilon_0)\}$}
	\psfrag{w8}[Bc][Bc]{$\{W_{2,P}(1),\dots,W_{2,P}(\Upsilon_2)\}$}
\includegraphics[height=5cm]{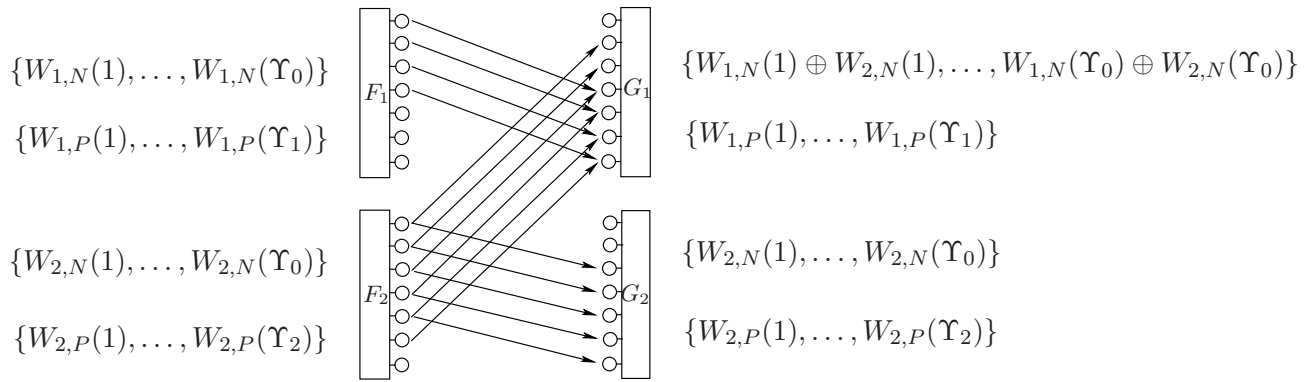}
\end{center}
\vspace{-4mm}
\caption{A deterministic $\Z$-neutralization network with the message demands.}
\label{fig:dz-model}
\vspace{-4mm}
\end{figure}

\begin{definition}
Consider a deterministic $\Z$ network, with gains $(n_{11},n_{12},n_{22})$.
as shown in Figure~\ref{fig:dz-model}. 
Each of the transmitters has a set of information 
bits to transmit to the receivers. This set for $F_i$ includes $\Upsilon_{i}$ private bits and 
$\Tz$ functional bits, namely, $\mathcal{W}_{i,P}=\{W_{i,P}(1),\dots,W_{i,P}(\Upsilon_i)\}$ and
$\mathcal{W}_{i,N}=\{W_{i,N}(1),\dots,X_{i,N}(\Tz)\}$. The second receiver wishes to receive all the
private and functional bits of $F_2$, while the first receiver is interested in receiving 
the private bits of $F_1$, and the \texttt{xor} of the functional bits of $F_1$ and $F_2$. 
More precisely, denoting by $\hat{\mathcal{W}}_i$ the set of bits $G_i$ is interested in, 
we have
\begin{align*}
 \hat{\mathcal{W}}_1 &= \mathcal{W}_{1,P} \cup \{\tilde{W}_{1,N}(j) \triangleq W_{1,N}(j) \oplus W_{2,N}(j): j=1,\dots,\Tz\},\nonumber\\
 \hat{\mathcal{W}}_2 &= \mathcal{W}_{2,P} \cup \mathcal{W}_{2,N}. 
\end{align*}
We term this network with the described decoding demands as deterministic $\Z$-neutralization network.
The goal is the characterize the set achievable tuples $(\Tz,\To,\Tt)$. 
\label{def:det-Z-neut}
\end{definition}

The following lemma gives an achievable rate region for the deterministic 
$\Z$-neutralization network. The proof of this lemma can be found in 
Appendix~\ref{sec:app-lm}.

\begin{lemma}
Consider the deterministic $\Z$-neutralization network defined in 
Definition~\ref{def:det-Z-neut} with channel gains $(n_{11},n_{12},n_{22})$ (see Figure~\ref{fig:dz-model}).
Any rate tuple  $(\Tz,\To,\Tt)$ satisfying
\begin{align}
\Tz &\leq \lambda\triangleq \min\{n_{11},n_{12},n_{22}\},\\
\Tz+\To &\leq n_{11},\\
\Tz+\Tt &\leq n_{22},\\
\Tz+\To+\Tt &\leq \mu\triangleq\max\{n_{11},n_{12},n_{22}, n_{11}+n_{22}-n_{12} \}.
\end{align}
is achievable for this network.
\label{lm:dz-neut}
\end{lemma}

Now, having an achievable rate region for the deterministic $\Z$-neutralization network, we are ready to 
present the coding scheme and analyze its rate region for the $\ZZ$ network. 

Recall that the $\ZZ$ network
consists of two cascaded $\Z$ network. In first layer, the source nodes split their message into private 
and functional parts. They can send these parts to the relays as long as their rates belong to the
achievable rate region of the first layer given in Lemma~\ref{lm:dz-neut}. Once the relays receive these 
sub-messages, forward them to the destination nodes using the same scheme for the private and 
functional  sub-messages. This can be done if the rate tuple for the sub-messages satisfy the 
corresponding inequalities for the second layer as well. Note that functional bits received at the 
destination are $\tilde{W}_{1,N}(j) \oplus W_{2,N}(j)=[W_{1,N}(j)\oplus W_{2,N}(j)]\oplus W_{2,N}(j)=W_{1,N}(j)$.
Therefore, the interference of these bits get neutralized, and pure information bits will be received 
at the destination. 

The achievable rate region of this scheme is given by 
\begin{align}
  \R{\DZZ}{ach}=\Big\{(R_1,R_2):& \exists \Tz,\To,\Tt \geq 0, \nonumber\\
  & R_1=\Tz+\To,\nonumber\\
  & R_2=\Tz+\Tt,\nonumber\\
  & \Tz \leq \min \{\lambda_m,\lambda_n\},\nonumber\\
  & \Tz+\To \leq \min \{m_{11},n_{11}\},\nonumber\\
  & \Tz+\Tt \leq \min \{m_{22},n_{22}\},\nonumber\\
  & \Tz+\To +\Tt\leq \min \{\mu_{m},\mu_{n}\}\Big\}
\end{align}
Here we used subscripts $m$ and $n$ to denote $\lambda$ and  $\mu$ parameters of the 
first and the second layer of the network, respectively. Applying Fourier-Motzkin elimination 
on this set to project it on the $(R_1,R_2)$ plane, gives us the rate region claimed in the theorem.



\section{Discussion}
\label{sec:disc}

Interference management is perhaps the most fundamental open problem
in wireless networks. The recent progress in (approximate)
characterization of the interference channel capacity and the utility
of the deterministic approach inspired the questions studied in this
paper. Even though the interference-relay networks studied in this
work were special, they revealed several new features needed for
information transmission. In particular, the interference
neutralization and network flow decomposition techniques were
uncovered through the study of $\ZZ$ and $\ZS$ networks. We also saw
the importance of using structured lattice codes for interference
neutralization. Moreover, we believe that the neutralization technique
is robust to channel uncertainties and one could get partial
neutralization in such situations. This is a topic of ongoing work on
this topic. We also believe that the outer bounding techniques
developed in this work could have more general applicability in the
wireless multiple-unicast problem. The two-unicast problem in
arbitrary layered wireless networks would be a natural next step
arising out of our work. The deterministic approach for this problem
has already provided some interesting new techniques
\cite{MDFT:ITW09}. In summary we believe that the deterministic
approach is a promising methodology to make progress on the wireless
multiple-unicast problem.

\appendices
\def\thesubsectiondis{\Alph{section}.\arabic{subsection}}
\renewcommand{\theequation}{\Alph{section}.\arabic{equation}}

\section{The Gaussian $\ZS$ Network}
\label{sec:gzs}
\setcounter{equation}{0}

\subsection{The Outer Bound}
\label{subsec:gzs-conv}

\begin{figure}[h!]
 \centering
      	\psfrag{s1}[Bc][Bc]{$S_1$}
	\psfrag{s2}[Bc][Bc]{$S_2$}
	\psfrag{r1}[Bc][Bc]{$A$}
	\psfrag{r2}[Bc][Bc]{$B$}
	\psfrag{d1}[Bc][Bc]{$D_1$}
	\psfrag{d2}[Bc][Bc]{$D_2$}
	\psfrag{a}[Bc][Bc]{$\sqrt{g_{11}}$}
	\psfrag{b}[Bc][Bc]{$\sqrt{h_{11}}$}
	\psfrag{c}[Bc][Bc]{$\sqrt{g_{22}}$}
	\psfrag{d}[Bc][Bc]{$\sqrt{h_{22}}$}
	\psfrag{f}[Bc][Bc]{$\sqrt{g_{12}}$}
	\psfrag{g}[Bc][Bc]{$\sqrt{h_{21}}$}
 	\psfrag{X_1}[Bc][Bc]{$x_1$}
	\psfrag{X_2}[Bc][Bc]{$x_2$}
	\psfrag{Y'_1}[Bc][Bc]{$y'_1$}
	\psfrag{Y'_2}[Bc][Bc]{$y'_2$}
	\psfrag{X'_1}[Bc][Bc]{$x'_1$}
	\psfrag{X'_2}[Bc][Bc]{$x'_2$}
	\psfrag{Y_1}[Bc][Bc]{$y_1$}
	\psfrag{Y_2}[Bc][Bc]{$y_2$}
	\psfrag{Z'_1}[Bc][Bc]{$z'_1$}
	\psfrag{Z'_2}[Bc][Bc]{$z'_2$}
	\psfrag{Z_1}[Bc][Bc]{$z_1$}
	\psfrag{Z_2}[Bc][Bc]{$z_2$}
	\psfrag{t_1}[Bc][Bc]{$\ $}
	\psfrag{t_2}[Bc][Bc]{$\ $}
\includegraphics[width=9cm]{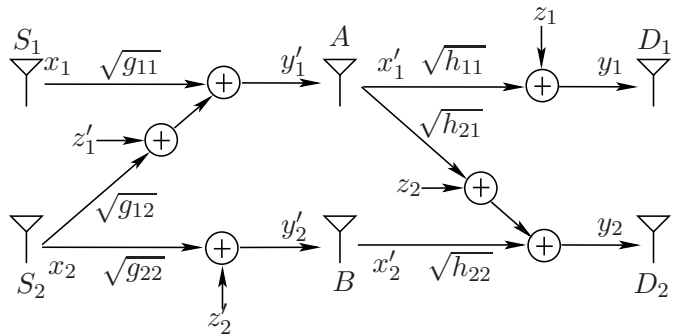}
\vspace{-4mm}
\caption{The Gaussian $\ZS$ network.}
\vspace{-4mm}
\label{fig:GZS:2}
\end{figure}

In the following we will prove each of the inequalities in (\ref{gzs:s1})-(\ref{gzs:d1d2}), separately. We will use
the notation as shown in Figure~\ref{fig:GZS:2}, and 
assume that the rate pair $(R_1,R_2)$ can be achieved with small enough decoding error probability using a code 
of length $\ell$.

\begin{lemma}
Any achievable rate pair $(R_1,R_2)$  satisfies
\begin{align}
\ell R_1 &\leq I(x_1^{\ell}; y_1^{\ell}) +\ell \e_{\ell},\label{eq:Fano:1}\\
\ell R_2 &\leq I(x_2^{\ell}; y_2^{\ell}) +\ell \e_{\ell},\label{eq:Fano:2}\\
\ell (R_1 +R_2) &\leq I(x_1^{\ell}, x_2^{\ell}; y_1^{\ell} , y_2^{\ell}) + \ell \e_{\ell}.\label{eq:Fano:12}
\end{align}
Note that $\e_{\ell}\rightarrow 0$ as $\ell$ grows.
\label{lm:gzs-fano}
\end{lemma}
This lemma is a consequence of the Fano's lemma combined with the decodability requirements imposed by the problem,
and its proof is given in  Appendix~\ref{sec:app-lm}. 

Most of the inequalities in (\ref{gzs:s1})-(\ref{gzs:d1d2}) are  cut-set type bounds, although the proof presented here 
are slightly different than the standard argument. However, the sum-rate bounds in \eqref{gzs:s1s2} and \eqref{gzs:d1d2}
are different from the well known cut-set bounds. These two bounds are in general tighter than the cut values 
for the corresponding cuts. This is because the decoders are inherently allowed to cooperate in deriving a cut-set bound, 
while individual decoding abilities are imposed in this problem. In the following we first present the proofs of 
\eqref{gzs:s1s2} and \eqref{gzs:d1d2}, which are more involved, and then prove the cut-set type bounds. 

\paragraph{The proofs of non-cut-set type bounds}
\begin{itemize}
 
\item (\ref{gzs:s1s2}) $\  R_1 +R_2 < \frac{1}{2}\log(1+ g_{11} + g_{12}) + \frac{1}{2}\log \left(1+\frac{g_{22}}{g_{12}}\right)$:
We start with Lemma~\ref{lm:gzs-fano} for the sum-rate which implies
\begin{align}
\ell (R_1 +R_2) & \leq I(x^{\ell}_1, x^{\ell}_2 ; y^{\ell}_1 , y^{\ell}_2) +\ell \e_{\ell} \nonumber\\
  &\leq  I(x^{\ell}_1, x^{\ell}_2 ; y^{'\ell}_1 , y^{'\ell}_2) +\ell \e_{\ell}\label{gzs:s1s2-0}\\
  &= I(x^{\ell}_1, x^{\ell}_2 ; y^{'\ell}_1 )+ I(x^{\ell}_1, x^{\ell}_2 ; y^{'\ell}_2 | y^{'\ell}_1) +\ell \e_{\ell}\nonumber\\ 
&\leq \frac{\ell}{2}\log(1+ g_{11} + g_{12}) + h(y^{'\ell}_2 | y^{'\ell}_1 ) 
  -  h(y^{'\ell}_2 | y^{'\ell}_1,x^{\ell}_1, x^{\ell}_2 )+\ell \e_{\ell},
\label{gzs:s1s2-1}
\end{align}
where \eqref{gzs:s1s2-0} follows from  the data processing inequality. 
Now, note that 
\begin{align*}
h(y^{'\ell}_2, W_1 | y^{'\ell}_1) &= h(y^{'\ell}_2 | y^{'\ell}_1 ) +H(W_1 | y^{'\ell}_1 , y^{'\ell}_2)\\
&= H(W_1|y^{'\ell}_1) + h(y^{'\ell}_2 | W_1, y^{'\ell}_1 )\nonumber\\
&\leq H(W_1|y^{\ell}_1) + h(y^{'\ell}_2 | W_1, y^{'\ell}_1 ).
\end{align*}
Therefore,
\begin{align}
h(y^{'\ell}_2 | y^{'\ell}_1 ) 
&\leq  h(y^{'\ell}_2 | W_1, y^{'\ell}_1 ) +\ell \e_{\ell}\nonumber\\
&\leq h(y^{'\ell}_2 | x^{\ell}_1, y^{'\ell}_1 ) +\ell \e_{\ell} \label{gzs:s1s2-2}  \\
&= h(y^{'\ell}_2 | x^{\ell}_1,  \sqrt{g_{12}} x^{\ell}_2 + z^{'\ell}_1 ) +\ell \e_{\ell} \label{gzs:s1s2-3}\\
&\leq h( \sqrt{g_{22}} x^{\ell}_2 + z^{'\ell}_2  |  \sqrt{g_{12}} x^{\ell}_2 + z^{'\ell}_1 ) +\ell \e_{\ell}\nonumber\\
&= h( \sqrt{g_{22}} x^{\ell}_2 + z^{'\ell}_2 
  - \frac{\sqrt{g_{22}}}{\sqrt{g_{12}}} (\sqrt{g_{12}} x^{\ell}_2 + z^{'\ell}_1) |  \sqrt{g_{12}} x^{\ell}_2 
  + z^{'\ell}_1 ) +\ell \e_{\ell}\nonumber\\
&\leq h( z^{'\ell}_2 - \frac{\sqrt{g_{22}}}{\sqrt{g_{12}}}  z^{'\ell}_1)  +\ell \e_{\ell}\nonumber\\
&= \frac{\ell}{2}\log(2\pi e)\left(1+\frac{g_{22}}{g_{12}}\right)+\ell \e_{\ell}, \label{gzs:s1s2-4}
\end{align}
where \eqref{gzs:s1s2-2} holds since $x_1^{\ell}$ is a function of $W_1$, and in \eqref{gzs:s1s2-3}
we used the invertibility property of the function $y_1^{\ell}= \sqrt{g_{11}} x^{\ell}_1 + \sqrt{g_{12}} x^{\ell}_2 + z^{'\ell}_1$. 
Replacing $h(y^{'\ell}_2 | y^{'\ell}_1 )$ from (\ref{gzs:s1s2-4}) in (\ref{gzs:s1s2-1}), we get the desired bound. 

\item (\ref{gzs:d1d2}) $\ \ R_1 + R_2 <  \frac{1}{2}\log \left(1+\frac{h_{11}}{h_{21}}\right)  
+ \frac{\ell}{2}\log(1+ h_{21} + h_{22}+2\sqrt{h_{21}h_{22}})$:
The sum-rate can be upper bounded as in Lemma~\ref{lm:gzs-fano}. Next, we have
\begin{align}
\ell (R_1 +R_2) & \leq I(x^{\ell}_1, x^{\ell}_2 ; y^{\ell}_1 , y^{\ell}_2) +\ell \e_{\ell} \nonumber\\
& \leq  I(x^{'\ell}_1, x^{'\ell}_2 ; y^{\ell}_1 , y^{\ell}_2) +\ell \e_{\ell}\nonumber\\
&= I(x^{'\ell}_1, x^{'\ell}_2 ; y^{\ell}_2 ) + I(x^{'\ell}_2 ; y^{\ell}_1 | y^{\ell}_2)
+ I(x^{'\ell}_1 ; y^{\ell}_1 | x^{'\ell}_2, y^{\ell}_2) +\ell \e_{\ell}.\label{gzs:d1d2-1}
\end{align}
The first term in (\ref{gzs:d1d2-1}) can be simply upper bounded as
\begin{align}
I(x^{'\ell}_1, x^{'\ell}_2 ; y^{\ell}_2 )\leq \frac{\ell}{2}\log(1+ h_{21} + h_{22}+2\sqrt{h_{21}h_{22}}).\label{gzs:d1d2:2}
\end{align}
In order to bound the second term, we can use the fact that $W_2$ can be decoded from $y^{\ell}_2$, and write
\begin{align}
I(x^{'\ell}_2 ;y^{\ell}_1, W_2  | y^{\ell}_2 ) &= I(x^{'\ell}_2 ;y^{\ell}_1  | y^{\ell}_2 )
      +  I(x^{'\ell}_2 ; W_2  | y^{\ell}_1, y^{\ell}_2 )\nonumber\\
&= I(x^{'\ell}_2 ;y^{\ell}_1  | W_2, y^{\ell}_2 )+ I(x^{'\ell}_2 ; W_2  |  y^{\ell}_2 ) \nonumber\\
&\leq I(x^{'\ell}_2 ;y^{\ell}_1  | W_2, y^{\ell}_2 ) + H( W_2  |  y^{\ell}_2 )\nonumber\\
&\leq I(x^{'\ell}_2 ; y^{\ell}_1 | y^{\ell}_2, W_2) +\ell \e_{\ell}.\nonumber
\end{align}
Therefore, 
\begin{align}
I(x^{'\ell}_2 ; y^{\ell}_1 | y^{\ell}_2) &\leq I(x^{'\ell}_2 ; y^{\ell}_1 | y^{\ell}_2, W_2) +\ell \e_{\ell}
\leq I(y^{'\ell}_2 ; y^{\ell}_1 | y^{\ell}_2, W_2) +\ell \e_{\ell}
=\ell \e_{\ell},\label{gzs:d1d2:4}
\end{align}
where the second inequality follows from the fact that $x^{'\ell}_2$ is a function of $y^{'\ell}_2$, and \eqref{gzs:d1d2:4}
holds since $y^{'\ell}_2$ and $y^{\ell}_1$ are independent if $W_2$ is given.

Finally, we bound the last term as follows. 
\begin{align}
I(x^{'\ell}_1 ; y^{\ell}_1 | x^{'\ell}_2, y^{\ell}_2) &= 
    I(x^{'\ell}_1 ; y^{\ell}_1 | x^{'\ell}_2, \sqrt{h_{21}} x^{'\ell}_1+z^{\ell}_2)\nonumber\\
&=  h(\sqrt{h_{11}} x^{'\ell}_1+z^{\ell}_1 | x^{'\ell}_2, \sqrt{h_{21}} x^{'\ell}_1+z^{\ell}_2) 
    - h( y^{\ell}_1 | x^{'\ell}_2, \sqrt{h_{21}} x^{'\ell}_1 +z^{\ell}_2, x^{'\ell}_1)\nonumber\\
&\leq h(\sqrt{h_{11}} x^{'\ell}_1+z^{\ell}_1 
    - \frac{ \sqrt{h_{11}} }{ \sqrt{h_{21}} }(\sqrt{h_{21}} x^{'\ell}_1+z^{\ell}_2) ) -h(z^{\ell}_1)\nonumber\\
&\leq \frac{\ell}{2}\log\left(1+\frac{h_{11}}{h_{12}}\right).\label{gzs:d1d2:3}
\end{align}
 Replacing the bound derived for the three terms, 
(\ref{gzs:d1d2:2}), (\ref{gzs:d1d2:4}), and \eqref{gzs:d1d2:3} in (\ref{gzs:d1d2-1}), we get the desired bound. 

\end{itemize}

\paragraph{The proofs of cut-set type bounds}
\begin{itemize}

\item (\ref{gzs:s1}) $\ \ R_1 < \frac{1}{2}\log(1+ g_{11})$: 
We start by  Lemma~\ref{lm:gzs-fano}, and write
\begin{align}
\ell R_1 &= I(x_1^{\ell}; y_1^{\ell}) +\ell \e_{\ell} \nonumber\\
& \leq  I(x_1^{\ell}; y_1^{'\ell}) +\ell \e_{\ell} \label{pr:gzs:s1-1}\\
& \leq  I(x_1^{\ell}; x^{\ell}_2 , y_1^{'\ell}) +\ell \e_{\ell} \nonumber\\
&= I(x_1^{\ell}; x^{\ell}_2 )  + I(x_1^{\ell};  y_1^{'\ell} | x^{\ell}_2)  +\ell \e_{\ell} \label{pr:gzs:s1-2}\\
&\leq \frac{\ell}{2}\log(1+g_{11})+ \ell \e_{\ell}
\end{align}
where \eqref{pr:gzs:s1-1} follows from  the data-processing inequality for the Markov chain 
$ x_1^{\ell}\leftrightarrow y_1^{'\ell}\leftrightarrow x_1^{'\ell}\leftrightarrow y_1^{\ell}$, 
and in \eqref{pr:gzs:s1-2} we used the fact that $x_1^{\ell}$ and $x_2^{\ell}$ are independent.
It is worth mentioning that this inequality essentially bounds the maximum flow that can be transmitted through 
the cut  $\Omega_s=\{S_1\}$ and $\Omega_d=\{S_2, A, B, D_1, D_2\}$.

\item (\ref{gzs:s2}) $\ \ R_2 < \frac{1}{2}\log(1+ g_{12} + g_{22})$:
Again starting from Lemma~\ref{lm:gzs-fano}, we have
\begin{align}
\ell R_2 
&\leq  I(x^{\ell}_2; y^{\ell}_2)+\ell \e_{\ell}\nonumber\\
&\leq  I(x^{\ell}_2;  y^{'\ell}_1, y^{'\ell}_2)+\ell \e_{\ell}\label{gzs:s2:data}\\
&\leq  I(x^{\ell}_2; x^{\ell}_1, y^{'\ell}_1, y^{'\ell}_2)+\ell \e_{\ell}\nonumber\\
&= I(x^{\ell}_2; x^{\ell}_1) + I(x_2^{\ell} ; y^{'\ell}_1, y^{'\ell}_2 | x^{\ell}_1)+\ell \e_{\ell}\nonumber\\
&= h(y^{'\ell}_1, y^{'\ell}_2 | x^{\ell}_1) - h(y^{'\ell}_1, y^{'\ell}_2 | x^{\ell}_1, x^{\ell}_2)+\ell \e_{\ell}\nonumber\\
&\leq  h(\sqrt{g_{12}} x^{\ell}_2 + z^{'\ell}_1, \sqrt{g_{22}} x^{\ell}_2 + z^{'\ell}_2) 
  - h(z^{'\ell}_1, z^{'\ell}_2)+\ell \e_{\ell}\nonumber\\
&\leq \frac{\ell}{2}\log(1+g_{12}+g_{22})+ \ell \e_{\ell},
\end{align}
where the data processing inequality implies (\ref{gzs:s2:data}) for the Markov chain 
$ x_2^{\ell}\leftrightarrow (y_1^{'\ell},y_2^{'\ell}) \leftrightarrow (x_1^{'\ell}, x_2^{'\ell})\leftrightarrow y_1^{\ell}$.
Note that this bound is essentially the cut-set bound for the cut $\Omega_s=\{S_2\}$ and $\Omega_d=\{S_1, A, B, D_1, D_2\}$.

%

\item (\ref{gzs:s2r2}) $\ \ R_2 < \frac{1}{2}\log(1+ g_{12}) + \frac{1}{2}\log(1+ h_{22})$: 
Again we use Lemma~\ref{lm:gzs-fano} to upper bound $R_2$ as
\begin{align}
\ell R_2  &\leq I(x^{\ell}_2 ; y^{\ell}_2) +\ell \e_{\ell}\nonumber\\
    &\leq I(x^{'\ell}_2, x^{\ell}_2 ; x^{\ell}_1 , y^{'\ell}_1 , y^{\ell}_2 )+\ell \e_{\ell}\nonumber\\
&= I(x^{'\ell}_2, x^{\ell}_2 ; x^{\ell}_1 ) +  I(x^{'\ell}_2, x^{\ell}_2 ; y^{'\ell}_1 , y^{\ell}_2 | x^{\ell}_1)+\ell\e_{\ell}\nonumber\\
&= I(x^{'\ell}_2, x^{\ell}_2 ; y^{'\ell}_1 | x^{\ell}_1)+  I(x^{'\ell}_2, x^{\ell}_2 ; y^{\ell}_2 | x^{\ell}_1,y^{'\ell}_1)
    +\ell\e_{\ell}\nonumber\\
&=  I( x^{\ell}_2 ; y^{'\ell}_1 | x^{\ell}_1)+    I( x^{'\ell}_2 ; y^{'\ell}_1 | x^{\ell}_1, x^{\ell}_2) 
+I( x^{'\ell}_2 ; y^{\ell}_2 | x^{\ell}_1,y^{'\ell}_1) 
    + I( x^{\ell}_2 ; y^{\ell}_2 | x^{\ell}_1,y^{'\ell}_1, x^{'\ell}_2)+\ell\e_{\ell} \label{gza:s2r2-1}\\
&= I( x^{\ell}_2 ; y^{'\ell}_1 | x^{\ell}_1)+ I( x^{'\ell}_2 ; y^{\ell}_2 | x^{\ell}_1,y^{'\ell}_1) +\ell\e_{\ell}\nonumber\\
&\leq \frac{\ell}{2}\log(1+ g_{12}) + \frac{\ell}{2}\log(1+ h_{22})+\ell\e_{\ell}.
\end{align}
Note that we used the fact that the second and fourth terms in \eqref{gza:s2r2-1} are zero. This follows from
\begin{align}
 I( x^{'\ell}_2 ; y^{'\ell}_1 | x^{\ell}_1, x^{\ell}_2) \leq 
    I( x^{'\ell}_2 ; y^{'\ell}_1- \sqrt{g_{11}} x^{\ell}_1 - \sqrt{g_{12}} x^{\ell}_2 | x^{\ell}_1, x^{\ell}_2)
= I( x^{'\ell}_2 ; z^{'\ell}_1 | x^{\ell}_1, x^{\ell}_2)=0,\nonumber
\end{align}
and 
\begin{align}
 I( x^{\ell}_2 ; y^{\ell}_2 | x^{\ell}_1,y^{'\ell}_1, x^{'\ell}_2) 
&\leq I( x^{\ell}_2 ; y^{\ell}_2 | x^{\ell}_1 ,x^{'\ell}_1, x^{'\ell}_2) \nonumber\\
&\leq I( x^{\ell}_2 ; y^{\ell}_2 - \sqrt{h_{21}} x^{'\ell}_1 - \sqrt{h_{22}} x^{'\ell}_2| x^{\ell}_1 ,x^{'\ell}_1, x^{'\ell}_2)  \nonumber\\
&\leq I( x^{\ell}_2 ; z^{\ell}_2  | x^{\ell}_1 ,x^{'\ell}_1, x^{'\ell}_2) =0. \nonumber
\end{align}

\item (\ref{gzs:s1s2r1}) $\ \ R_1+R_2  <  \frac{1}{2} \log(1+g_{22}) + \frac{1}{2}\log(1+h_{11} +h_{21})$:\\ 
We start from Lemma~\ref{lm:gzs-fano} and write
\begin{align}
\ell( R_1+R_2) & \leq I( y^{\ell}_1, y^{\ell}_2 ; x^{\ell}_1, x^{\ell}_2) +\ell\e_{\ell}\label{gzs:s1s2r1-1}\\
    &\leq  I( y^{\ell}_1, y^{\ell}_2 ; x^{'\ell}_1, x^{\ell}_2) +\ell\e_{\ell}\nonumber\\
&\leq I( y^{\ell}_1, y^{\ell}_2 , y^{'\ell}_2; x^{'\ell}_1, x^{\ell}_2) +\ell\e_{\ell}\nonumber\\
&=  I( y^{'\ell}_2; x^{\ell}_2) + I(  y^{'\ell}_2; x^{'\ell}_1 | x^{\ell}_2) 
    + I( y^{\ell}_1, y^{\ell}_2 ; x^{'\ell}_1, x^{\ell}_2 | y^{'\ell}_2)+\ell\e_{\ell}\nonumber\\
&= I( y^{'\ell}_2; x^{\ell}_2) + I( y^{\ell}_1, y^{\ell}_2 ; x^{'\ell}_1, x^{\ell}_2 | x^{'\ell}_2)+\ell\e_{\ell}\nonumber\\
&=  I( y^{'\ell}_2; x^{\ell}_2) + h( y^{\ell}_1, y^{\ell}_2| x^{'\ell}_2) 
    - h( y^{\ell}_1, y^{\ell}_2 | x^{\ell}_2, x^{'\ell}_1 , x^{'\ell}_2)+\ell\e_{\ell}\nonumber\\
& \leq  I( y^{'\ell}_2; x^{\ell}_2) +  h( \sqrt{h_{11}} x^{'\ell}_1 + z^{\ell}_1, \sqrt{h_{21}} x^{'\ell}_1 + z^{\ell}_2) 
    - h( z^{\ell}_1, z^{\ell}_2) +\ell\e_{\ell}\nonumber\\
&\leq \frac{\ell}{2} \log(1+g_{22}) + \frac{\ell}{2}\log(1+h_{11} +h_{21})+\ell\e_{\ell}
\end{align}
where \eqref{gzs:s1s2r1-1} follows from the data processing inequality for the Markov chain 
$ (x_1^{\ell},x_2^{\ell}) \leftrightarrow (y_1^{'\ell},y_2^{'\ell}) \leftrightarrow (x_1^{'\ell}, x_2^{'\ell})
\leftrightarrow (y_1^{\ell},y_2^{\ell})$. Note that this bound essentially captures the maximum flow of information through 
the cut $\Omega_s=\{S_1, S_2, A\}$ and $\Omega_d=\{B,D_1,D_2\}$.

\item (\ref{gzs:s1s2r2}) $\ \ R_1+R_2  <  \frac{1}{2} \log(1+g_{11} +g_{12}) + \frac{1}{2}\log(1+h_{22} )$:\\ 
Similar to the previous bounds, we start from Lemma~\ref{lm:gzs-fano} and write
\begin{align}
 R_1+R_2 & \leq I( y^{\ell}_1, y^{\ell}_2 ; x^{\ell}_1, x^{\ell}_2) +\ell\e_{\ell}\nonumber\\
&\leq  I( y^{'\ell}_1, y^{\ell}_2 ; x^{\ell}_1, x^{\ell}_2) +\ell\e_{\ell}\label{gzs:s1s2r2-1}\\
&\leq I( y^{'\ell}_1, y^{\ell}_2 ; x^{\ell}_1, x^{\ell}_2, x^{'\ell}_2)+\ell\e_{\ell}\nonumber\\
&= I( y^{'\ell}_1; x^{\ell}_1, x^{\ell}_2) +I( y^{'\ell}_1 ; x^{'\ell}_2|  x^{\ell}_1, x^{\ell}_2)
+ I(  y^{\ell}_2 ; x^{'\ell}_2 | y^{'\ell}_1) 
  + I( y^{\ell}_2 ; x^{\ell}_1, x^{\ell}_2| y^{'\ell}_1, x^{'\ell}_2)+\ell\e_{\ell}\label{gzs:s1s2r2-2}\\
&= I( y^{'\ell}_1; x^{\ell}_1, x^{\ell}_2)+ I(  y^{\ell}_2 ; x^{'\ell}_2 | y^{'\ell}_1) +\ell\e_{\ell}\nonumber\\
&\leq \frac{\ell}{2} \log(1+g_{11}+g_{12}) +  I(  y^{\ell}_2 ; x^{'\ell}_2 | y^{'\ell}_1) +\ell\e_{\ell}. \label{gzs:s1s2r2-3}
\end{align}
Note that in \eqref{gzs:s1s2r2-1} we used the data processing inequality. An argument similar to that 
is used in the proof of \eqref{gzs:s2r2} shows that the second and fourth terms in \eqref{gzs:s1s2r2-2} are zero. 
Now, we have
\begin{align}
I(  y^{\ell}_2 ; x^{'\ell}_2 | y^{'\ell}_1) &= h(  y^{\ell}_2 | y^{'\ell}_1)- h(  y^{\ell}_2 | x^{'\ell}_2, y^{'\ell}_1) \nonumber\\
&\leq  h(  y^{\ell}_2 | x^{'\ell}_1)- h(  y^{\ell}_2 | x^{'\ell}_1, x^{'\ell}_2, y^{'\ell}_1) \nonumber\\
&=  h(  \sqrt{h_{22}} x^{'\ell}_2 + z_2^{\ell} | x^{'\ell}_1)- h(  z_2^{\ell} | x^{'\ell}_1, x^{'\ell}_2, y^{'\ell}_1) \nonumber\\
&\leq h(  \sqrt{h_{22}} x^{'\ell}_2 + z_2^{\ell}) -h(  z_2^{\ell})\nonumber\\
&\leq \frac{\ell}{2}\log(1+h_{22}) \label{gzs:s1s2r2-4}
\end{align}
Finally, we obtain the desired bound by replacing \eqref{gzs:s1s2r2-4} in \eqref{gzs:s1s2r2-3}. 
It is worth mentioning that this bound is the same as the cut-set bound for the cut  $\Omega_s=\{S_1, S_2, B\}$ 
and $\Omega_d=\{A,D_1,D_2\}$.

\item (\ref{gzs:d1}) $\ \ R_1 < \frac{1}{2}\log(1+ h_{11})$: 
Using Lemma~\ref{lm:gzs-fano} and the data processing inequality, we can write
\begin{align}
\ell R_1 & \leq I(x^{'\ell}_1; y^{\ell}_1)+\ell \e_{\ell}
\leq I(x^{'\ell}_1; y^{\ell}_1)+\ell \e_{\ell}
\leq \frac{\ell}{2}\log(1+h_{11})+ \ell \e_{\ell}.
\end{align}

\item (\ref{gzs:d2}) $\ \ R_2 < \frac{1}{2}\log(1+ h_{21}+h_{22} +2 \sqrt{h_{21}h_{22}})$: 
Starting from Lemma~\ref{lm:gzs-fano} and applying the data processing inequality for the Markov chain 
$x_2^{\ell} \leftrightarrow (y_1^{'\ell},y_2^{'\ell}) \leftrightarrow (x_1^{'\ell}, x_2^{'\ell})\leftrightarrow y_2^{\ell}$,
we have
\begin{align}
\ell R_2 &\leq I(x^{\ell}_2; y^{\ell}_2)+\ell \e_{\ell}
\leq I(x^{'\ell}_1,x^{'\ell}_2; y^{\ell}_2)+\ell \e_{\ell}
\leq \frac{\ell}{2}\log(1+h_{21}+h_{22}+  2 \sqrt{h_{21}h_{22}})+ \ell \e_{\ell}.
\end{align}
Note that $x^{'\ell}_1$ and $x^{'\ell}_2$ are not independent. However, their variance is upper bounded by 
$(\sqrt{h_{21}}+\sqrt{h_{22}})^2$.

\item (\ref{gzs:r2d2}) $\ \ R_2 < \frac{1}{2}\log(1+ g_{22}) + \frac{1}{2}\log(1+ h_{21})$:
Consider the cut which partitions the network into $\Omega_s=\{S_1,S_2,A,D_1\}$ and $\Omega_d=\{B,D_2\}$. We have
\begin{align}
\ell R_2  &\leq I(x^{\ell}_2 ; y^{\ell}_2) +\ell \e_{\ell}\nonumber\\
&\leq I(x^{'\ell}_1, x^{\ell}_2 ; y^{'\ell}_2 , y^{\ell}_2 )+\ell \e_{\ell}\nonumber\\
&=  I( x^{\ell}_2 ; y^{'\ell}_2 )+    I( x^{'\ell}_1 ; y^{'\ell}_2 | x^{\ell}_2)
+I( x^{'\ell}_1 ; y^{\ell}_2 | y^{'\ell}_2) + I( x^{\ell}_2 ; y^{\ell}_2 |y^{'\ell}_2, x^{'\ell}_1)+\ell \e_{\ell}\label{gzs:r2d2-1}\\
&= I( x^{\ell}_2 ; y^{'\ell}_2 )+ I( x^{'\ell}_1 ; y^{\ell}_2 | y^{'\ell}_2)+\ell \e_{\ell}\nonumber\\
&=  I( x^{\ell}_2 ; y^{'\ell}_2 )+ I( x^{'\ell}_1 ; y^{\ell}_2 | y^{'\ell}_2, x^{'\ell}_2)+\ell \e_{\ell}\nonumber\\
&\leq \frac{\ell}{2}\log(1+ g_{22}) + \frac{\ell}{2}\log(1+ h_{21})+\ell \e_{\ell}.
\end{align}
We again used an argument similar to that is used in proof of \eqref{gzs:s2r2} to show that the second and fourth 
terms in \eqref{gzs:r2d2-1} are zero.

This completes the proof of the outer bound in  Theorem~\ref{thm:gzs}.
\end{itemize}

\subsection{The Achievability Part}
\label{subsec:gzs-ach}
In this section we provide an encoding scheme for the Gaussian $\ZS$ network, and show that the
rate region that can be achieved using this scheme is only a constant bit gap away from 
the outer bound. 

\paragraph*{Large Channel Gains}
In this part, we assume that all channel gains are at least $1$, \emph{i.e.,} $g_{ij}\geq 1$, and
$h_{ij}\geq 1$. Note that if any of 
the gains are small, then either one of the rates are small (of the order of our 
constant bit gap), or the cross links are negligible. We will discuss these cases later. 

The encoding scheme proposed for the Gaussian $\ZS$ network consists of two separate parts. We first split the message 
of the second source nodes as $W_1=\Uoo$ and  $W_2=(\Uto,\Utt,\Utr)$, where $\Uto$ can be decoded at both relay nodes $A$ and $B$,
and $\Utt$ and $\Utr$ can be decoded only at $A$ and $B$, respectively (see Figure~\ref{fig:GZS:Z}). Denoting the rate of message 
$W_{i}^{(j)}$ by $\Upsilon_{i,j}$, the following rate constraints are imposed by this message splitting
\begin{align}
R_1&=\Too,\label{eq:gzs:sp-1-1}\\
R_2&=\Tto+\Ttt+\Ttr.\label{eq:gzs:sp-1-2}
\end{align}
An achievable rate region for this message splitting 
is given in Lemma~\ref{lm:GZS:Z}.

In the second layer of the network (see Figure~\ref{fig:GZS:S}), relay node $A$ further  splits its messages as follows: $W_1=\Uoo=\left(\Voo,\Vot\right)$, 
$\Uto=\left(\Vto,\Vtt\right)$, and $\Utt=\left( \Vtr,\Vtf\right)$ . A similar message splitting 
is also performed at node $B$ to obtain 
$\Uto=\left(\Vto,\Vtt\right)$ and $\Utr=\Vtv$. This message splitting imposes the following rate equations
\begin{align}
\Too &= \Qoo+\Qot, \label{eq:gzs:sp-2-1}\\
\Tto &= \Qto+\Qtt, \label{eq:gzs:sp-2-2}\\
\Ttt &= \Qtr+\Qtf , \label{eq:gzs:sp-2-3}\\
\Ttr &= \Qtv,  \label{eq:gzs:sp-2-4}
\end{align}
where $\Theta_{i,j}$ denotes the rate of the message $V_{i}^{(j)}$.
Next, the relay nodes have to convey the messages to the destination nodes such that $D_1$ can decode $\Voo$, $\Vot$, $\Vto$ and $\Vtr$, 
and $D_2$ be able to decode $\Voo$, $\Vto$, $\Vtt$,
$\Vtr$, $\Vtf$ and $\Vtv$. An achievable rate region for this transmission scenario is given in Lemma~\ref{lm:GZS:S}. 

Putting the rate constraints in Lemma~\ref{lm:GZS:Z} and Lemma~\ref{lm:GZS:S} together with the equations in \eqref{eq:gzs:sp-1-1}-\eqref{eq:gzs:sp-1-2}
and \eqref{eq:gzs:sp-2-1}-\eqref{eq:gzs:sp-2-4}, we obtain the following achievable rate region for the Gaussian $\ZS$ network. 
\begin{align}
\R{\GZS}{ach}=\Big\{(R_1,R_2): \exists &\Too, \Tto,\Ttt,\Ttr, \Qoo,\Qot,\Qto,\Qtt, \Qtr,\Qtf,\Qtv \geq 0,\\
&R_1=\Too,\nonumber\\
&R_2=\Tto+\Ttt+\Ttr,\nonumber\\
&\Too = \Qoo+\Qot, \nonumber\\
&\Tto = \Qto+\Qtt, \nonumber\\
&\Ttt = \Qtr+\Qtf, \nonumber\\
&\Ttr = \Qtv,  \nonumber\\
&\Too \leq \+{\CGN{1+g_{11}}-\frac{1}{2}},\nonumber\\
&\Ttt \leq  \+{\CGN{1+\frac{g_{12}}{g_{22}} }-\frac{1}{2}},\nonumber\\
&\Tto+\Ttt \leq  \+{\CGN{1+g_{12}}-\frac{1}{2}},\nonumber\\
&\Too +\Tto+\Ttt  \leq \+{\CGN{1+g_{11}+g_{12}}-\frac{1}{2}},\nonumber\\
&\Ttr \leq \+{\CGN{1+\frac{g_{22}}{g_{12}} }-\frac{1}{2}},\nonumber\\
&\Tto+\Ttr  \leq \+{\CGN{1+g_{22}}-\frac{1}{2}},\nonumber\\
&\Qoo+\Qot+\Qto+\Qtr \leq \+{\CGN{1+h_{11}} -\frac{1}{2}},\nonumber\\
&\Qot \leq \+{\CGN{1+\frac{h_{11}}{h_{12}} }-\frac{1}{2}},\nonumber\\
&\Qtf  \leq \+{\CGN{1+\frac{h_{21}} {h_{11}}}-\frac{1}{2}},\nonumber\\
&\Qoo+\Qtr+\Qtf  \leq \+{\CGN{1+h_{21}}-\frac{1}{2}},\nonumber\\
&\Qtv \leq \+{\CGN{1+h_{22}}-\frac{1}{2}},\nonumber\\
&\Qoo+\Qto+\Qtt+\Qtr+\Qtf+\Qtv  \leq \+{\CGN{1+h_{21}+h_{22}}-\frac{1}{2}}\Big\}.\nonumber
\end{align}

We  apply the Fourier-Motzkin elimination on this region, to project it on the coordinated $R_1$ and $R_2$, 
and obtain the following rate region. After some simplifications, we get 

\begin{align}
\R{\GZS}{ach}=\Big\{(R_1,R_2): 
&R_1 \leq \+{\CGN{g_{11}} -\frac{1}{2}},\nonumber\\
&R_2 \leq \+{\CGN{ g_{12} + g_{22} }-\frac{1}{2}},\nonumber\\
&R_1 + R_2 \leq  \+{\CGN{g_{11} + g_{12}} + \CGN{ \frac{g_{22}}{g_{12}} }-\frac{1}{2}},\nonumber\\
 &R_2 \leq \+{\CGN{ g_{12}} + \CGN {h_{22}}-\frac{1}{2}},\nonumber\\
 &R_1+R_2  \leq  \+{\CGN{ g_{22} } + \CGN{ h_{11} +h_{21}} -\frac{1}{2}},\nonumber\\
 &R_1+R_2 \leq  \+{\CGN{ g_{11}+g_{12}} + \CGN{ h_{22} } -\frac{1}{2}},\nonumber\\
&R_1 \leq \+{\CGN{ h_{11}}-\frac{1}{2}},\nonumber\\
 &R_2 \leq \+{\CGN{  h_{21} + h_{22} } -\frac{1}{2}},\nonumber\\
 &R_2 \leq  \+{\CGN{  g_{22} } + \CGN{  h_{21} }-\frac{1}{2}},\nonumber\\
&R_1 + R_2 \leq  \+{\CGN{ h_{21} + h_{22}}+ \CGN{ \frac{h_{11}}{h_{21}}}-\frac{1}{2} }  
\Big\}.\nonumber
\end{align}
Note that this rate region is characterized by a set of constraints which are similar to the
inequalities in the definition of $\R{\GZS}{}$, except for the additive constants, and the fact 
that $\log(1+x)$ is replaced by $\log(x)$. Note that since $x\geq 1$, we have 
\begin{align}
\frac{1}{2}\log (1+x ) - \frac{1}{2}\log (x ) \leq \frac{1}{2}.
\end{align}
Hence, the difference between the RHS's of two sets of inequalities 
do not exceed $1$ for $R_1$, and $3/2$ for $R_2$ and $R_1+R_2$.  
Therefore, for any rate pair $(R_1,R_2)\in \R{\GZS}{}$, we have 
$(R_1-1,R_2-1.5)\in \R{\GZS}{ach}$.
This completes the proof.

\paragraph*{Small Channel Gains}
We will show in this part that if any of the channel gains are small, then the outer bound in Theorem~\ref{thm:gzs}
is still within a constant bit gap of an achievable rate region. This argument is based on the analysis of the same network, 
in which all the links with gain smaller than $1$ are removed. One can show that the capacity region of this modified network is within a
constant gap from that of the original one. On the other hand, we can argue that the gap between the achievable rate pairs of the modified network 
and the outer bound in Theorem~\ref{thm:gzs} is bounded by a constant. Therefore, we can conclude that if $(R_1,R_2)\in\R{\GZS}{}$ then 
$(R_1-\delta_1,R_2-\delta_2)$ is achievable for the original network, where $\delta_1=1$ and $\delta_2=1.5$. 

The main intuition behind this argument is the fact that 
since all the nodes are assumed to have power constraint equal to $1$, the flow of information through a link 
with gain not exceeding $1$ is upper bounded by $\frac{1}{2}\log(1+\mathsf{SNR}) \leq \frac{1}{2}\log(1+1)=\frac{1}{2}$ bit. Therefore, 
by removing such links from the network, the achievable rates change by at most $\frac{1}{2}$ bit. On the other hand, the incoming signals over small 
channel gains may act as an interference on the original network, which cause a total noise power not exceeding $1$. Therefore, by doubling the noise 
variances of the original network, we guarantee that capacity region of the modified network is always smaller than that of the original one. 

The advantage of analyzing the modified network instead of the original one is that some of the links are removed in the modified 
network, which convert it to simpler network to analyze. 

A precise  analysis of the modified networks requires considering several cases separately. However, similar techniques and ideas 
will be used for all cases. In the following we present one illustrating example, and skip the details for the other cases. 

\begin{example}
 Consider the Gaussian $\ZS$ network in Figure~\ref{fig:GZS}, and assume that $g_{12}=0$. Therefore, the first layer of the network 
 would be two parallel links as shown in Figure~\ref{fig:GZS:3}, where $\E[\tilde{z}^{'2}_1]=2$. 
\begin{figure}[h!]
 \centering
 	\psfrag{s1}[Bc][Bc]{$S_1$}
	\psfrag{s2}[Bc][Bc]{$S_2$}
	\psfrag{r1}[Bc][Bc]{$A$}
	\psfrag{r2}[Bc][Bc]{$B$}
	\psfrag{d1}[Bc][Bc]{$D_1$}
	\psfrag{d2}[Bc][Bc]{$D_2$}
	\psfrag{a}[Bc][Bc]{$\sqrt{g_{11}}$}
	\psfrag{b}[Bc][Bc]{$\sqrt{h_{11}}$}
	\psfrag{c}[Bc][Bc]{$\sqrt{g_{22}}$}
	\psfrag{d}[Bc][Bc]{$\sqrt{h_{22}}$}
	\psfrag{f}[Bc][Bc]{$\sqrt{g_{12}}$}
	\psfrag{g}[Bc][Bc]{$\sqrt{h_{21}}$}
 	\psfrag{X_1}[Bc][Bc]{$x_1$}
	\psfrag{X_2}[Bc][Bc]{$x_2$}
	\psfrag{Y'_1}[Bc][Bc]{$y'_1$}
	\psfrag{Y'_2}[Bc][Bc]{$y'_2$}
	\psfrag{X'_1}[Bc][Bc]{$x'_1$}
	\psfrag{X'_2}[Bc][Bc]{$x'_2$}
	\psfrag{Y_1}[Bc][Bc]{$y_1$}
	\psfrag{Y_2}[Bc][Bc]{$y_2$}
	\psfrag{Z'_1}[Bc][Bc]{$\tilde{z}'_1$}
	\psfrag{Z'_2}[Bc][Bc]{$z'_2$}
	\psfrag{Z_1}[Bc][Bc]{$z_1$}
	\psfrag{Z_2}[Bc][Bc]{$z_2$}
	\psfrag{t_1}[Bc][Bc]{$\ $}
	\psfrag{t_2}[Bc][Bc]{$\ $}
\includegraphics[width=9cm]{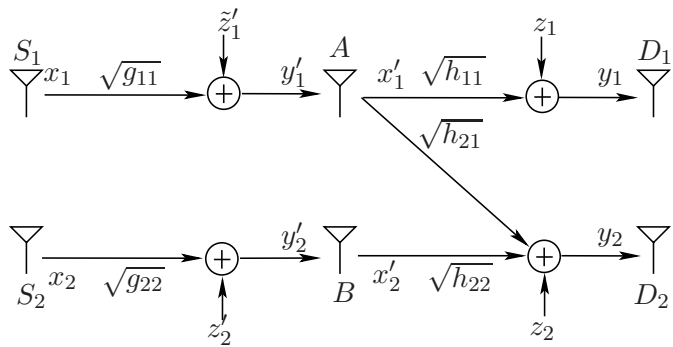}
\vspace{-5mm}
\caption{A modified $\ZS$ network obtained assuming $g_{12}=0$.}
\label{fig:GZS:3}
\vspace{-5mm}
\end{figure}
 Moreover, the rate region in \eqref{gzs:s1}-\eqref{gzs:d1d2} will be 
 reduced to 
 \begin{align}
R_1 &\leq \frac{1}{2}\log(1+ g_{11})\label{gzs-example-1}\\
R_2 &\leq \frac{1}{2}\log(1+ g_{12} )\label{gzs-example-2}\\
R_1 &\leq \frac{1}{2}\log(1+ h_{11})\label{gzs-example-3}\\
R_2 &\leq \frac{1}{2}\log(1+ h_{22})\label{gzs-example-4}\\
R_1 + R_2 &\leq  \frac{1}{2}\log(1+ h_{21} + h_{22}+2\sqrt{h_{21}h_{22}})+ \frac{1}{2}\log \left(1+\frac{h_{11}}{h_{21}}\right).  
\label{gzs-example-5}
\end{align}
The encoding strategy for this network is fairly simple. Let $(R_1,R_2)$ be a 
rate pair satisfying \eqref{gzs-example-1}-\eqref{gzs-example-5}. The goal is 
to show that $(R_1-1,R_2-1)$ is achievable. Since $(R_1-1,R_2-1)$ satisfies 
\eqref{gzs-example-1} and \eqref{gzs-example-2}, transmission 
over the first layer of the network from the source nodes to the relays is simply done using random Gaussian codes.

The second layer of the network is a Gaussian $\S$ network.  Once the relays decode the messages received from the 
first layer of the network, they encode them using an encoding strategy similar to that of the $\Z$ network in Example~\ref{ex:GZ}
in Section~\ref{sec:examples}. Note that the sum-rate bounds  in \eqref{gzs-example-5} and the outer bound of the $\S$ network 
are slightly different. However, their difference is upper bounded by 
\begin{align}
 \frac{1}{2}\log(1+ h_{21} + h_{22}+2\sqrt{h_{21}h_{22}}) -  \frac{1}{2}\log(1+ h_{21} + h_{22})
 &= \frac{1}{2}\log\left(1+ \frac{2\sqrt{h_{21}h_{22}} } {1+ h_{21} + h_{22} } \right) \nonumber\\
 &< \frac{1}{2}\log (1+1) =\frac{1}{2}.
\end{align}
Therefore, the loss caused by this difference is at most $\frac{1}{2}$ bit, and $(R_1-1,R_2-1)$ would be achievable. 
On the other hand, as we argued before, the capacity of the modified network is an inner bound for the original one, and hence, 
$(R_1-1,R_2-1)$ is achievable for the $\ZS$ network as well.

\end{example}

\section{The Gaussian $\ZZ$ Network}
\label{sec:gzz}
\setcounter{equation}{0}

\subsection{The Outer Bound}
\label{subsec:gzz-conv}

\begin{figure}[h!]
 \centering
 	\psfrag{s1}[Bc][Bc]{$S_1$}
	\psfrag{s2}[Bc][Bc]{$S_2$}
	\psfrag{r1}[Bc][Bc]{$A$}
	\psfrag{r2}[Bc][Bc]{$B$}
	\psfrag{d1}[Bc][Bc]{$D_1$}
	\psfrag{d2}[Bc][Bc]{$D_2$}
	\psfrag{a}[Bc][Bc]{$\sqrt{g_{11}}$}
	\psfrag{b}[Bc][Bc]{$\sqrt{h_{11}}$}
	\psfrag{c}[Bc][Bc]{$\sqrt{g_{22}}$}
	\psfrag{d}[Bc][Bc]{$\sqrt{h_{22}}$}
	\psfrag{f}[Bc][Bc]{$\sqrt{g_{12}}$}
	\psfrag{g}[Bc][Bc]{$\sqrt{h_{12}}$}
 	\psfrag{X_1}[Bc][Bc]{$x_1$}
	\psfrag{X_2}[Bc][Bc]{$x_2$}
	\psfrag{Y'_1}[Bc][Bc]{$y'_1$}
	\psfrag{Y'_2}[Bc][Bc]{$y'_2$}
	\psfrag{X'_1}[Bc][Bc]{$x'_1$}
	\psfrag{X'_2}[Bc][Bc]{$x'_2$}
	\psfrag{Y_1}[Bc][Bc]{$y_1$}
	\psfrag{Y_2}[Bc][Bc]{$y_2$}
	\psfrag{Z'_1}[Bc][Bc]{$z'_1$}
	\psfrag{Z'_2}[Bc][Bc]{$z'_2$}
	\psfrag{Z_1}[Bc][Bc]{$z_1$}
	\psfrag{Z_2}[Bc][Bc]{$z_2$}
	\psfrag{t_1}[Bc][Bc]{$\gamma_1$}
	\psfrag{t_2}[Bc][Bc]{$\gamma_2$}
\includegraphics[width=9cm]{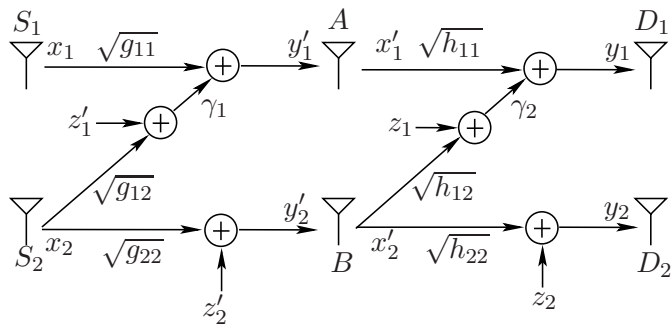}
\vspace{-4mm}
\caption{The Gaussian $\ZZ$ network.}
\label{fig:GZZ-2}
\vspace{-4mm}
\end{figure}

In the following we present the proof for each of the inequalities in
(\ref{gzz:s1-o})-(\ref{gzz:d1d2-o}), separately.  We again present the
Gaussian $\ZZ$ network in Figure~\ref{fig:GZZ-2}, to clarify the
notation used in the proof. In particular, we use two variables,
which are the noisy signals received at $A$ and $D_1$ through the
cross links assuming the direct links were absent, namely,
\begin{align}
\gamma_1=\sqrt{g_{12}} x_2 + z'_1,\nonumber\\
\gamma_2=\sqrt{h_{12}} x'_2 + z_1.\nonumber
\end{align}
Note that $y'_1=\sqrt{g_{11}} x_{1}+\gamma_1$ and $y_1=\sqrt{h_{11}} x'_{1}+\gamma_2$.

Suppose that the rate pair $(R_1,R_2)$ is achieved with a small decoding
error probability $\e_{\ell}$ using a code of length $\ell$. The following chains
of inequalities provide upper bounds on the individual rates as well
as the sum-rate. We again use Lemma~\ref{lm:gzs-fano}, which
essentially captures the decodability requirements of the network.

The individual rate bounds in \eqref{gzz:s1-o}-\eqref{gzz:d2-o} have
the same structure as the cut-set bound, although we derive them
through a slightly different argument. However, the two sum-rate
bounds in \eqref{gzz:s1s2-o} and \eqref{gzz:d1d2-o} are conceptually
different than the cut-set bounds. These two bounds which are tighter
than cut-set bounds are derived through a genie-aided argument; that
is, we assume that the signal sent over the cross link of one layer is
given by a genie to the receiver of the other layer (relay node $A$ in
layer $1$ and destination node $D_1$ in layer $2$).  Therefore, we
present the proofs of \eqref{gzz:s1s2-o} and \eqref{gzz:d1d2-o}
first. The more standard cut-set type bounds are provided later for
completeness.

\paragraph*{a) The proof of the genie-aided bounds}
\begin{itemize}
 
\item (\ref{gzz:s1s2-o}) $\ \ R_1+R_2 \leq \frac{1}{2} \log(1+g_{11}+g_{12}) + \frac{1}{2}\log\left(1+\frac{g_{22}}{g_{12}}\right) + \frac{1}{2}\log(1+h_{12})$:
We start with the sum-rate inequality in Lemma~\ref{lm:gzs-fano}, and write  
\begin{align}
\ell(R_1+R_2) & \leq I( y^{\ell}_1, y^{\ell}_2 ; x^{\ell}_1, x^{\ell}_2) +\ell \e_{\ell}\nonumber\\
&\leq I( y^{'\ell}_1, y^{'\ell}_2 ; x^{\ell}_1, x^{\ell}_2) +\ell \e_{\ell}\nonumber\\
&\leq I( y^{'\ell}_1, y^{'\ell}_2, \gamma_2^{\ell} ; x^{\ell}_1, x^{\ell}_2) +\ell \e_{\ell}\nonumber\\
&= I( y^{'\ell}_1, \gamma_2^{\ell} ; x^{\ell}_1, x^{\ell}_2) 
    + I( y^{'\ell}_2 ; x^{\ell}_1, x^{\ell}_2| y^{'\ell}_1, \gamma_2^{\ell} ) +\ell \e_{\ell}\nonumber\\
&\leq I( y^{'\ell}_1, \gamma_2^{\ell} ; x^{\ell}_1, x^{\ell}_2) 
    + I( y^{'\ell}_2, \gamma_1^{\ell} ; x^{\ell}_1, x^{\ell}_2| y^{'\ell}_1, \gamma_2^{\ell} ) +\ell \e_{\ell}\nonumber\\
&= I( y^{'\ell}_1, \gamma_2^{\ell} ; x^{\ell}_1, x^{\ell}_2) 
    + I( \gamma_1^{\ell} ; x^{\ell}_1, x^{\ell}_2| y^{'\ell}_1, \gamma_2^{\ell} ) 
    +  I( y^{'\ell}_2 ; x^{\ell}_1, x^{\ell}_2| y^{'\ell}_1, \gamma_1^{\ell}, \gamma_2^{\ell} ) +\ell \e_{\ell}\nonumber\\
&= I( y^{'\ell}_1, \gamma_2^{\ell} ; x^{\ell}_1, x^{\ell}_2) 
    + I( \gamma_1^{\ell} ; x^{\ell}_1, x^{\ell}_2| y^{'\ell}_1, \gamma_2^{\ell} )  
+   I( y^{'\ell}_2 ; x^{\ell}_2| y^{'\ell}_1, \gamma_1^{\ell}, \gamma_2^{\ell} )  
    + I( y^{'\ell}_2 ; x^{\ell}_1|  x^{\ell}_2 , y^{'\ell}_1, \gamma_1^{\ell}, \gamma_2^{\ell} ) +\ell \e_{\ell}.
 \label{gzz:s1s2_1}
\end{align}
Each of the terms in (\ref{gzz:s1s2_1}) can be bounded as follows. In
order to bound the first term, we can simply write
\begin{align}
 I( y^{'\ell}_1, \gamma_2^{\ell} ; x^{\ell}_1, x^{\ell}_2) 
&= I( \gamma^{\ell}_2; x^{\ell}_1, x^{\ell}_2 ) + I(y^{'\ell}_1 ; x^{\ell}_1, x^{\ell}_2|\gamma^{\ell}_2 )\nonumber\\
&=I( \gamma^{\ell}_2; x^{\ell}_1, x^{\ell}_2 ) + h(y^{'\ell}_1 |\gamma^{\ell}_2 ) 
    - h(y^{'\ell}_1 | x^{\ell}_1, x^{\ell}_2, \gamma^{\ell}_2 )\nonumber\\
&\leq I( \gamma^{\ell}_2; x^{\ell}_1, x^{\ell}_2 ) + h(y^{'\ell}_1 )  
    - h(y^{'\ell}_1 | x^{\ell}_1, x^{\ell}_2 )\label{gzz:s1s2:cond}\\
&= I(\gamma_2^{\ell} ; x^{\ell}_1, x^{\ell}_2) + I( y^{'\ell}_1; x^{\ell}_1, x^{\ell}_2 ) \nonumber\\
&\leq I(\gamma_2^{\ell} ; x^{'\ell}_2) + I( y^{'\ell}_1; x^{\ell}_1, x^{\ell}_2 ) \label{gzz:s1s2:mrk}\\
&= \frac{\ell}{2}\log(1+h_{12})+ \frac{\ell}{2}  \log(1+g_{11}+g_{12}), \label{gzz:s1s2_2}
\end{align}
where in (\ref{gzz:s1s2:cond}) we have used the fact that conditioning
decreases the entropy, and the Markov chain $\gamma^{\ell}_2
\leftrightarrow x^{'\ell}_2 \leftrightarrow y^{'\ell}_2
\leftrightarrow (x^{\ell}_1,x^{\ell}_2) \leftrightarrow
y^{'\ell}_1$. Also (\ref{gzz:s1s2:mrk}) follows from the same Markov
chain.

For the second term, we can write
\begin{align}
 I( \gamma_1^{\ell} ; x^{\ell}_1, x^{\ell}_2| y^{'\ell}_1, \gamma_2^{\ell} )&= 
 	 I(  y^{'\ell}_1-\gamma_1^{\ell} ; x^{\ell}_1, x^{\ell}_2| y^{'\ell}_1, \gamma_2^{\ell} )\nonumber\\
&=I (\sqrt{g_{11}} x^{\ell}_1 ; x^{\ell}_1, x^{\ell}_2| y^{'\ell}_1, \gamma_2^{\ell} )\nonumber\\
&\leq I (W_1 ; x^{\ell}_1, x^{\ell}_2| y^{'\ell}_1, \gamma_2^{\ell} )\nonumber\\
&\leq H(W_1 |y^{'\ell}_1, \gamma_2^{\ell} )\nonumber\\ 
&\leq H(W_1 | y^{\ell}_1)\leq \ell \e_{\ell}, \label{gzz:s1s2_3}
\end{align}
where the last inequality holds since $y^{\ell}_1=\sqrt{h_{11}}
x^{'\ell}_1+\gamma_2^{\ell} = f_1(y^{'\ell}_1) + \gamma_2^{\ell}
=f_2(y^{'\ell}_1, \gamma_2^{\ell})$.

In order to bound the third term in (\ref{gzz:s1s2_1}) we can write 
\begin{align}
  I( y^{'\ell}_2 ; x^{\ell}_2| y^{'\ell}_1, \gamma_1^{\ell}, \gamma_2^{\ell} )
&= h( x^{\ell}_2 | y^{'\ell}_1, \gamma_1^{\ell}, \gamma_2^{\ell} )
      -h( x^{\ell}_2 | y^{'\ell}_1, y^{'\ell}_2, \gamma_1^{\ell}, \gamma_2^{\ell} )\nonumber\\
&\leq h( x^{\ell}_2 |  \gamma_1^{\ell} )
      -h( x^{\ell}_2 | y^{'\ell}_1, y^{'\ell}_2, \gamma_1^{\ell}, \gamma_2^{\ell} )\label{gzz:s1s2:cond_2}\\
  &= h( x^{\ell}_2 |  \gamma_1^{\ell} )
      -[h( x^{\ell}_2, t^{\ell}_1 | y^{'\ell}_1, y^{'\ell}_2, \gamma_2^{\ell} ) 
      - h(t^{\ell}_1 | y^{'\ell}_1, y^{'\ell}_2, \gamma_2^{\ell})]\nonumber\\
  &= h( x^{\ell}_2 |  \gamma_1^{\ell} )-[h( x^{\ell}_2, t^{\ell}_1 | y^{'\ell}_1, y^{'\ell}_2 ) 
      - h(t^{\ell}_1 | y^{'\ell}_1, y^{'\ell}_2)]\label{gzz:s1s2:mrk_1}\\
  &= h( x^{\ell}_2 |  \gamma_1^{\ell} )-h( x^{\ell}_2| y^{'\ell}_1, y^{'\ell}_2, \gamma_1^{\ell} ) \nonumber\\
  &= h( x^{\ell}_2 | \gamma_1^{\ell} )-h(  x^{\ell}_2 | x^{\ell}_1, y^{'\ell}_2,  \gamma_1^{\ell} )\label{gzz:s1s2:cond_3}\\
  &= h( x^{\ell}_2 |  \gamma_1^{\ell} )-[h( x^{\ell}_2| y^{'\ell}_2, \gamma_1^{\ell} )
      + h( x^{\ell}_1| x^{\ell}_2, y^{'\ell}_2, \gamma_1^{\ell} ) - h(x^{\ell}_1 | y^{'\ell}_2, \gamma_1^{\ell})]\nonumber\\
  &= h( x^{\ell}_2 |  \gamma_1^{\ell} )-h( x^{\ell}_2| y^{'\ell}_2, \gamma_1^{\ell} )\label{gzz:s1s2:cond_4}\\
  &= I( y^{'\ell}_2 ; x^{\ell}_2| \gamma_1^{\ell} )\nonumber\\
  &=h( y^{'\ell}_2 | \gamma_1^{\ell} ) - h( y^{'\ell}_2 |  \gamma_1^{\ell}, x^{\ell}_2 )\nonumber\\
  &= h( y^{'\ell}_2 | \gamma_1^{\ell} ) - h( y^{'\ell}_2 | x^{\ell}_2 )\label{gzz:s1s2:mrk_2}\\
  &= h( y^{'\ell}_2 - \sqrt{\frac{g_{22}}{g_{12}}} \gamma_1^{\ell} | \gamma_1^{\ell} ) 
      - h( y^{'\ell}_2 -\sqrt{g_{22}} x^{\ell}_2| x^{\ell}_2 )\nonumber\\
  &= h( z^{'\ell}_2 - \sqrt{\frac{g_{22}}{g_{12}}} z^{'\ell}_1 | \gamma_1^{\ell} ) - h(  z^{'\ell}_2  | x^{\ell}_2 )\nonumber\\
  &\leq h( z^{'\ell}_2 - \sqrt{\frac{g_{22}}{g_{12}}} z^{'\ell}_1 ) - h(  z^{'\ell}_2 )\label{gzz:s1s2:rem_cond}\\
  &=
  \frac{\ell}{2}\log\left(1+\frac{g_{22}}{g_{12}}\right),\label{gzz:s1s2_4}
\end{align}
where in (\ref{gzz:s1s2:cond_2}) we have used the fact that
conditioning reduces the differential entropy, and
\eqref{gzz:s1s2:mrk_1} holds due to the Markov chain
$(x^{\ell}_2,t^{\ell}_1) \leftrightarrow
(y^{'\ell}_1,y^{'\ell}_2)\leftrightarrow \gamma_2^{\ell} $. Then in
(\ref{gzz:s1s2:cond_3}) we replaced $(y^{'\ell}_1, \gamma_1^{\ell} )$
by $(x^{\ell}_1, \gamma_1^{\ell} )$ since there is an one-to-one map,
$y^{'\ell}_1=\sqrt{g_{11}} x^{\ell}_1 +t^{\ell}_1$, between these
joint variables, and in (\ref{gzz:s1s2:cond_4}) we used the fact that
$x^{\ell}_1$ is independent of $(x^{\ell}_2, y^{'\ell}_2,t^{\ell}_1)$
to conclude $h( x^{\ell}_1| x^{\ell}_2, y^{'\ell}_2, \gamma_1^{\ell} )
= h(x^{\ell}_1 | y^{'\ell}_2, \gamma_1^{\ell})=h(x)$. Also
(\ref{gzz:s1s2:mrk_2}) holds due to the Markov chain $y^{'\ell}_2
\leftrightarrow x^{\ell}_2 \leftrightarrow \gamma_1^{\ell} $. Finally,
(\ref{gzz:s1s2:rem_cond}) is true due to removing conditioning and the
fact that $z'_2$ is independent of $x_2$.

Finally for the last term in (\ref{gzz:s1s2_1}) we have

\begin{align}
I( y^{'\ell}_2 ; x^{\ell}_1|  x^{\ell}_2 , y^{'\ell}_1, \gamma_1^{\ell}, \gamma_2^{\ell} ) 
&\leq I( y^{'\ell}_2 ; W_1|  x^{\ell}_2 , y^{'\ell}_1, \gamma_1^{\ell}, \gamma_2^{\ell} )\nonumber\\
&\leq H( W_1|  x^{\ell}_2 , y^{'\ell}_1, \gamma_1^{\ell}, \gamma_2^{\ell} )\nonumber\\
&\leq H( W_1|  y^{'\ell}_1,\gamma_2^{\ell} )\nonumber\\
&\leq H( W_1|  y^{\ell}_1 )\label{gzz:s1s2:func}\\
&\leq \ell \e_{\ell}, \label{gzz:s1s2_5}
\end{align}
where \eqref{gzz:s1s2:func} is due to the fact that
$y^{\ell}_1=\sqrt{h_{11}} x^{'\ell}_1+\gamma_2^{\ell} =
f_1(y^{'\ell}_1) + \gamma_2^{\ell} =f_2(y^{'\ell}_1, \gamma_2^{\ell})$
is a function of $(y^{'\ell}_1,\gamma_2^{\ell})$, and
\eqref{gzz:s1s2_5} is just the Fano's inequality.

Replacing (\ref{gzz:s1s2_2}), (\ref{gzz:s1s2_3}), (\ref{gzz:s1s2_4}),
and (\ref{gzz:s1s2_5}) in (\ref{gzz:s1s2_1}), we get

\begin{align}
  R_1+R_2 \leq \frac{1}{2} \log(1+g_{11}+g_{12}) +
  \frac{1}{2}\log\left(1+\frac{g_{22}}{g_{12}}\right) +
  \frac{1}{2}\log(1+h_{12}) +3\ell \e_{\ell}.
\end{align}

\item (\ref{gzz:d1d2-o}) Before proving this inequality, we
  present a lemma which will be used in this proof.  We will present
  the proof of this lemma later in Appendix~\ref{sec:app-lm}.

\begin{lemma}
  Let $X_1$ and $X_2$ be two (arbitrarily correlated) random variables
  with variance constraints $\E[X^2_1]=\sigma_1^2$ and
  $\E[X^2_2]=\sigma_2^2$, which form a Markov chain $X_1
  \leftrightarrow \Gamma \leftrightarrow X_2$ for some random variable
  $\Gamma$. Also assume that $Z$ is a zero-mean unit variance Gaussian
  random variable independent of $X_1$, $X_2$ and $\Gamma$. Then the
  conditional differential entropy of $Y=X_1+X_2+Z$ is upper bounded
  by
\begin{align}
h(Y|\Gamma) \leq \frac{1}{2}\log 2\pi e(1+\sigma_1^2+\sigma_2^2).
\end{align}
\label{lm:zzg:indep-mac}
\end{lemma}

Now, in order to prove \eqref{gzz:s1s2-o},  we start with Lemma~\ref{lm:gzs-fano}. 
%
\begin{align}
  \ell(R_1+R_2) &\leq  I( y_1^{\ell}, y_2^{\ell} ; x_1^{\ell},x_2^{\ell}) + \ell \e_{\ell}\nonumber\\
  & \leq I( y^{\ell}_1, y^{\ell}_2, \gamma_1^{\ell} ; x^{\ell}_1, x^{\ell}_2)+ \ell \e_{\ell}\nonumber\\
  &= I( y^{\ell}_1, \gamma_1^{\ell} ; x^{\ell}_1, x^{\ell}_2) + I( y^{\ell}_2 ; x^{\ell}_1, x^{\ell}_2| y^{\ell}_1, \gamma_1^{\ell} )+ \ell \e_{\ell}\nonumber\\
  &= I( \gamma_1^{\ell} ; x^{\ell}_1, x^{\ell}_2) +I(y_1^{\ell};
  x_1^{\ell}, x_2^{\ell} | \gamma_1^{\ell}) + I( y^{\ell}_2 ;
  x^{\ell}_1, x^{\ell}_2| y^{\ell}_1, \gamma_1^{\ell} )+ \ell
  \e_{\ell}.\label{gzz:d1d2_0}
\end{align}
Since $\gamma_1^{\ell}$ is independent of $x_1^{\ell}$, the first term
can be simply bounded as
\begin{align}
  I( \gamma_1^{\ell} ; x^{\ell}_1, x^{\ell}_2) &= I(\gamma_1^{\ell} ; x_2^{\ell}) + I(t^{\ell}; x_1^{\ell} | x^{\ell}_2) 
  =  I(\gamma_1^{\ell} ; x_2^{\ell}) + I(z^{'\ell}_1; x_1^{\ell} | x^{\ell}_2)
  \leq \frac{\ell}{2} \log (1+g_{12}).
\label{gzz:d1d2_1}
\end{align}

For the second term we can write
\begin{align}
  I(y_1^{\ell}; x_1^{\ell}, x_2^{\ell} | \gamma_1^{\ell}) &= h(y_1^{\ell} | \gamma_1^{\ell}) - h(y_1^{\ell} | x_1^{\ell}, x_2^{\ell}, \gamma_1^{\ell})\nonumber\\
  &\leq h(y_1^{\ell} | \gamma_1^{\ell}) - h(y_1^{\ell} | x^{'\ell}_1, x^{'\ell}_2)\label{gzz:d1d2:a}\\
  &=h(y_1^{\ell} | \gamma_1^{\ell}) - h(z_1^{\ell} | x^{'\ell}_1, x^{'\ell}_2)\nonumber\\
  &=h(\sqrt{h_{11}}x^{'\ell}_1 + \sqrt{h_{12}}x^{'\ell}_2 + z^{\ell}_1 | \gamma_1^{\ell}) - h(z_1^{\ell})\nonumber\\
  &\leq \frac{\ell}{2}\log \Big( 2\pi e (1+h_{11}+h_{12})\Big)- \frac{\ell}{2}\log 2\pi e \label{gzz:d1d2:b}\\
  &=\frac{\ell}{2}\log (1+h_{11}+h_{12}),\label{gzz:d1d2_2}
\end{align}
where $(\ref{gzz:d1d2:a})$ follows from the Markov chain $ y_1^{\ell}
\leftrightarrow (x^{'\ell}_1,x^{'\ell}_2)\leftrightarrow (x_1^{\ell},
x_2^{\ell} , \gamma_1^{\ell}) $. In $(\ref{gzz:d1d2:b})$ we have used
Lemma~\ref{lm:zzg:indep-mac} for $t^{\ell}_1$, $x^{'\ell}_1$ and
$x^{'\ell}_2$ which form a Markov chain, since
\begin{align*}
  I(x^{'\ell}_1;x^{'\ell}_2|t^{\ell}_1) \leq
  I(y^{'\ell}_1;y^{'\ell}_2|t^{\ell}_1)
  =I(\frac{y^{'\ell}_1-t^{\ell}_1}{\sqrt{g_{11}}};y^{'\ell}_2|t^{\ell}_1)
  = I(x^{\ell}_1 ; y^{'\ell}_2 |
  t^{\ell}_1)=h(x^{\ell}_1|t^{\ell}_1)-h(x^{\ell}_1 | t^{\ell}_1,
  y^{'\ell}_2)=0.
\end{align*}

The third term can be further upper bounded by
\begin{align}
  I( y^{\ell}_2  ; x^{\ell}_1, x^{\ell}_2| y^{\ell}_1, \gamma_1^{\ell} )
  &= h( y^{\ell}_2 | y^{\ell}_1, \gamma_1^{\ell}) - h(y^{\ell}_2 | x^{\ell}_1, x^{\ell}_2, y^{\ell}_1, \gamma_1^{\ell})\nonumber\\
  &\leq h( y^{\ell}_2 | y^{\ell}_1, \gamma_1^{\ell}) - h(y^{\ell}_2 | x^{'\ell}_2)\label{gzz:d1d2:c}\\
  &\leq h( y^{\ell}_2 | y^{\ell}_1, \gamma_1^{\ell}) - h(y^{\ell}_2 | x^{'\ell}_2,x^{'\ell}_1, y^{\ell}_1, \gamma_1^{\ell})\label{gzz:d1d2:d}\\
  &=I( y^{\ell}_2  ; x^{'\ell}_1, x^{'\ell}_2| y^{\ell}_1, \gamma_1^{\ell} )\nonumber\\
  &\leq I( y^{\ell}_2, \gamma_2^{\ell}  ; x^{'\ell}_1, x^{'\ell}_2| y^{\ell}_1, \gamma_1^{\ell} )\nonumber\\
  &= I( \gamma_2^{\ell} ; x^{'\ell}_1, x^{'\ell}_2| y^{\ell}_1,
  \gamma_1^{\ell} ) + I( y^{\ell}_2 ; x^{'\ell}_1, x^{'\ell}_2|
  y^{\ell}_1, \gamma_1^{\ell},\gamma_2^{\ell} ),
\end{align}
where both \eqref{gzz:d1d2:c} and \eqref{gzz:d1d2:d} follow from the
Markov chain $ y_2^{\ell} \leftrightarrow x^{'\ell}_2 \leftrightarrow
(x_1^{\ell} , y^{\ell}_1, \gamma_1^{\ell}) $. Now, we have

\begin{align}
  I( \gamma_2^{\ell} ; x^{'\ell}_1, x^{'\ell}_2| y^{\ell}_1, \gamma_1^{\ell} )
  &=  I(  y^{\ell}_1-\gamma_2^{\ell} ; x^{'\ell}_1, x^{'\ell}_2| y^{\ell}_1, \gamma_1^{\ell} )\nonumber\\
  &=I (\sqrt{h_{11}} x^{'\ell}_1 ; x^{'\ell}_1, x^{'\ell}_2| y^{\ell}_1, \gamma_1^{\ell} )\nonumber\\
  &\leq I ( y^{'\ell}_1 ; x^{'\ell}_1, x^{'\ell}_2| y^{\ell}_1, \gamma_1^{\ell} )\label{gzz:d1d2:e}\\
  &\leq I ( y^{'\ell}_1-t^{\ell}_1 ; x^{'\ell}_1, x^{'\ell}_2| y^{\ell}_1, \gamma_1^{\ell} )\nonumber\\
  &= I ( \sqrt{g_{11}} x^{\ell}_1 ; x^{'\ell}_1, x^{'\ell}_2| y^{\ell}_1, \gamma_1^{\ell} )\nonumber\\
  &\leq I (W_1 ; x^{'\ell}_1, x^{'\ell}_2| y^{\ell}_1, \gamma_1^{\ell} )\nonumber\\
  &\leq H(W_1 | y^{\ell}_1)\leq \ell \e_{\ell}, \label{gzz:d1d2_3}
\end{align}
where \eqref{gzz:d1d2:e} follows from the fact that $x^{'\ell}_1$ is a function of $y^{'\ell}_1$. Finally, 

\begin{align}
  I( y^{\ell}_2 ; x^{'\ell}_1, x^{'\ell}_2| y^{\ell}_1, \gamma_1^{\ell}, \gamma_2^{\ell} )
  &= h( y^{\ell}_2 | y^{\ell}_1, \gamma_1^{\ell}, \gamma_2^{\ell} )
      -h( y^{\ell}_2 | y^{\ell}_1, \gamma_1^{\ell}, \gamma_2^{\ell}, x^{'\ell}_1, x^{'\ell}_2)\nonumber\\
  &=h(y^{\ell}_2 - \sqrt{\frac{h_{22}}{h_{12}}}\gamma_2^{\ell}| y^{\ell}_1, \gamma_1^{\ell}, \gamma_2^{\ell} )
      - h( y^{\ell}_2 -\sqrt{h_{22}} x^{'\ell}_2 | y^{\ell}_1, \gamma_1^{\ell}, \gamma_2^{\ell}, x^{'\ell}_1, x^{'\ell}_2)\nonumber\\
  &= h(z^{\ell}_2-\sqrt{\frac{h_{22}}{h_{12}}} z^{\ell}_1 |  y^{\ell}_1, \gamma_1^{\ell}, \gamma_2^{\ell}) 
      -  h( z^{\ell}_2 | y^{\ell}_1, \gamma_1^{\ell}, \gamma_2^{\ell}, x^{'\ell}_1, x^{'\ell}_2)\nonumber\\
  &\leq h(z^{\ell}_2-\sqrt{\frac{h_{22}}{h_{12}}} z^{\ell}_1) - h( z^{\ell}_2)\label{gzz:d1d2:f}\\
  &=
  \frac{\ell}{2}\log\left(1+\frac{h_{22}}{h_{12}}\right).\label{gzz:d1d2_4}
\end{align}
Here, in \eqref{gzz:d1d2:f} we have used the fact that conditioning
decreases the differential entropy, and the fact that $z^{\ell}_2$ is
independent of $(y^{\ell}_1, \gamma_1^{\ell}, \gamma_2^{\ell},
x^{'\ell}_1, x^{'\ell}_2)$.  Replacing (\ref{gzz:d1d2_1}),
(\ref{gzz:d1d2_2}), (\ref{gzz:d1d2_3}), and (\ref{gzz:d1d2_4}) in
(\ref{gzz:d1d2_0}), we will obtain the desired inequality.

\end{itemize}

\paragraph*{b) The proofs of cut-set type bounds}

\begin{itemize}

\item (\ref{gzz:s1-o}) $\ \ R_1 \leq \frac{1}{2}\log(1+ g_{11})$:
The individual rate bound can be simply obtained from 
\begin{align}
  \ell R_1 &= I(x_1^{\ell}; y_1^{\ell}) +\ell \e_{\ell} \nonumber\\
  & \leq  I(x_1^{\ell}; y_1^{'\ell} ,y_2^{'\ell}) +\ell \e_{\ell} \label{pr:gzz:s1-1}\\
  &=  I(x_1^{\ell}; y_1^{'\ell} | y_2^{'\ell})  +  I(x_1^{\ell}; y_2^{'\ell}) +\ell \e_{\ell} \nonumber\\
  &= h(y_1^{'\ell} | y_2^{'\ell})  - h(y_1^{'\ell} | x_1^{\ell} , y_2^{'\ell})  \label{pr:gzz:s1-2}\\
  &\leq h(y_1^{'\ell} | x_2^{\ell})  - h(z_1^{'\ell} | x_1^{\ell} , y_2^{'\ell}) \nonumber\\
  &= h(\sqrt{g_{11}} x^{\ell}_1 +z^{'\ell}_1)-h(z^{'\ell}_1)+\ell \e_{\ell}\nonumber\\
  &\leq \frac{\ell}{2}\log(1+g_{11})+ \ell \e_{\ell},\nonumber
\end{align}
where \eqref{pr:gzz:s1-1} follows from the data processing inequality
for the Markov chain $ x_1^{\ell}\leftrightarrow (y_1^{'\ell}
,y_2^{'\ell})\leftrightarrow y_1^{\ell}$, and \eqref{pr:gzz:s1-2}
follows from the Markov chain $ y_1^{'\ell} \leftrightarrow x_2^{\ell}
\leftrightarrow y_2^{'\ell}) $.  Note that $\e_{\ell}\rightarrow 0$ as
$\ell$ grows.  It is worth mentioning that this bound is similar to
the cut-set bound for the cut $\Omega_s=\{S_1\}$ and $\Omega_d=\{S_2,
A, B, D_1, D_2\}$.

\item (\ref{gzz:s2-o}) $\ \ R_2 \leq \frac{1}{2}\log(1+ g_{22})$: 

  For the second rate bound, we can start with Lemma~\ref{lm:gzs-fano}
  and write
\begin{align}
\ell R_2  &\leq I(x_2^{\ell} ; y^{\ell}_2 )  + \ell \e_{\ell} 
\leq I(x_2^{\ell} ; y^{'\ell}_2 ) + \ell \e_{\ell}
\leq  \frac{\ell}{2}\log(1+g_{22})+  \ell \e_{\ell},\nonumber
\end{align}
where we have used the data processing inequality and the Markov chain
$ x_2^{\ell}\leftrightarrow y_2^{'\ell} \leftrightarrow x_2^{'\ell}
\leftrightarrow y_2^{\ell}$ in the second inequality. Note that this bound
captures the maximum flow of information through the cut specified by
$\Omega_s=\{S_2\}$ and $\Omega_d=\{S_1,A, B, D_1, D_2\}$.

\item (\ref{gzz:d1-o}) $\ \ R_1 \leq \frac{1}{2}\log(1+ h_{11})$:
  In order to prove this upper bound, we use the cut-set bound for the
  cut $\Omega_s=\{S_1, S_2, A, B, D_2\}$ and $\Omega_d=\{D_1\}$.
\begin{align}
\ell R_1 &\leq I(x^{'\ell}_1 ; y^{\ell}_1 | x^{'\ell}_2) +  \ell \e_{\ell}\nonumber\\
&= h(y^{\ell}_1| x^{'\ell}_2)- h(y^{\ell}_1| x^{'\ell}_1, x^{'\ell}_2)+  \ell \e_{\ell}\nonumber\\
&= h(\sqrt{h_{11}}x^{'\ell}_1+z^{\ell}_1| x^{'\ell}_2)- h(z^{\ell}_1| x^{'\ell}_1, x^{'\ell}_2)+  \ell \e_{\ell}\nonumber\\
&\leq h(\sqrt{h_{11}}x^{'\ell}_1+z^{\ell}_1)- h(z^{\ell}_1)+  \ell \e_{\ell}\nonumber\\
&\leq \frac{\ell}{2}\log(1+ h_{11})+  \ell \e_{\ell}.\nonumber
\end{align}

\item (\ref{gzz:d2-o}) $\ \ R_2 \leq \frac{1}{2}\log(1+ h_{22})$:
Starting from Lemma~\ref{lm:gzs-fano}, we can write
\begin{align}
\ell R_2 &\leq I(x^{'\ell}_2 ; y^{\ell}_2) +  \ell \e_{\ell}
\leq I(x^{'\ell}_2 ; y^{\ell}_2) +  \ell \e_{\ell}
\leq \frac{1}{2}\log(1+ h_{22}) +  \ell \e_{\ell},\nonumber
\end{align}
where the second inequality follows from the data processing inequality for the Markov chain 
$ x_2^{\ell}\leftrightarrow y_2^{'\ell} \leftrightarrow x_2^{'\ell} \leftrightarrow y_2^{\ell}$.

\end{itemize}
This shows that the rate region in Theorem~\ref{thm:gzz} is an outer
bound for the achievable region region of the Gaussian $\ZZ$ network.

\subsection{The Achievability Part}
\label{subsec:gzz-ach}

In this section we present an encoding/decoding scheme, and derive an 
achieve rate region for this strategy. We then show that the gap between 
the boundary of this achievable rate region and that of the outer bound 
presented  in Theorem~\ref{thm:gzz} is upper bounded by a constant. 

Similar to the Gaussian $\ZS$ network, we only consider the large channel gain case, where we assume that all the channel gains are 
lower bounded by $1$. A similar argument to that we used for the $\ZS$ network shows that for small channel gain cases the network is reduced to 
a simple one and its gap analysis is fairly simple. 

We
essentially use the result of Lemma~\ref{lm:GZZ:Z} as an achievable
rate region for the $\Z$-neutralization network. We use notation
$(\lambda_{g}, \mu_{g})$ and $(\lambda_{h}, \mu_{h})$ to
distinguish between  $\lambda$ and $\mu$ parameters of the first and
the second layers of the network.

In the first layer of the network, each source node splits its message
into two parts, namely, functional and private parts, $W_1=\left(
  U_1^{(0)}, U_1^{(1)}\right)$ and $W_2=\left( U_2^{(0)},
  U_2^{(1)}\right)$, where the functional parts, have the same rate,
\emph{i.e.,} $\Upsilon_{1,0}=\Upsilon_{2,0}=\Tz$. 
  Both transmitters use a common
lattice code to encode their functional sub-messages into
$\x_{1,0}=\psi(U_1^{(0)})$ and $\x_{2,0}=\psi(U_2^{(0)})$, where
$\psi$ is the one-to-one encoding map induced by the lattice code. We
define the partial-invertible function by
\begin{align}
  \phi\left(U_1^{(0)}, U_2^{(0)}\right) =\psi^{-1}\left (
    \psi(U_1^{(0)}) + \psi(U_2^{(0)}) \right) = \psi^{-1}(\x_{1,0} +
  \x_{2,0}).
\end{align}

We denote the rates of the private 
sub-messages by $\To$ and $\Tt$, where $\Upsilon_i=R_i-\Tz$, for $i=1,2$.
The goal is to encode and forward messages to $A$ and $B$ in such a
way that $A$ can decode $U_1^{(1)}$ and $\phi(U_1^{(0)},U_2^{(0)})$,
and $B$ can decode $U_2^{(0)}$ and $U_2^{(1)}$. Based on
Lemma~\ref{lm:GZZ:Z}, this can be done provided that
\begin{align*}
\Tz &\leq \+{\CGN{\lambda_{g}}-\frac{1}{2}},\\
\Tz+\To &\leq \+{\CGN{g_{11}}-1},\\
\Tz +\Tt &\leq \+{\CGN{g_{22}}-1},\\
\Tz +\To+\Tt &\leq \+{\CGN{\mu_{g}}-\frac{3}{2}}.
\end{align*}

The second layer of the network is another $\Z$-neutralization
network with transmitters $A$ and $B$, and receivers $D_1$ and $D_2$. 
We use $V_1^{(0)}=\psi^{-1}\left ( \psi(U_1^{(0)} ) + \psi(
  U_2^{(0)} ) \right)$, $V_1^{(1)}= U_{1}^{(1)}$ as the functional and
private messages of the first relay node, and $V_2^{(0)}= \psi^{-1}
(-\x_{2,0}) = \psi^{-1} (-\psi (U_2^{(0)} ))$ and $V_2^{(1)}=
U_2^{(1)}$ for the functional and private messages of second
relay. Denoting the corresponding rates by $\Qz$, $\Qo$, and $\Qt$, we have
\begin{align}
\Theta_i&=\Upsilon_i, \qquad i=0,1,2.
\end{align}
The goal is to encode and send these messages to the destinations,
such that $D_1$ can decode $\phi(V_1^{(0)}, V_2^{(0)}) $ and
$V_1^{(1)}$, and $D_2$ can decode $V_2^{(0)}$ and $V_2^{(1)}$. Again
we use the achievable rate region proposed in Lemma~\ref{lm:GZZ:Z}.
\begin{align*}
\Qz &\leq \+{\CGN{\lambda_{h}}-\frac{1}{2}},\\
\Qz+\Qo &\leq \+{\CGN{h_{11}}-1},\\
\Qz+\Qt &\leq \+{\CGN{h_{22}}-1},\\
\Qz+\Qo+\Qt &\leq \+{\CGN{\mu_{h}}-\frac{3}{2}}.
\end{align*}
Note that the first destination observes $\phi(V_1^{(0)}, V_2^{(0)})$, 
which is equivalent to 
\begin{align}
\phi\left(V_1^{(0)}, V_2^{(0)}\right) &= \psi^{-1}\left( \psi( V_1^{(0)} ) + \psi( V_2^{(0)}  ) \right)\nonumber\\
&= \psi^{-1}\left( \psi(U_1^{(0)} ) + \psi( U_2^{(0)} )  -\psi (U_2^{(0)}) \right)\\
 &= U_1^{(0)}.
\end{align}
Therefore, combining it with $V_1^{(1)}=U_{1}^{(1)}$, the first destination 
node can decode $W_1$. The second destination node $D_2$ has
$V_2^{(0)}$ and $V_2^{(1)} =U_2^{(1)}$, and can compute
\begin{align}
\psi^{-1} \left(- \psi(V_2^{(0)})\right) = U_2^{(0)},
\end{align}
and hence it decodes $W_2$.

This scheme can reliably transmit the messages with rate pair in
\begin{align}
\R{\GZZ}{ach}=\Big\{(R_1,R_2): &\exists \ \Tz,\To,\Tt, \Qz, \Qo,\Qt \geq 0,\\
& R_1=\Qz+\Qo,\nonumber\\
& R_2=\Qz+\Qt, \nonumber\\
& \Theta_{i}=\Upsilon_i,\qquad i=0,1,2,\nonumber\\
&\Tz \leq \+{\CGN{\lambda_{g}}-\frac{1}{2}},\nonumber\\
&\Tz+\To \leq \+{\CGN{g_{11}}-1},\nonumber\\
&\Tz +\Tt \leq \+{\CGN{g_{22}}-1},\nonumber\\
&\Tz +\To+\Tt \leq \+{\CGN{\mu_{g}}-\frac{3}{2}},\nonumber\\
&\Qz \leq \+{\CGN{\lambda_{h}}-\frac{1}{2}},\nonumber\\
&\Qz+\Qo \leq \+{\CGN{h_{11}}-1},\nonumber\\
&\Qz+\Qt \leq \+{\CGN{h_{22}}-1},\nonumber\\
&\Qz+\Qo+\Qt \leq \+{\CGN{\mu_{h}}-\frac{3}{2}}\Big\}.
\end{align}

It only remains to apply Fourier-Motzkin elimination to project
this region onto $(R_1,R_2)$. This gives us
\begin{align}
\R{\GZZ}{ach}=\Big\{(R_1,R_2): 
&R_1\leq \+{\CGN{g_{11}}-1},\nonumber\\
&R_1 \leq \+{\CGN{h_{11}}-1},\nonumber\\
&R_2 \leq \+{\CGN{g_{22}}-1},\nonumber\\
&R_2 \leq \+{\CGN{h_{22}}-1},\nonumber\\
&R_1 + R_2 \leq  \+{\CGN{\mu_{g}}+ \CGN{\lambda_{h}}-\frac{3}{2}},\nonumber\\
&R_1 + R_2 \leq  \+{\CGN{\mu_{h}}+ \CGN{\lambda_{g}}-\frac{3}{2}}\Big\}.
\nonumber
\end{align}
Note that the RHS's of the sum-rate bounds depend on the order of the
channel gains. For most of possible orderings, these two inequalities
would be consequences of the individual rate bounds. For example, if
$\lambda_{h}= h_{22}$, then the last bound is implied by the first
and fourth bounds, since $\mu_{g}\geq g_{11}$. It can be shown in
general that $\R{\GZZ}{ach}$ is equivalent to
\begin{align}
\R{\GZZ}{ach}=\Big\{(R_1,R_2): 
&R_1\leq \+{\frac{1}{2}\log (g_{11})-1},\nonumber\\
&R_1 \leq \+{\frac{1}{2}\log (g_{22})-1},\nonumber\\
&R_2 \leq \+{\frac{1}{2}\log (h_{11})-1},\nonumber\\
&R_2 \leq \+{\frac{1}{2}\log (h_{22})-1},\nonumber\\
&R_1 + R_2 \leq  \+{\frac{1}{2}\log(\mu_{g})+ \frac{1}{2}\log(h_{12})-\frac{3}{2}},\nonumber\\
&R_1 + R_2 \leq  \+{\frac{1}{2}\log(\mu_{h})+ \frac{1}{2}\log(g_{12})-\frac{3}{2}}\Big\}.
\nonumber
\end{align}

Now, note that $g_{11}\geq 1$, $g_{12}\geq 1$, and $g_{22}\geq 1$. These
imply
\begin{align}
\frac{1}{2} \log(1+g_{11}+g_{12}) + \frac{1}{2}\log\left(1+\frac{g_{22}}{g_{12}}\right) 
&\leq \frac{1}{2} \log (3\max\{g_{11},g_{12}\}) +  \frac{1}{2}\log\left(\frac{2\max\{g_{12},g_{22}\} }{g_{12}}\right)\nonumber\\
&\leq \CG{\max\{g_{11},g_{12}\}\cdot \max\{g_{22},g_{12}\} }{g_{12}} + \frac{1}{2} \log 6\nonumber\\
&\leq \frac{1}{2}\log(\mu_{h,\Z})+ \frac{1}{2} \log 6.\label{eq:gzs-gap1}
\end{align}
We also have 
\begin{align}
  \frac{1}{2}\log (1+ x) \leq \frac{1}{2}\log (x) +\frac{1}{2}
\label{eq:gzs-gap2}
\end{align}
for all $x\geq 1$.  Applying \eqref{eq:gzs-gap1} and
\eqref{eq:gzs-gap2}, we obtain the following achievable rate region,
which is a subset of $\R{\GZZ}{ach}$.
\begin{align*}
\R{\GZZ}{ach,2}=\Big\{(R_1,R_2):
&R_1 \leq \+{\frac{1}{2}\log(1+ g_{11}) -\frac{3}{2}},\\
&R_2 \leq \+{\frac{1}{2}\log(1+ g_{22}) -\frac{3}{2}},\\
&R_1 \leq \+{\frac{1}{2}\log(1+ h_{11}) -\frac{3}{2}},\\
&R_2 \leq \+{\frac{1}{2}\log(1+ h_{22}) -\frac{3}{2}},\\
&R_1+R_2 \leq \+{  \frac{1}{2} \log(1+g_{11}+g_{12}) + \frac{1}{2}\log\left(1+\frac{g_{22}}{g_{12}}\right) 
  + \frac{1}{2}\log(1+h_{12}) - \frac{7}{2}} ,\\
&R_1+R_2  \leq  \+{\frac{1}{2} \log(1+h_{11}+h_{12}) + \frac{1}{2}\log\left(1+\frac{h_{22}}{h_{12}}\right) 
  + \frac{1}{2}\log(1+g_{12}) - \frac{7}{2}}  \Big\} 
\end{align*}
Therefore, for any rate pair $(R_1,R_2)\in \R{\GZZ}{}$, the rate pair
$(R_1-\frac{1}{4}\log 12, R_2-\frac{1}{4}\log 12 )$ belongs to
$\R{\GZZ}{ach,2}$, and therefore can be achieved using the proposed
encoding scheme.

\section{Proof of Lemmas}
\label{sec:app-lm}
\setcounter{equation}{0}

\begin{IEEEproof}[Discussion of Example~\ref{ex:GZ} in Section~\ref{sec:examples}]
The converse proof is fairly simple and follows from a similar argument we used to prove \eqref{gzs:s1}, \eqref{gzs:s2}, and \eqref{gzs:s1s2}
in Appendix~\ref{sec:gzs}. 

In the following we will present an encoding strategy which guarantees to achieve rate pair $(R_1-\frac{1}{2},R_2-\frac{1}{2})$,
provided that  $(R_1,R_2)\in\R{\Z}{}$. This gives us an approximate capacity characterization for the Gaussian $\Z$ network. In 
order to do this, we consider the following two cases.  

\noindent\emph{Case A: $g_{12}\geq g_{22}$}

Assume $(R_1,R_2)$ be an achievable rate pair.  Then, the first receiver $G_1$ is able to decode $W_1$ 
sent at rate $R_1$, and remove the signal associated to $W_1$ from its received signal. 
The remaining signal provides a higher $\SNR$ to decode $W_2$ than the signal received at $G_2$. 
Therefore, in this particular regime, the first receive would be able to decode both messages. 
Hence, we have a Gaussian multiple access channel from  $F_1$ and $F_2$ to $G_1$, combined with a 
line network from $F_2$ to $G_2$. Therefore, the intersection of the rate regions of the Gaussian 
MAC and the line networks is simply achievable. That is
\begin{align}
\R{\Z}{ach,A}=&\left\{ (R_1,R_2):  R_1 \leq \CGN{1+g_{11}},
		 R_2 \leq \CGN{1+g_{12}},
		 R_1+R_2 \leq \CGN{1+g_{11}+g_{12}}\right\}\nonumber\\
& \bigcap \nonumber\\
&\left\{ (R_1,R_2):  R_2 \leq \CGN{1+g_{22}}\right\}\nonumber\\
=&\Big\{ (R_1,R_2):  R_1 \leq \CGN{1+g_{11}},
		 R_2 \leq \CGN{1+g_{22}},
		 R_1+R_2 \leq \CGN{1+g_{11}+g_{12}}\Big\}.
\end{align}
Note that the individual rate bounds in $\R{\Z}{}$ and $\R{\Z}{ach,A}$ are the same. 
Moreover, the difference between the sum rate bounds is bounded by 
\begin{align}
\CGN{1+\frac{g_{22}}{g_{12}}} \leq \CGN{1+1} =\frac{1}{2}.
\end{align}
Therefore, the gap between each boundary point of $\R{\Z}{}$ and $\R{\Z}{ach,A}$ is at most $\frac{1}{2}$ bit.

\noindent\emph{Case B: $g_{12}\leq g_{22}$:}

The encoding scheme we introduce for this case is similar to Han-Kobayashi's scheme for $2$-user 
interference channel. We first split the second message $W_2$ into the common and private 
parts, $W_2=(W_2^c,W_2^p)$, with rates $R_2^c$ and $R_2^p$, respectively, where $W_2^c$ can 
be decoded at both receivers and $W_2^p$ is only decodable at $G_2$. Sub-messages $W_1$,
$W_2^c$, and $W_2^p$ are encoded by corresponding randomly generated Gaussian codes to 
$\x_1$, $\x_2^c$ and $\x_2^p$, and the resulting codewords are sent over the channel. 

We allocate $\alpha_p=1/g_{12}$ fraction of the transmission power available at $F_2$ to $W_2^p$, 
 and the remaining power $\alpha_c=1-\alpha_p$ is allocated to $W_2^c$. Therefore, we have
\begin{align*}
\x_2=\sqrt{\alpha_c}\x_2^p + \sqrt{\alpha_c}\x_2^p.
\end{align*} 
 
The first receiver, $G_1$, decodes $W_1$ and $W_2^c$ treating $W_2^p$ as noise. Therefore, 
the effective noise power received at $G_1$ would be $\E[\sqrt{g_{12} \alpha_p} x_p +z_1]^2=2$. According 
to the capacity region of Gaussian multiple access channel, this can be done provided that 
\begin{align}
\begin{array}{rl}
R_1 &\leq \CGN{1+\frac{g_{11}}{2}},\\
R_2^c &\leq \CGN{\frac{1+g_{12}}{2}},\\
R_1 +R_2^c &\leq \CGN{\frac{1+g_{11}+g_{12}}{2}}.
\end{array}
\label{eq:ex:GZ:MAC}
\end{align} 
The second decoder first decodes $W_2^c$ treating $W_2^p$ as noise.  It then removes the 
corresponding codeword from the received signal, and decodes $W_2^p$. This can be done as long as
\begin{align}
\begin{array}{rl}
R_2^c &\leq \CGN{\frac{1+g_{22}}{1+g_{22}/g_{12}}},\\
R_2^p &\leq \CGN{1+\frac{g_{22}}{g_{12}}}.
\end{array}
\label{eq:ex:GZ:line}
\end{align}
Note that we have two upper bounds for $R_2^c$. However, it is easy to show that $\frac{1+g_{22}}{1+g_{22}/g_{12}} \geq {\frac{1+g_{12}}{2} }$, 
for $1\leq g_{12}\leq g_{22}$, and therefore, the first bound dominates the second one. 
 Using Fourier-Motzkin elimination to write the achievable region in terms of $R_1$ and 
$R_2=R_2^c+R_2^p$, and after some simplification, we get that the region
\begin{align}
\R{\Z}{ach,B}=\Big\{ (R_1,R_2): & R_1 \leq \CGN{1+g_{11}}-\frac{1}{2},\\
			  & R_2 \leq \CGN{1+g_{22}}-\frac{1}{2},\\
			  & R_1+R_2 \leq \CGN{1+g_{11}+g_{12}}+\CGN{1+\frac{g_{22}}{g_{12}}}-\frac{1}{2}.
\Big\}. 
\end{align}
is achievable. Therefore, if $(R_1,R_2)\in\R{\Z}{}$, then $(R_1-\frac{1}{2},R_2-\frac{1}{2})$ is achievable. 
\end{IEEEproof}

\begin{IEEEproof}[Proof of Lemma~\ref{lm:GZS:Z} in Section~\ref{sec:pr-outline}]
The following achievability scheme simply uses 
 superposition encoding of sub-messages at $F_2$, and a successively decode and cancel strategy at $G_1$ and $G_2$. 
 We use a random codebook with a proper number of codewords, generated according to a zero-mean unit-variance Gaussian 
 distribution for each message.  A proper 
 power allocation for the messages at the transmitters allow the decoders to apply 
a decode and cancel strategy. We denote the codeword corresponding to the message $U_{i}^{(j)}$ by $\x_{i,j}$, and the 
power allocated to this message by $\alpha_{i,j}$. 

The available power at $F_2$ can be arbitrarily allocated to its sub-messages. In particular, we choose the power coefficients so that 
they satisfy $\att\leq 1/g_{22}$,  $\atr \leq 1/g_{12}$, and $\ato=1-\att-\atr$. 
In the decoding part,  $G_1$ and $G_2$ treat $\Utr$ and $\Utt$, respectively,  as noise. Therefore, the total noise  at $G_1$ and $G_2$ 
would be $\tilde{\z}_1=\sqrt{g_{12} \atr} \x_{2,3} +\z_1$ and $\tilde{\z}_2=\sqrt{g_{22} \att} \x_{2,2} +\z_2$. However, the effective 
noise power cannot exceed $2$ since
$\E[g_{12} \atr  + 1] \leq 2 $ and $\E[ g_{22} \att +1] \leq 2 $. 


The receiver $F_1$ observes a Gaussian multiple 
access channel (with noise power upper bounded by $2$), where $\Uoo$ is sent by one user, and $(\Uto,\Utt)$ is
sent by the other user.  The bounds in \eqref{eq:GZ-1}-\eqref{eq:GZ-4} guarantee that these rates are achievable over the 
multiple access channel. 

On the other hand, the channel from $F_2$ to $G_2$ is Gaussian point-to-point channel with modified additive noise. Therefore, any total rate 
not exceeding its capacity can be reliably transmitted. This is condition is fulfilled here since $\Tto+ \Ttr$ satisfies \eqref{eq:GZ-6}. Finally, the
bound on the power allocated to $\Utr$ upper bounds its rate as in \eqref{eq:GZ-5}.

\end{IEEEproof}

\begin{IEEEproof}[ Proof of Lemma~\ref{lm:GZS:S} in Section~\ref{sec:pr-outline}]
  Again, the achievability scheme we propose for the Gaussian $\S$ interference
  network (illustrated in Figure \ref{fig:GZS:S}) is based on 
  superposition coding, and a successively decode and cancel decoding
  strategy, such that the requirements of the problem are fulfilled. A
  proper power allocation is required to guarantee achievability of
  the rate tuples mentioned in this lemma.

Note that $G_1$ does not decode $\Vtt$ and $\Vtf$, and treats them as noise. 
We choose the total fraction of power allocated to $\Vtt$ and $\Vtf$ to be at most $1/h_{11}$, that is $\att+\atf\leq1/h_{11}$. Therefore, the total noise power
received at $G_1$ is upper bounded as $\E[h_{11} (\att +\atf )+ 1] \leq 2$. 

Similarly, $\Vot$ is treated as noise at $G_2$. By bounding the fraction of power allocated to this sub-message, we can upper bound the effective noise
power observed at $G_2$ by $\E[h_{11} (\att+\atf) + 1] \leq 2$.

The point-to-point Gaussian channel from $F_1$ to $G_1$ can support any sum-rate below its capacity as in \eqref{eq:GS-1}. Moreover,  $\Qot$ 
is bounded above since its allocated power does not exceed $1/h_{12}$. 

On the other hand, we have a Gaussian multiple access channel from $F_1$ and $F_2$ to $G_2$, with total noise power not exceeding $2$. 
The bounds in \eqref{eq:GS-3}-\eqref{eq:GS-6} guarantee that the desired rates belong to the capacity region of this  channel, and therefore 
they are achievable. We skip the details of  power allocation here, but we point out that the achievability of the region is a consequence 
of the Gaussian multiple access rate region achievability. 

  

\end{IEEEproof}


\begin{IEEEproof}[Proof of Lemma~\ref{lm:GZZ:Z} in Section~\ref{sec:pr-outline}]
In this part we show that any rate tuple satisfying \eqref{gzz:z:1}-\eqref{gzz:z:4} is achievable. 
The main idea of this proof can be summarized as follows.
\begin{itemize}
\item  Use a common codebook with group structure, such as lattice codes, for $W_1^{(0)}$ and $W_2^{(0)}$, 
which maps them to $\x_{1,0}$ and $\x_{2,0}$
\item Choose a proper power allocation for $\x_{1,0}$ and $\x_{2,0}$ such that they get received at $G_1$ at the same power level; 
More precisely, denoting their power allocation by $\az$ and $\bz$, they should satisfy 
$g_{11}\az= g_{12}\bz$. This condition guarantees that the two lattice points get scaled by the same factor, 
and therefore the result is still a lattice point on the scaled lattice and can be decoded as long as enough signal to noise ratio
is provided. 
\item Use random Gaussian codebooks to encode the private sub-messages to $\x_{1,1}$ and $\x_{1,2}$, and use proper power allocation, 
$\aoq$ and $\beta_1$.
\end{itemize}

The first receiver $G_1$ needs to decode the partial-invertible $\phi$ which we define as
\begin{align}
 \phi \left(W_1^{(0)}, W_2^{(0)}\right) &= \psi^{-1} \left( \psi \left( W_1^{(0)}\right) + \psi \left( W_1^{(0)}\right) \right)\nonumber\\
&= \psi^{-1} \left( \x_{1,0} + \x_{2,0} \right)\nonumber
\end{align}
where $\psi$ is the one-to-one encoding function which maps the functional messages to the common lattice codebook. Note that 
the group structure of the code impels that  $\x_{1,0} + \x_{2,0}$ is still a valid codeword. It is easy to check that this function is 
partial-invertible. 

Let us define 
\begin{align}
 \eta=\min\left\{g_{11},g_{12},g_{22},\frac{g_{11}g_{22}}{g_{12}}\right\}.
\end{align}
Depending on the minimizer in $\eta$, we identify four cases. In each case, the achievable rate region is a polytope, 
with a certain number of corner points. It suffices to show the achievability only for the cornet points, since 
a standard time-sharing argument guarantees achievability for the rest of the region. 

The proof details for each corner point includes message splitting, and power allocation for sub-messages such that the 
decoders be able to decode corresponding messages. In the following we describe this strategy in details for the case 
where $\eta=g_{11}$. The extension of this method for other cases is straight-forward, and therefore we skip it here to 
sake of brevity.
%
 
\paragraph*{Case I. $\eta=g_{11}$}

It is clear from the definition of $\eta$ that in this case $g_{11}\leq g_{12} \leq g_{22}$, and therefore $\lambda=g_{11}$ and 
$\mu=g_{22}$. Hence, the desired region is characterized by all non-negative rate tuples $(P_0,P_{1},P_{2})$ satisfying 
\begin{align*}
P_0+P_1 &\leq \CGN{g_{11}},\\
P_0+P_1+P_2 &\leq \CGN{g_{22}}.
\end{align*}
This rate region is illustrated in Figure~\ref{fig:GZZ-Z-1}. It suffices to show that the corner points $A$, $B$ and $C$ are achievable, 
since the points $D$ and $E$ are degenerated from $B$ and $C$, respectively. 

\begin{figure}[h!]
\begin{center}
 	\psfrag{t0}[Bc][Bc]{$P_0$}
	\psfrag{t1}[Bc][Bc]{$P_1$}
	\psfrag{t2}[Bc][Bc]{$P_2$}
	\psfrag{a}[Bc][Bc]{$A$}
	\psfrag{b}[Bc][Bc]{$B$}
	\psfrag{c}[Bc][Bc]{$C$}
	\psfrag{d}[Bc][Bc]{$D$}
	\psfrag{e}[Bc][Bc]{$E$}
	\includegraphics[width=6cm]{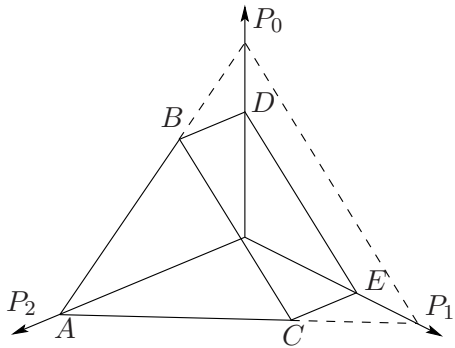}
\end{center}
\caption{Achievable rate region of the $\Z$-neutralization network when $\eta=g_{11}$.}
\label{fig:GZZ-Z-1}
\end{figure}

\begin{itemize}
 \item $A: (P_0,P_1,P_2)=\left(0,0,\CGN{g_{22}}-\frac{3}{2}\right)$\\
 The encoding strategy for this corner point is fairly simple. The second transmitter uses all its available power to send $W_2^{(1)}$,
while the first transmitter keeps silent. That is, $\x_1=0$ and $\x_2=\x_{2,1}$. The first decoder has nothing  to decode, and the 
second one can decode $\x_2$ from $\y_2$ as long as $P_2\leq \CGN{1+g_{22}}$. It is clear that in particular $P_2=\CGN{g_{22}}-\frac{3}{2}$ 
is achievable. 

\item $B: (P_0,P_1,P_2)=\left(\CGN{g_{11}}-1,0,\CGN{g_{22}}-\CGN{g_{11}}-\frac{1}{2}\right)$\\
The first encoder sends its lattice codeword with power allocation $\az=(g_{11}-1)/g_{11}$. The second encoder splits its private message 
into $W_2^{(1)}=(W_2^{(1,1)}, W_2^{(1,2)})$ of rates $P_{2,1}$ and $P_{2,2}$ where $P_{2}=P_{2,1}+P_{2,2}$. 
Then it sends 
\begin{align*}
\x_2&=\sqrt{\boo} \x_{2,1,1} + \sqrt{\bz} \x_{2,0} + \sqrt{\bot} \x_{2,1,2}
\end{align*}
where the power allocation coefficients are fixed to be $\bot=1/g_{12}$, 
$\bz=(g_{11}-1)/g_{12}$, and $\boo=1-\bz-\bot$. The signal received at the destinations are
\begin{align}
\y_1 &= \sqrt{g_{11}} \x_1 + \sqrt{g_{12}} \x_2 + \z_1,\nonumber\\
&= \sqrt{g_{12}-g_{11}} \x_{2,1,1} + \sqrt{g_{11}-1} [\x_{1,0}+ \x_{2,0}] +  \x_{2,1,2} + \z_1,\\
\y_2 &= \sqrt{g_{22}} \x_2 + \z_2,\nonumber\\
&= \sqrt{\frac{g_{22}(g_{12}-g_{11})}{g_{12}}} \x_{2,1,1} + \sqrt{\frac{g_{22}(g_{11}-1)}{g_{12}}} \x_{2,0} 
  +  \sqrt{\frac{g_{22}}{g_{12}}}\x_{2,1,2} + \z_2.
\end{align}
The first node decode and cancel $\x_{2,1,1}$, $\tilde{\x}_0=\x_{1,0}+ \x_{2,0}$, and $\x_{2,1,2} $ in order, while the second one 
performs the same decoding for  $\x_{2,1,1}$, $\x_{2,0}$, and $\x_{2,1,2} $. It is easy to show that the rates 
$P_{2,1}=\CGN{g_{12}/g_{11}}-0.5$, $P_0=\CGN{g_{11}}-1$, and $P_{2,2}=\CGN{g_{22}/g_{12}}$ are achievable, which implies
the private rates  $P_{2}=P_{2,1}+P_{2,2}=\CGN{g_{22}/g_{11}}-\frac{1}{2}$ for the second transmitter.

\item $C: (P_0,P_1,P_2)=\left(0,\CGN{g_{11}}-1,\CGN{g_{22}}-\CGN{g_{11}}-\frac{1}{2}\right)$\\
For this rate tuple, the rate of the functional message is zero. 
The second transmitter splits its private message similar to that of corner point $B$. The transmission power is distributed between 
among the sub-message as $\az=0$, $\aoq=1$, $\bot=1/g_{12}$, $\bz=0$, and $\boo=1-\bot$. A similar argument to that of corner point $B$
shows that the rates $P_{2,1}=\CGN{g_{12}/g_{11}}-0.5$, $P_{1}=\CGN{g_{11}}-1$, and $P_{2,2}=\CGN{g_{22}/g_{12}}$ are achievable,
which implies the 
achievability of the rate point $C$.
\end{itemize}

\end{IEEEproof}

\begin{IEEEproof}[Proof of Lemma~\ref{prop:dzs} in Section~\ref{subsec:DZS-Ach}]
Let $(R_1,R_2)\in\R{\DZS}{}$ be an arbitrary rate pair which satisfies (\ref{dzs:s1})-(\ref{dzs:d1d2}). 
In particular $R_1\leq \min\{m_{11},n_{11}\}$.
We claim that $(R_1,R_2)\in \R{\DZS}{1}(t) \times \R{\DZS}{2}(t)$ for $t=R_1$, and therefore $(R_1,R_2)$ is 
achievable using network decomposition.  In order to do this we have to show that any $R_2$ satisfying 
\eqref{dzs:s1}-\eqref{dzs:d1d2}, fulfills the constraints in the definition of $\R{\DZS}{2}(R_1)$.

Using \eqref{dzs:s2} and \eqref{dzs:s1s2}, we have
\begin{align}
R_2 &\leq \min\big(\max(m_{11},m_{12}) + (m_{22}-m_{12})^+ - R_1, \max(m_{12}, m_{22}) \big)\nonumber\\
&= \min\left(\max(m_{11},m_{12}) -R_1, m_{12}\right)+(m_{22}-m_{12})^+\nonumber\\
&= m'_{12}(R_1)+(m_{22}-m_{12})^+\nonumber\\
& \leq m'_{12}(R_1)+(m'_{22}(R_1)-m'_{12}(R_1))^+ \label{lm:zs1-pr-1}\\
&= \max(m'_{12}(R_1), m'_{22}(R_1)),\label{pr:prop-zsd-1}
\end{align}
where in (\ref{lm:zs1-pr-1}) we have used the fact that 
\[\left(\min(a,b)-\min(c,d)\right)^+\geq \min\left((a-c)^+,(b-d)^+\right).\]

Moreover, since $R_2$ satisfies \eqref{dzs:s1s2}, \eqref{dzs:s1s2r1}, and \eqref{dzs:d2}, we have
\begin{align}
R_2 & \leq \min \big( \max(m_{11},m_{12})+(m_{22}-m_{12})^+ -R_1, m_{22}+\max(n_{11},n_{21}) -R_1, m_{22}+n_{21} \big) \nonumber\\
&\leq \min \big( \max(m_{11},m_{12})+(m_{22}-m_{12})^+ -R_1, m_{22} \big)  + \min \big( \max(n_{11},n_{21}) -R_1, n_{21}\big)\label{lm:zs1-pr-2}\\
&= m'_{22}(R_1)+n'_{21}(R_1),\label{pr:prop-zsd-2}
\end{align}
where (\ref{lm:zs1-pr-2}) holds since 
\[\min (a,b) +\min (c,d)\geq \min (a, b+c,b+d),\]
for non-negative $a$, $b$, $c$, and $d$.

In order to show that the third constraint is satisfied, we can start with \eqref{dzs:s2r2}, \eqref{dzs:s1s2r2}, and \eqref{dzs:d1d2}. 
\begin{align}
R_2 & \leq \min \big( \max(m_{11},m_{12})+n_{22}+ -R_1, m_{12}+n_{22}, \max(n_{11},n_{21}) +(n_{22}-n_{21})^+ -R_1\big)\nonumber\\
& \leq \min\big( \max(m_{11},m_{12})+ -R_1, m_{12} \big)  + \min \big( \max(n_{11},n_{21})+(n_{22}-n_{21})^+ -R_1, n_{22}\big)\nonumber\\
&=m'_{12}(r_1)+n'_{22}(r_1).\label{pr:prop-zsd-3}
\end{align}

Finally, using \eqref{dzs:d2} and \eqref{dzs:d1d2}, we have

\begin{align}
R_2 & \leq \min \big( \max(n_{11},n_{21})+(n_{22}-n_{21})^+ -r_1, \max(n_{21},n_{22}) \big)\nonumber\\
& = \min\big( \max(n_{11},n_{21}) -r_1, n_{21}\big)+(n_{22}-n_{21})^+\nonumber\\
&= n'_{21}(r_1)+(n_{22}-n_{21})^+\nonumber\\
&\leq n'_{21}(r_1)+(n'_{22}(r_1)-n'_{21}(r_1))^+\nonumber\\
&= \max(n'_{21}(r_1), n'_{22}(r_1)).\label{pr:prop-zsd-4}
\end{align}
Putting inequalities in \eqref{pr:prop-zsd-1} and \eqref{pr:prop-zsd-2}-\eqref{pr:prop-zsd-4} together shows that $R_2\in \R{\DZS}{2}(R_1)$, 
and completes the proof.
\end{IEEEproof}

\begin{IEEEproof}[Proof of Lemma~\ref{lm:dz-neut} in Section~\ref{subsec:dzz:ach}]

The coding strategy we present here is based a network decomposition, where the 
sub-nodes and the links of the deterministic $\Z$-interference network are partitioned
into two disjoint sets. We analyze the rate region of each network, and derive an achievable 
rate region for the original network based on this analysis.

We just point out here that in this coding strategy, the second sender $F_2$, never sends 
a bit on a sub-node which is not received at $G_2$, even if $n_{12}>n_{22}$.

The first partition of the network $\N_1$, consists of those sub-nodes in $G_1$ which are connected 
to one of the top $m_{11}$ sub-nodes of $F_1$ \emph{and} one of the top $m_{22}$ sub-nodes of $F_2$. 
All the sub-nodes in the network which are related to (see Definition~\ref{def:node-related}) any 
of these sub-nodes also belong to the first network partition. The remaining nodes and link form the
second part of the network $\N_2$. It is clear that these two networks are node-disjoint, and do not cause 
interference on each other.

We first characterize the number sub-nodes in $G_1$ which belong to $\N_1$, by determining whether each of them 
can receive a bit from $F_1$, $F_2$, or both of them. We denote the number of levels 
in $G_1$ which are only connected to a transmitting level in $F_1$ by $k_1$. Similarly, the 
number of those only connected to a a transmitting level (the top $\min(n_{12},n_{22})$) in $F_2$ by $k_2$. Finally, 
$k_0$ denotes the number of levels which are connected to transmitting levels of both $F_1$
and $F_2$ (see Figure~\ref{fig:dz}).

\begin{figure}
\begin{center}
 	\psfrag{s1}[Bc][Bc]{$F_1$}
	\psfrag{s2}[Bc][Bc]{$F_2$}
	\psfrag{r1}[Bc][Bc]{$G_1$}
	\psfrag{r2}[Bc][Bc]{$G_2$}
	\psfrag{k0}[Bc][Bc]{$k_0=3$}
	\psfrag{k1}[Bc][Bc]{$k_1=1$}
	\psfrag{k2}[Bc][Bc]{$k_2=2$}
\includegraphics[height=6cm]{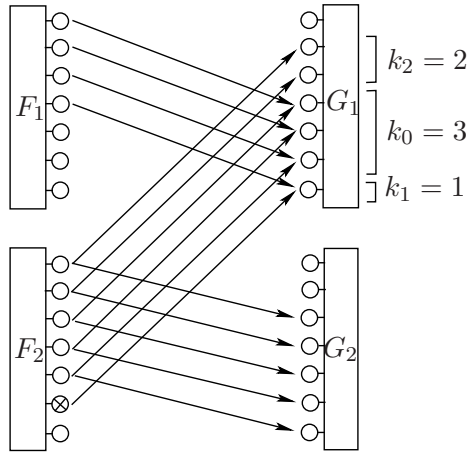}
\end{center}
\caption{A deterministic $\Z$-neutralization network. The upper $2$ sub-nodes in $G_1$ are only 
connected to $F_2$, and therefore $k_2=2$. The next $3$ sub-nodes receive information from both $F_1$ and
$F_2$, and hence $k_0=3$. Although the lowest sub-node is also connected to both transmitters, it only 
receives information from $F_1$ since $F_2$
keeps silent on its sub-nodes below $n_{22}$.}
\label{fig:dz}
\end{figure}

First, we derive $k_0$. Enumerate the levels of $G_1$ from $1$ (for the highest) to $q$ 
(for the lowest). Let $j$ be the index of a sub-node in $G_1$ belong to $\N_1$, \emph{i.e.,} it receives bits from both $F_1$
and $F_2$. Its neighbors in $F_1$ and $F_2$ (if there is any) are indexed by $j+n_{11}-q$ and $j+n_{12}-q$, respectively.
Therefore, $j$ belongs to $\N_1$ if and only if $1\leq j+n_{11}-q \leq n_{11}$ and $1\leq j+n_{12}-q\leq \min(n_{12},n_{22})$.
Therefore, the number of such sub-nodes is given by 
\begin{align}
 k_0&= \left[\min\{q,q-n_{12}+n_{22}\}- \max\{q-n_{11},q-n_{12}\}\right]^+\nonumber\\
&=\min\{n_{11},n_{12},n_{22},(n_{11}+n_{22}-n_{12})^+\}.
\end{align}

It is clear from the definition of $k_0$ that the remaining $n_{11}-k_0$ lowest levels of $G_1$
are only connected to sub-nodes of $F_1$, and hence, $k_1=n_{11}-k_0$. Similarly, 
$\min\{n_{12},n_{22}\}$ sub-nodes in $G_1$ are receiving information from $F_2$, where $k_0$ of them 
are also connected to $F_1$. Therefore, the remaining sub-nodes are only connected to $G_2$. Thus, 
$k_2=\min\{n_{12},n_{22}\}-k_0$. 

We partition the network into two parts: The first part consists of the $k_0$ sub-nodes of $G_1$ connected
to both $F_1$ and $F_2$, and sub-nodes connected to them. The remaining sub-nodes form the second partition 
of the network. We characterize the achievable tuples for each, denoted by $(Q'_{0},Q'_{1},Q'_{2})$ and 
$(Q''_{0},Q''_{1},Q''_{2})$, respectively. The fact that these two partitions are isolated allows us to 
conclude that the summation of such achievable tuples is also achievable for the original network.

Consider the first partition of the network. 
It is clear that any of the $k_0$ levels of $G_1$ connected to both $F_1$ and $F_2$ and can be used to communicate
a functional bit, since $G_1$ naturally receives the \texttt{xor} of the transmitting bits. On the other hand, 
such sub-node can be used to communicate one private bit from any of $F_1$ or $F_2$ to $G_1$ by keeping the other 
one silent. Therefore, any rate tuple satisfying 
\begin{align}
Q'_0+Q'_1+Q'_2 \leq k_0
\label{eq:DZ-ach-1} 
\end{align}
is achievable.  

The non-interfered links of the second partition of the network can be used to send private bits from the 
transmitters to $G_1$ simultaneously. Moreover, each 
transmitter can use one of its non-interfering sub-nodes to send a functional bit to $G_1$, and then, $G_1$ 
computes their \texttt{xor}, after receiving them separately. This can provide up to $\min\{k_1,k_2\}$ new 
functional bits for $G_1$. Moreover, the lower $(n_{22}-n_{12})^+$ sub-nodes of $F_2$ which are connected to 
$G_2$ but not to $G_1$ can be used
to send private bits to $G_2$ without causing any interference at $G_1$. 

Hence, this strategy can transmit any rate tuple satisfying 
\begin{align}
Q''_0 &\leq \min\{k_1,k_2\},\nonumber\\
Q''_0 +Q''_1&\leq k_1,\nonumber\\
Q''_0 +Q''_2&\leq k_2+(n_{22}-n_{12})^+.
\label{eq:DZ-ach-2}
\end{align}

Summing up the rates achieved on each partition of the network, we have arrive at $Q_i=Q'_i+Q''_i$ for 
$i=0,1,2$, where $(Q'_0,Q'_1,Q'_2)$'s and $(Q''_0,Q''_1,Q''_2)$ satisfy \eqref{eq:DZ-ach-1} and \eqref{eq:DZ-ach-2}, 
respectively.  It only remains to apply the Fourier-Motzkin elimination to project the rate region on the 
$(Q_0,Q_1,Q_2)$ space. This gives us 
\begin{align}
Q_0 &\leq k_0+\min\{k_1,k_2\},\nonumber\\
Q_0 +Q_1&\leq k_0+k_1,\nonumber\\
Q_0 +Q_2&\leq k_0+k_2+(n_{22}-n_{12})^+, \nonumber\\
Q_0 +Q_1+Q_2&\leq k_0+k_1+k_2+(n_{22}-n_{12})^+.
\label{eq:DZ-ach-3}
\end{align}
Some simple manipulations show that the RHS's of the inequalities in \eqref{eq:DZ-ach-3} are the same 
as that claimed in the lemma. 

\end{IEEEproof}

\begin{IEEEproof}[Proof of Lemma~\ref{lm:gzs-fano} in Appendix~\ref{subsec:gzs-conv}]
As mentioned before, we will use the Fano's inequality in order to prove this lemma. 
We have
\begin{align}
\ell R_1 &= H(W_1) = I(W_1; y_1^{\ell}) + H(W_1| y_1^{\ell})\nonumber\\
&\leq  I(W_1; y_1^{\ell}) + \ell e_{\ell} \label{pr:fano:1-1}\\
&\leq I(x_1^{\ell} ; y_1^{\ell}) + \ell e_{\ell},\label{pr:fano:1-2}
\end{align}
where \eqref{pr:fano:1-1} is implied by the Fano's inequality, and in \eqref{pr:fano:1-2} 
we used the data processing inequality for the Markov chain $W_1 \leftrightarrow x_1^{\ell} \leftrightarrow y_1^{\ell}$.
Note that where $\e_{\ell}\rightarrow 0$ as $\ell$ grows.
The proofs of the other two inequalities follow the same lines, and we skip them to sake of brevity.
\end{IEEEproof}

\begin{IEEEproof}[Proof of Lemma~\ref{lm:zzg:indep-mac} in Appendix~\ref{subsec:gzz-conv}]
Note that $Z$ is independent of everything else, and $X_1$ and $X_2$ are conditionally independent. Without loss of generality 
we can also assume that $\mu_i(\gamma)=\E[X_i|\Gamma=\gamma]=0$ for $\forall \gamma$ (otherwise for any given $\Gamma=\gamma$, 
we can shift $X_i$ by $\mu_i(\gamma)$, while the entropy does not change). Let $\E[X^2_i | \Gamma=\gamma]=\sigma_i^2(\gamma)$ 
for $i=1,2$. Therefore the conditional variance of $Y$ can be bounded as
\begin{align}
\E[Y^2|\Gamma=\gamma]&=\E[(X_1+X_2+Z)^2|\Gamma=\gamma]= \sigma_1^2(\gamma) +\sigma_2^2(\gamma) +1. 
\end{align}
Therefore,
\begin{align}
h(Y|\Gamma)&=\E_\Gamma[h(Y|\Gamma=\gamma)]=\E_T[h(X_1+X_2+Z|\Gamma=\gamma)]\nonumber\\
&\leq \E_\Gamma[\log 2\pi e (\sigma_1^2(\gamma)+\sigma_2^2(\gamma)+1)]\label{gzz:lm:gs}\\
&\leq \log 2\pi e (\E_\Gamma[\sigma_1^2(\gamma)+\sigma_2^2(\gamma)+1])\label{gzz:lm:conc}\\
&=\log 2\pi e (\sigma_1^2+\sigma_2^2+1),\label{gzz:lm:tower}
\end{align}
where in (\ref{gzz:lm:gs}) we have used the fact that Gaussian random variable has the maximum differential entropy among 
all random variables with the same variance, and (\ref{gzz:lm:conc}) follows from the concavity of the function $\log(\cdot)$. 
Finally, \eqref{gzz:lm:tower} is just the tower property, $\E_\Gamma[\E[X_i^2|\Gamma]]=\E[X_i^2]$.
\end{IEEEproof}

\bibliography{RelayInterference_arxiv}

\begin{thebibliography}{10}
\providecommand{\url}[1]{#1}
\csname url@rmstyle\endcsname
\providecommand{\newblock}{\relax}
\providecommand{\bibinfo}[2]{#2}
\providecommand\BIBentrySTDinterwordspacing{\spaceskip=0pt\relax}
\providecommand\BIBentryALTinterwordstretchfactor{4}
\providecommand\BIBentryALTinterwordspacing{\spaceskip=\fontdimen2\font plus
\BIBentryALTinterwordstretchfactor\fontdimen3\font minus
  \fontdimen4\font\relax}
\providecommand\BIBforeignlanguage[2]{{%
\expandafter\ifx\csname l@#1\endcsname\relax
\typeout{** WARNING: IEEEtran.bst: No hyphenation pattern has been}%
\typeout{** loaded for the language `#1'. Using the pattern for}%
\typeout{** the default language instead.}%
\else
\language=\csname l@#1\endcsname
\fi
#2}}

\bibitem{Schrijver}
A.~Schrijver, \emph{Theory of Linear and Integer Programming}.\hskip 1em plus
  0.5em minus 0.4em\relax New York: Wiley, 1998.

\bibitem{HK:81}
T.~S. Han and K.~Kobayashi, ``A new achievable rate region for the interference
  channel,'' \emph{IEEE Transactions on Information Theory}, vol.~27, pp.
  49--60, January 1981.

\bibitem{ETW:08}
R.~H. Etkin, D.~Tse, , and H.~Wang, ``Gaussian interference channel capacity to
  within one bit,'' \emph{IEEE Transactions on Information Theory}, vol.~54,
  no.~12, pp. 5534--5562, Dec. 2008.

\bibitem{ADT07a}
A.~Avestimehr, S.~Diggavi, and D.~Tse, ``A deterministic approach to wireless
  relay networks,'' in \emph{Proceedings of Allerton Conference on
  Communication, Control, and Computing}, Illinois, USA, Sept. 2007, see:
  http://licos.epfl.ch/index.php?p=research\_projWNC.

\bibitem{GJ09}
K.~Gomadam and S.~A. Jafar, ``The effect of noise correlation in
  amplifyand-forward relay networks,'' \emph{IEEE Transactions on Information
  Theory}, vol.~55, no.~2, pp. 731--745, Feb. 2009.

\bibitem{GK00}
P.~Gupta and P.~Kumar, ``The capacity of wireless networks,'' \emph{IEEE
  Transactions on Information Theory}, vol.~46, no.~2, pp. 388--404, Mar. 2000.

\bibitem{OLT07}
A.~Ozgur, O.~Leveque, and D.~Tse, ``Hierarchical cooperation achieves optimal
  capacity scaling in ad hoc networks,'' \emph{IEEE Transactions on Information
  Theory}, vol.~53, no.~10, pp. 3549--3572, Oct. 2007.

\bibitem{Massimo09}
M.~Franceschetti, M.~Migliore, and P.~Minero, ``The capacity of wireless
  networks: Information-theoretic and physical limits,'' \emph{IEEE
  Transactions on Information Theory}, vol.~55, no.~8, pp. 3413--3424, Aug.
  2009.

\bibitem{PV09}
V.~Prabhakaran and P.~Viswanath, ``Interference channel with destination
  cooperation,'' in \emph{IEEE International Symposium on Information Theory
  (ISIT)}, Seoul, Korea, June 2009.

\bibitem{SuhTse09}
C.~Suh and D.~Tse, ``Symmetric feedback capacity of the gaussian interference
  channel to within one bit,'' in \emph{IEEE International Symposium on
  Information Theory (ISIT)}, Seoul, Korea, June 2009.

\bibitem{BT08}
G.~Bresler and D.~Tse, ``{The Two-User Gaussian Interference Channel: A
  Deterministic View},'' \emph{{European Transactions in Telecommunications}},
  vol.~19, pp. 333--354, June 2008.

\bibitem{CJ08}
V.~Cadambe and S.~Jafar, ``{Interference Alignment and Degrees of Freedom of
  the K-User Interference Channel},'' \emph{{IEEE Trans. Information Theory}},
  vol.~54, no.~8, pp. 3425--3441, August 2008, {}.

\bibitem{MMK08}
M.~Maddah-Ali, A.~Motahari, and A.~Khandani, ``{Communication Over MIMO X
  Channels: Interference Alignment, Decomposition, and Performance Analysis},''
  \emph{{IEEE Trans. Information Theory}}, vol.~54, no.~8, pp. 3457--3470,
  August 2008.

\bibitem{CT91}
T.~M. Cover and J.~Thomas, \emph{Elements of Information Theory}.\hskip 1em
  plus 0.5em minus 0.4em\relax New York: Wiley, 1991.

\bibitem{MDFT:Al08}
S.~Mohajer, S.~N. Diggavi, C.~Fragouli, and D.~N.~C. Tse, ``Transmission
  techniques for relay-interference networks,'' in \emph{Proceedings of
  Allerton Conference on Communication, Control, and Computing}, Illinois, USA,
  Sept. 2008.

\bibitem{ADT07b}
A.~Avestimehr, S.~Diggavi, and D.~Tse, ``Wireless network information flow,''
  in \emph{Proceedings of Allerton Conference on Communication, Control, and
  Computing}, Illinois, USA, Sept. 2007, see:
  http://licos.epfl.ch/index.php?p=research\_projWNC.

\bibitem{MDFT:ITW09}
S.~Mohajer, S.~Diggavi, C.~Fragouli, and D.~N.~C. Tse, ``Capacity of
  deterministic z-chain relay-interference network,'' in \emph{Proceedings of
  IEEE Information Theory Workshop}, Volos, Greece, 2009, pp. 331--335.

\end{thebibliography}

\end{document}